\documentclass[10pt,journal,twocolumn,compsoc]{IEEEtran}

\usepackage[utf8]{inputenc}
\usepackage{textgreek}
\usepackage{graphicx}
\usepackage{amsmath}
\usepackage{amssymb}
\usepackage{amsfonts}
\usepackage{authblk}
\usepackage{calc}
\usepackage{url}

\usepackage[backend=biber,style=numeric-comp,sorting=none,maxnames=2]{biblatex}

\DeclareSourcemap{
  \maps[datatype=bibtex]{
    \map[overwrite]{
      \step[fieldsource=doi, final]
      \step[fieldset=urn, null]
      \step[fieldset=url, null]
      \step[fieldset=eprint, null]
      \step[fieldset=issn, null]
      \step[fieldset=isbn, null]
    }  
  }
}

\usepackage{xcolor}
\usepackage{hyperref}
\hypersetup{
    pdftoolbar=true,
    pdfmenubar=true,
    pdffitwindow=false,
    pdfstartview={FitH},
    pdftitle={Deflectometry for specular surfaces: an overview},
    pdfauthor={Jan Burke, Alexey Pak, Sebastian Höfer, Mathias Ziebarth, Masoud Roschani, Jürgen Beyerer},
    pdfsubject={An overview of deflectometry},
    pdfcreator={Jan Burke},
    pdfproducer={Jan Burke},
    pdfkeywords={deflectometry, machine vision, industrial inspection, metrology, phase shifting, surface reconstruction, surface inspection, sensor calibration},
    pdfnewwindow=true,
    linkcolor=red,
    citecolor={0 0.9 0},
    urlcolor={0 0.9 0.9},
}

\usepackage[font=small]{caption}
\usepackage{soul}
\usepackage{siunitx}
\usepackage{multirow}
\usepackage{tikz}

\newcommand{\orcidicon}[1]{
    \href
    {https://orcid.org/#1}
    {\includegraphics[height=\heightof{B}]{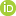}}
}

\begin{document}

\title{Deflectometry for specular surfaces: an overview}

\author{Jan Burke$^1 \orcidicon{0000-0002-0608-4765}$, Alexey Pak$^1$, Sebastian H{\"o}fer$^1$, Mathias Ziebarth$^1$, Masoud Roschani$^1$, J{\"u}rgen Beyerer$^1,{^2}$
  \thanks{$^*$ jan.burke@iosb.fraunhofer.de} \\
  $^1$Fraunhofer Institute of Optronics, System Technologies
    and Image Exploitation IOSB \\
  Fraunhoferstra{\ss}e 1, 76131 Karlsruhe, Germany \\
  $^2$Karlsruhe Institute of Technology – KIT,
    Department of Informatics, Vision and Fusion Laboratory (IES) \\
    c/o Technologiefabrik, Haid-und-Neu-Stra{\ss}e 7, 76131 Karlsruhe, Germany 
}
\maketitle

\begin{abstract}
Deflectometry as a technical approach to assessing reflective surfaces has now existed for almost 40 years. Different aspects and variations of the method have been studied in multiple theses and research articles, and reviews are also becoming available for certain subtopics. Still a field of active development with many unsolved problems, deflectometry now encompasses a large variety of application domains, hardware setup types, and processing workflows designed for different purposes, and spans a range from qualitative defect inspection of large vehicles to precision measurements of microscopic optics. Over these years, many exciting developments have accumulated in the underlying theory, in the systems design, and in the implementation specifics. This diversity of topics is difficult to grasp for experts and non-experts alike and may present an obstacle to a wider acceptance of deflectometry as a useful tool in other research fields and in the industry.

This paper presents an attempt to summarize the status of deflectometry, and to map relations between its notable "spin-off" branches. The intention of the paper is to provide a common communication basis for practitioners and at the same time to offer a convenient entry point for those interested in learning and using the method. The list of references is extensive but definitely not exhaustive, introducing some prominent trends and established research groups in order to facilitate further self-directed exploration by the reader.
\end{abstract}
\vspace{-0.2cm}

\begin{center}
\resizebox{0.72\columnwidth}{!}{
    \begin{tabular}{p{1cm}|p{6cm}}
        \multicolumn{2}{c}{\textsf{\textbf{Glossary}}} \\
        \hline
        ANN     &   artificial neural network\\
        CMOS    &   complementary metal–oxide–semiconductor\\
        DFT     &   discrete Fourier transform\\
        DM      &   deflectometry\\
        E-ELT   &   European Extremely Large Telescope\\
        EMVA    &   European Machine Vision Association\\
        FP      &   fringe projection\\
        GPU     &   graphics processing unit\\
        GU      &   gloss units\\
        HD      &   high definition\\
        IPS     &   in-plane switching\\
        IRDM    &   infrared deflectometry\\
        LC      &   liquid crystal\\
        LWIR    &   long-wave infrared\\
        MTF     &   modulation transfer function\\
        OF      &   optical flow\\
        PDE     &   partial differential equation\\
        QA      &   quality assurance\\
        RMS     &   root-mean-square\\
        SF      &   specular flow\\
        SFS     &   shape from shading\\
        SFSF    &   shape from specular flow\\
        SNR     &   signal-to-noise ratio\\
        UVDM    &   ultraviolet deflectometry\\
    \end{tabular}
}
\end{center}

\section{Introduction}
\label{sec:introduction}

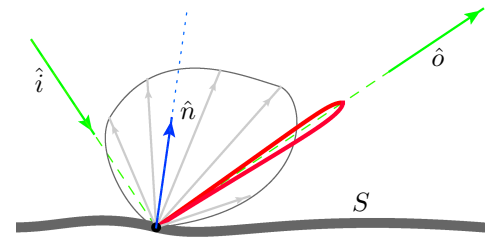
\begin{figure}
    \centering
    \includegraphics[width=0.7\columnwidth]{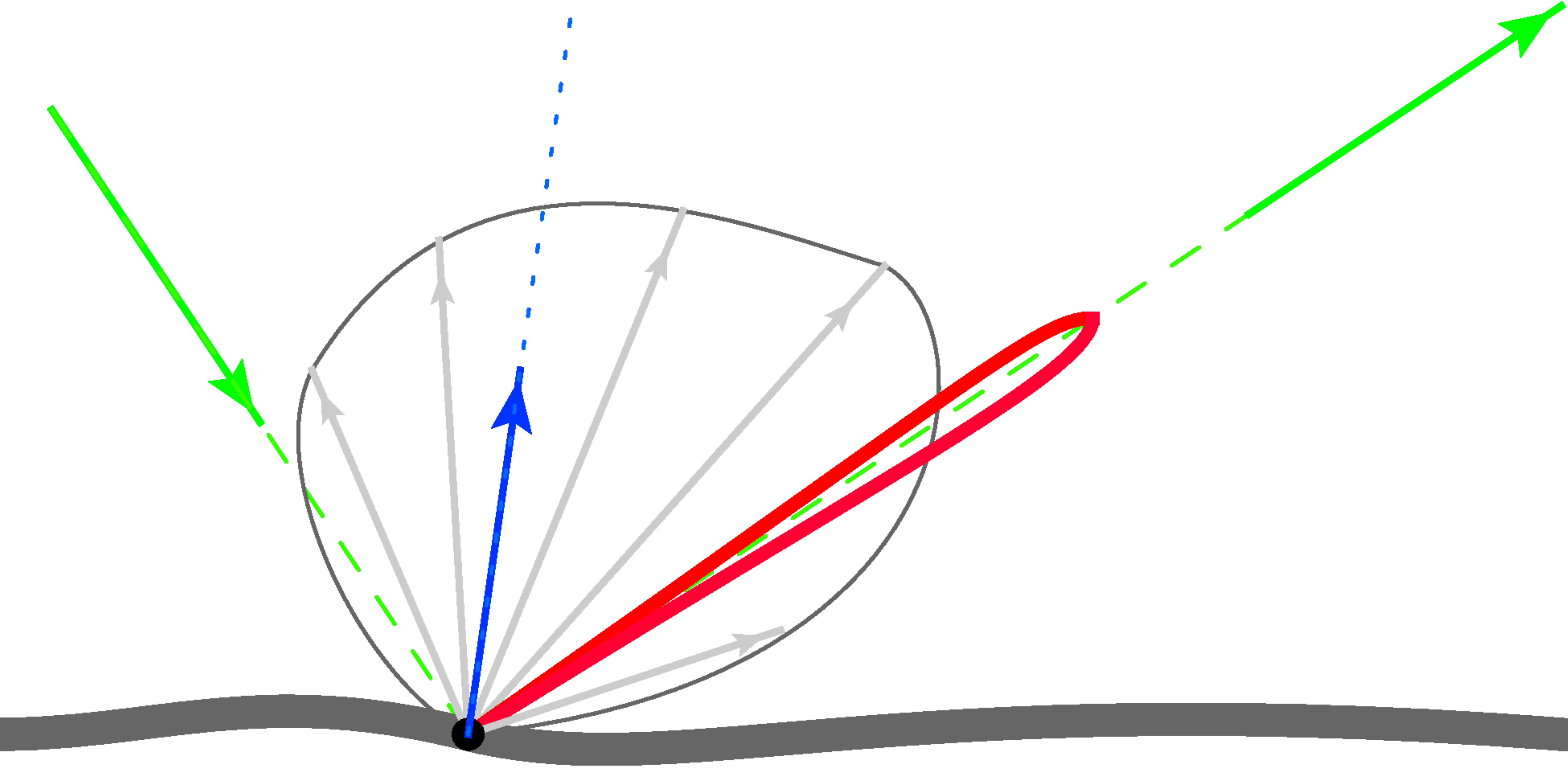}
    \put(-115, 45){$\hat{n}$}
    \put(-170, 55){$\hat{i}$}
    \put(-20, 65){$\hat{o}$}
    \put(-50, 10){$S$}
    \caption{Diffuse and specular reflection from a surface. A ray of light is incident along a direction $\hat{i}$ at some point on the surface $S$ where the unit normal vector is $\hat{n}$. In the distribution of the reflected light one can identify a diffuse component (relatively broad and smooth angular distribution indicated by the gray shape and arrows) and a specular component (thick red shape) that is collimated along the direction $\hat{o} = \hat{i} - 2\hat{n}~ (\hat{i}^T\hat{n})$ (reflection law).}
    \label{fig:reflection_components}
    \vspace{-0.5cm}
\end{figure}

If a ray of light reflects from a smooth surface, the resulting angular distribution can in general be separated into two components. Diffusely reflected light spans a relatively broad range of angles, while the specular component is to the first approximation perfectly collimated along the direction prescribed by the reflection law (Fig.~\ref{fig:reflection_components}). The higher the surface quality (i.e., the smaller the scale of the residual roughness), the more prominent the specular component becomes. For mirrors, the predominantly specular reflection is the primary purpose of polishing; for other technical surfaces, it may be a by-product of precision manufacturing. Measuring high-quality polished surfaces with low slope or shape uncertainty is an important problem in metrology.

Since such surfaces typically require delicate handling, when solving this problem it is natural to give preference to non-contact optical measurement techniques. Of the latter, methods relying on diffuse reflection (e.g. fringe projection, laser triangulation) are inapplicable, and one is left with interferometry (accurate but expensive for large areas or complex surface shapes) or one must embrace specular reflection and wield the power of deflectometry. 

\begin{figure}
    \includegraphics[height=3.4cm]{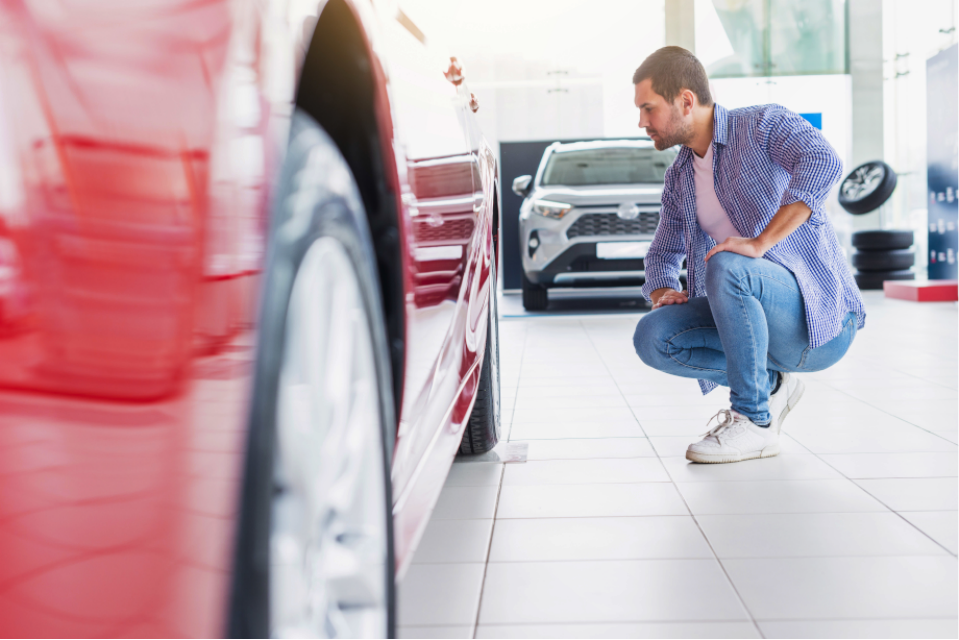}
    \put(-15, 5){\colorbox{white}{(a)}}
    \hfill
    \includegraphics[height=3.4cm]{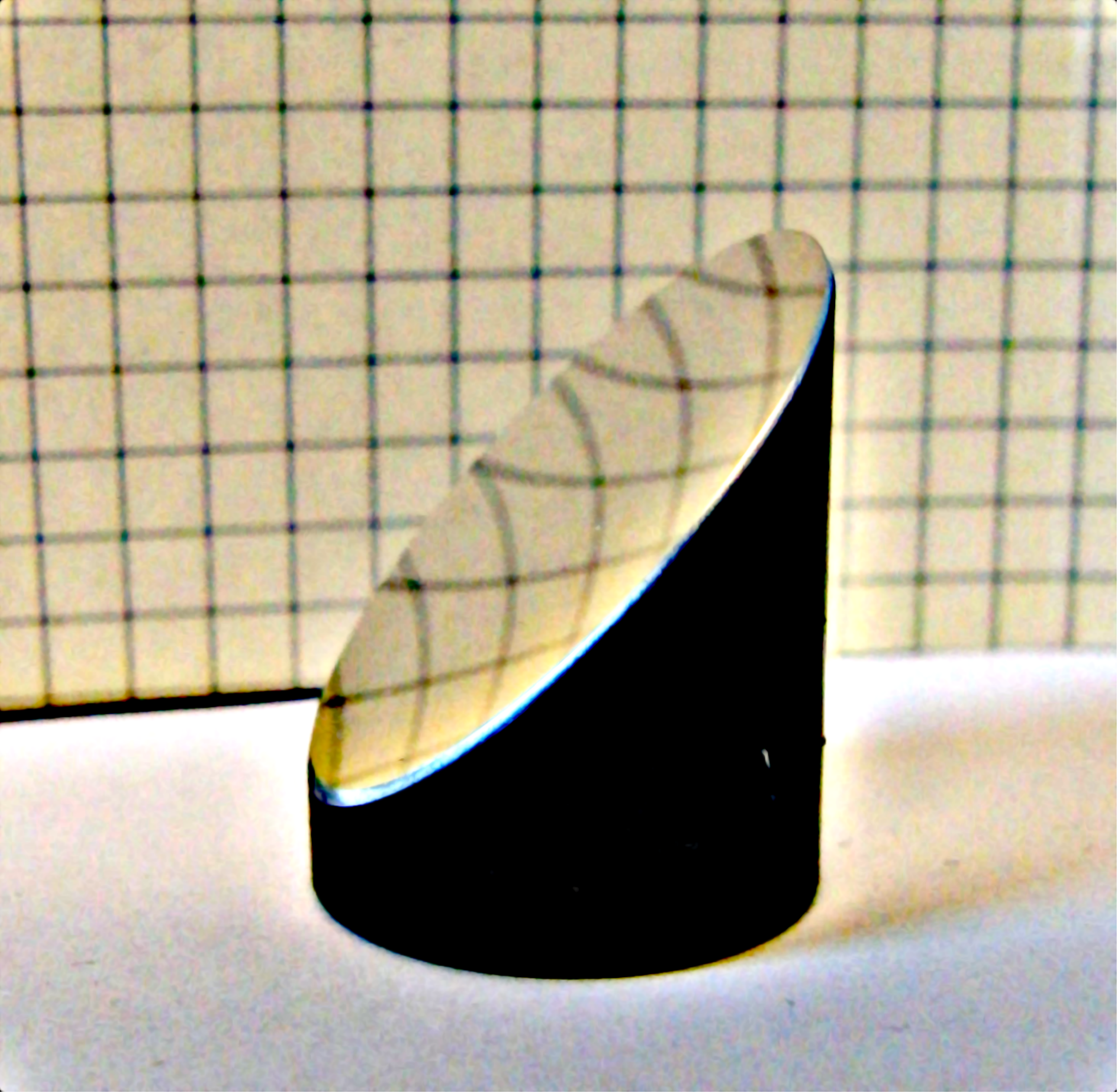}
    \put(-15, 5){\colorbox{white}{(b)}}
    \caption{Examples of surfaces amenable to deflectometric inspection, (a) manual: car inspection during purchase (photo designed by Freepik), and (b) automated: a high-quality off-axis parabolic mirror (photo by Marc Sandner).}
    \label{fig:examples}
    \vspace{-0.7cm}
\end{figure}

Humans seem to know intuitively how to exploit specularities when inspecting mirrors. Given a glossy surface such as in Fig.~\ref{fig:examples}(a), we adjust the viewing angle in order to observe virtual (reflected) images of contrast-rich objects in the environment. An imperfection in the surface shape leads to a visible distortion in the virtual image. In order to enhance the signal, we may slightly move our head or the studied object (thus creating specular flow) and observe how the reflection changes. The largest changes then would typically be associated with higher curvatures of the reflecting surface -- i.e., defects or non-uniformities. This simple recipe allows even untrained observers to detect slope variations at the level of mrad (Sec.~\ref{sec:aesthetics}). Automated industrial deflectometry employs similar principles in order to measure objects such as in Fig.~\ref{fig:examples}(b), where the target slope uncertainties may reach a few \si{\micro\radian}.

Historically, deflectometry has been studied in at least two different research communities, with rather different objectives. In computer vision, the problem of reconstructing mirror shapes from images is known as "shape from specular reflection" or "shape from specularities", whereas in optical metrology the terms "reflection grating method" and "deflectometry" are more common. Some sources mention "reflectometry"; we would like to discourage the use of this term as it encompasses much more than deflectometry and lacks a precise definition. In what follows, "deflectometry" (DM) will refer to any method that relies on specular reflection and treats diffusely reflected light as noise.

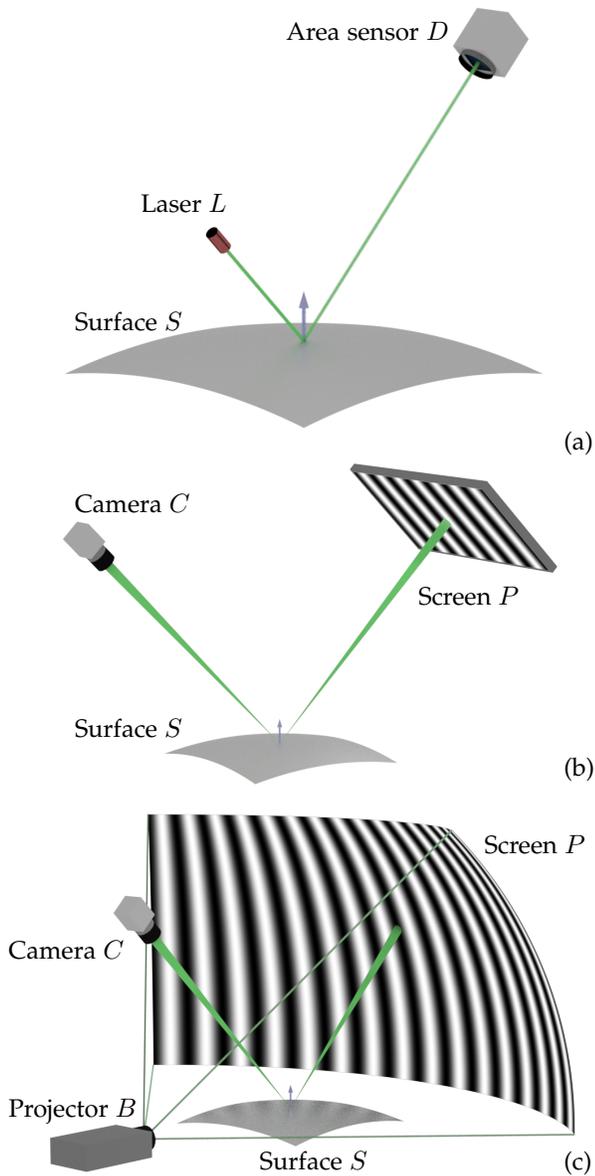
\begin{figure}
    \centering
    \includegraphics[bb=0 0 1000 950, clip, width=0.8\columnwidth]{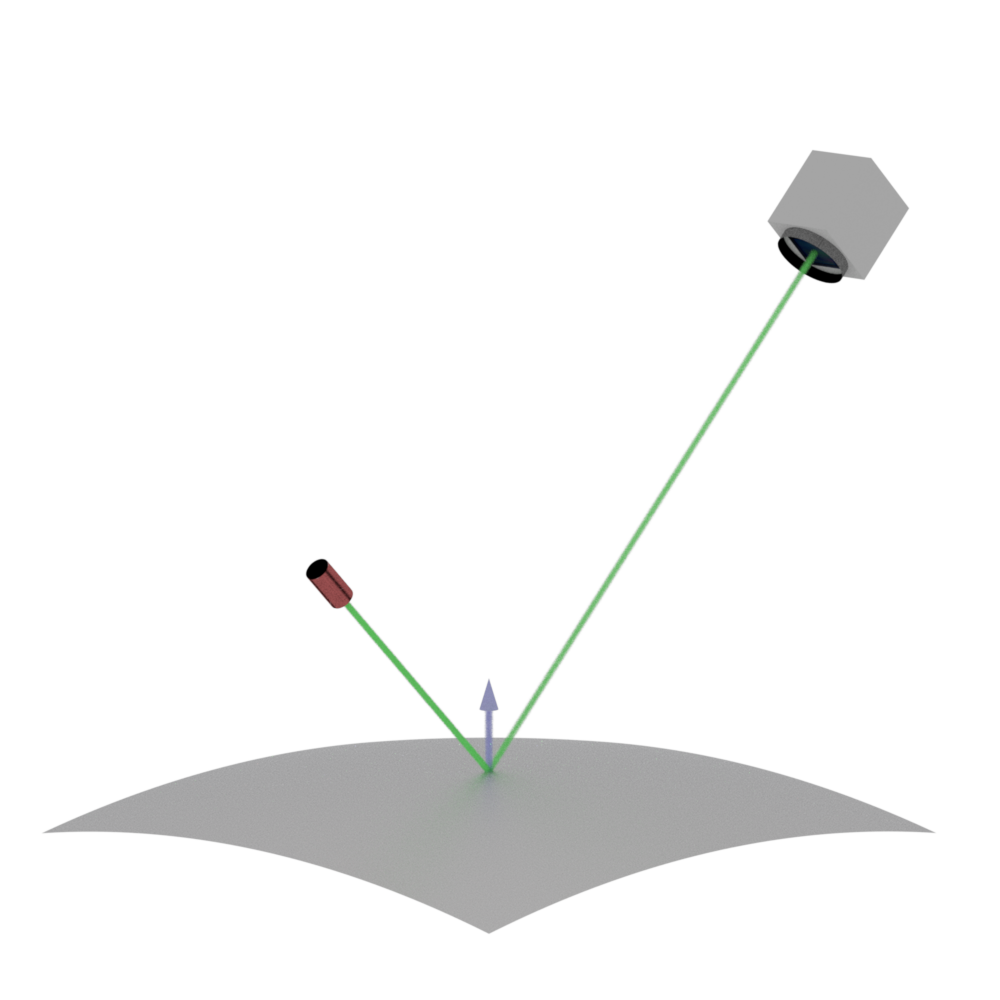}
    \put(-190,50){Surface $S$}
    \put(-165,95){Laser $L$}
    \put(-110,160){Area sensor $D$}
    \put(-5,5){(a)}
    \vspace{-0.1cm}
    \\
    \includegraphics[bb=0 0 1200 750, clip, width=0.8\columnwidth]{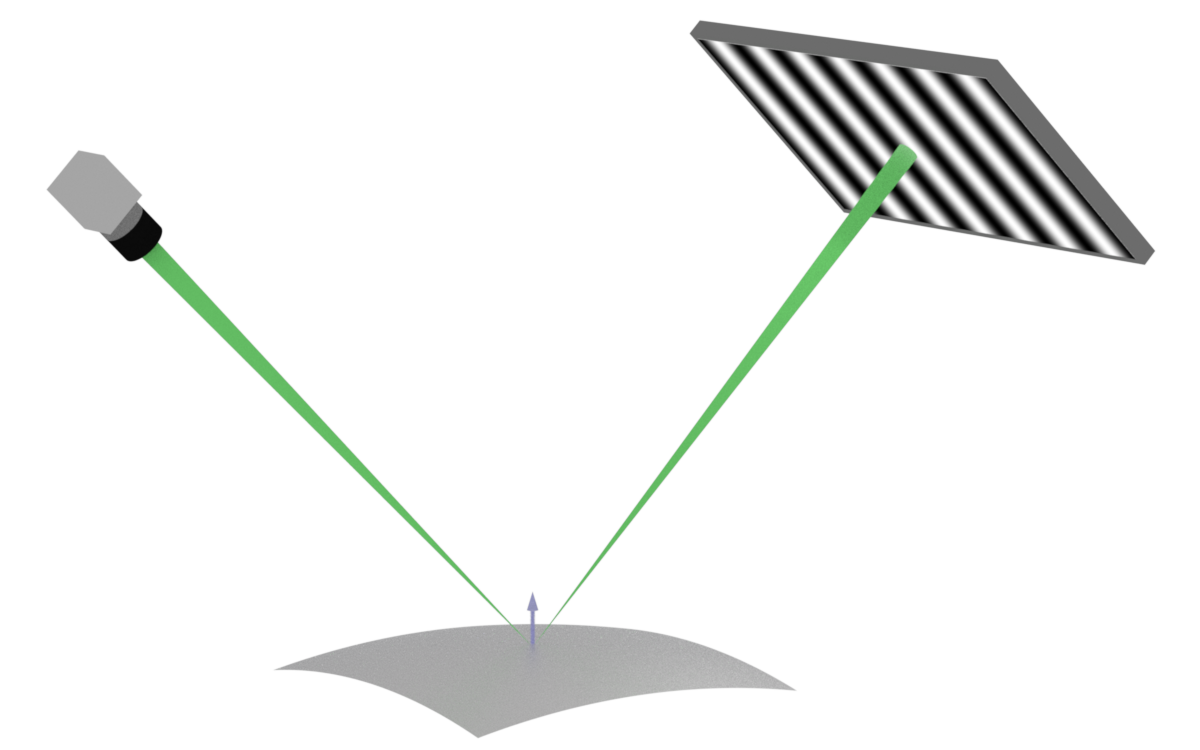}
    \put(-190,105){Camera $C$}
    \put(-60,70){Screen $P$}
    \put(-190,20){Surface $S$}
    \put(-5,5){(b)}
    \vspace{0.2cm}
    \\
    \includegraphics[bb=0 0 1200 850, clip, width=0.8\columnwidth]{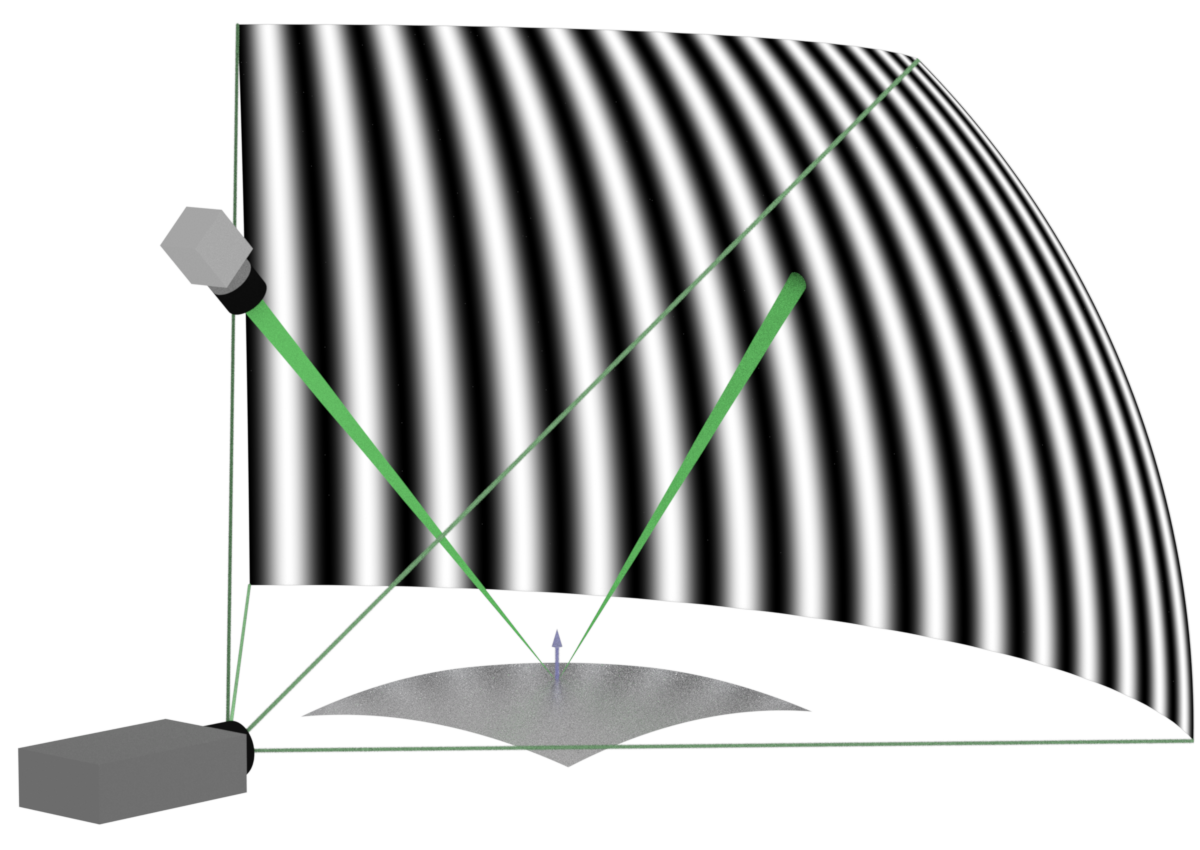}
    \put(-215,85){Camera $C$}
    \put(-35,125){Screen $P$}
    \put(-120,5){Surface $S$}
    \put(-215,25){Projector $B$}
    \put(-5,5){(c)}
    \caption{Three main deflectometric schemes: (a) "direct": a collimated light source $L$ illuminates the surface $S$, the reflected ray is detected by a sensor $D$. (b) "inverse" or "basic": a pixel on a flat screen $P$ emits un-directed light, a camera $C$ registers its reflection in the surface. (c) "inverse" with curved screen; a projector $B$ illuminates a screen that provides the reference pattern and allows coverage of a larger solid angle around the object $S$. Note how in (b) and (c) the camera is focused on the object; this gives best lateral resolution of the surface but comes at a sacrifice in angular resolution (see~\ref{sec:focus}).}
    \label{fig:dm_setup}
    \vspace{-0.5cm}
\end{figure}

For the purposes of the following discussion, let us introduce three specific implementations of DM (Fig.~\ref{fig:dm_setup}). In the "direct" scheme, a collimated (e.g. laser) beam reflects from the surface and arrives as a (possibly blurred) dot on e.g. a CMOS matrix; in order to assess an extended object, one has to scan it mechanically while tracking the position of the reflected beam. The "inverse" approach employs inexpensive flat screens and digital cameras and may inspect relatively large object patches without movement. If we assume that the position of the emitting pixel and the direction of the respective camera's "view ray" are known, then the symmetry of the reflection law guarantees the optical equivalence to the "direct" scheme. In reality, this technology becomes viable only when coupled with coded pattern sequences. Without them, pixels on the screen must be turned on and off one by one (scanning the screen surface) so that only a single point emits light as the camera makes a shot. This would lead to extremely lengthy measurements. Instead, a handful of special patterns can be displayed on the screen such that the position of each screen pixel is unambiguously encoded in the sequence of gray-scale values and can later be identified based on the recorded camera frames (Sec.~\ref{sec:processing}). Since it is the most popular scheme in applications, we will also refer to the setup in Fig.~\ref{fig:dm_setup}(b) as to "basic". Also note that the camera cannot be focused on the entire surface, since the viewing will necessarily be oblique to some extent and the depth of focus plays a significant role in the measurement; this issue has been investigated in detail in~\cite{Kammel.2004}. Finally, if an even larger coding screen is needed, or if a non-flat shape suits the object better, we can illuminate a wall or a screen of pre-designed shape with a projector. The price to pay for the extra components is significant additional effort for the geometrical calibration if required. While an LC display can be approximated as a regular dot pattern on a plane, the projector variant must account for the much larger uncertainties of the projector and the screen. An intermediate approach may use a curved monitor screen~\cite{Liu.2021} but still requires thorough characterization of the geometry.

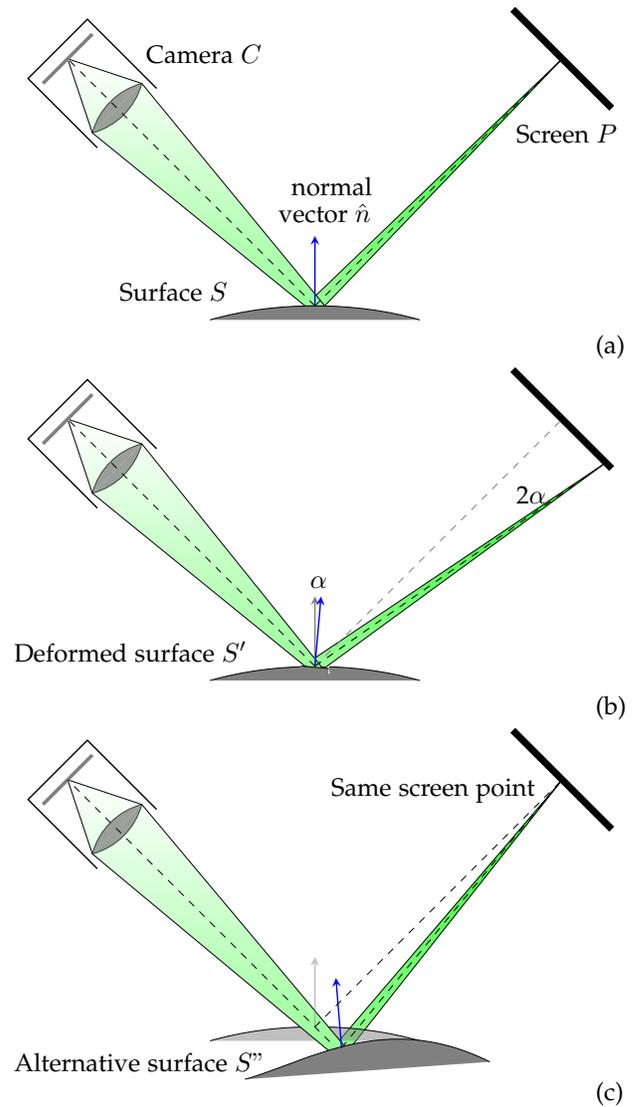
\begin{figure}
    \centering
    \includegraphics[width=0.9\columnwidth]{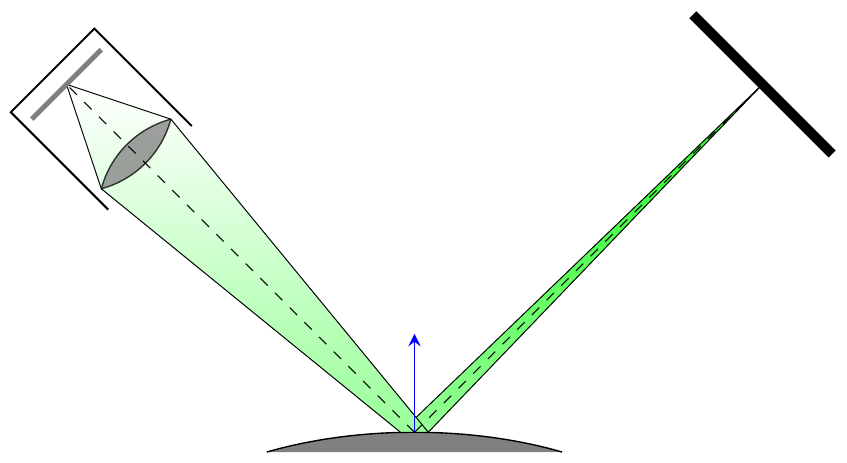}
    \put(-10,-10){(a)}
    \put(-180,100){Camera $C$}
    \put(-125,50){normal}
    \put(-130,40){vector $\hat{n}$}
    \put(-40,70){Screen $P$}
    \put(-190,10){Surface $S$}
    \vspace{0.0cm}
    \\
    \centering
    \includegraphics[width=0.9\columnwidth]{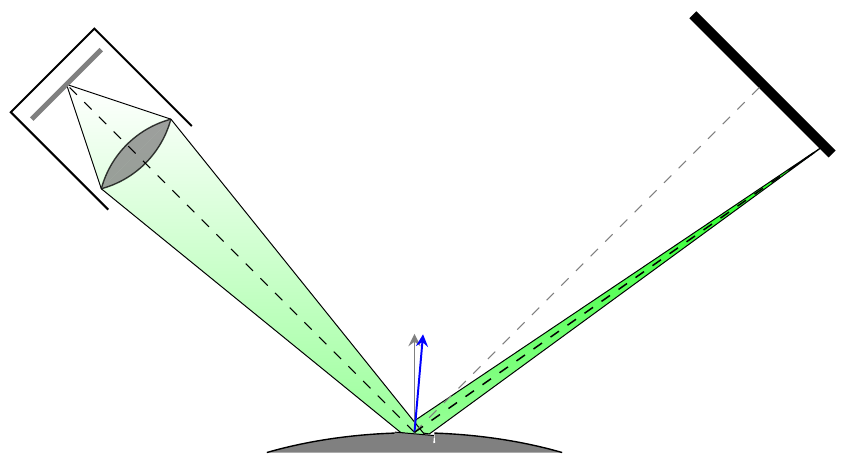}
    \put(-10,-10){(b)}
    \put(-230,10){Deformed surface $S'$}
    \put(-118,38){$\alpha$}
    \put(-40,70){2$\alpha$}
    \vspace{0.0cm}
    \\
    \centering
    \includegraphics[width=0.9\columnwidth]{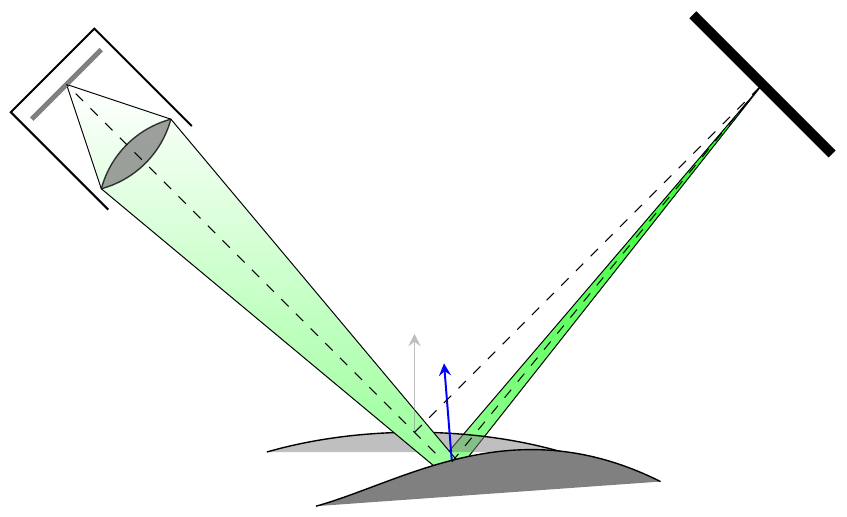}
    \put(-10,-5){(c)}
    \put(-110,110){Same screen point}
    \put(-230,6){Alternative surface $S"$}
    \vspace{0.0cm}
    \caption{(a) Schematic ray paths in the "basic" DM setup, notations are as in Fig.~\ref{fig:dm_setup}(b). The lens is focused on the screen. (b) If the surface tilts by an angle $\alpha$ in the incidence plane, the direction of the reflected ray changes by $\theta = 2\alpha$. The detected displacement on the screen grows linearly with the distance between the surface $S$ and the screen $P$, while the measurement uncertainty of the surface inclination angle decreases. (c) Point-wise deflectometric observations are inherently ambiguous: for any detection there exists a continuum of possible surface positions and shapes that are consistent with the recorded reflections. Here the camera is focused on the screen, which gives best angular resolution – but at the cost of the object surface being less finely resolved and its curvatures modifying the observation geometry (see~\ref{sec:focus}).}
    \label{fig:sensitivity}
    \vspace{-0.5cm}
\end{figure}

We know from everyday experience that specularities are very sensitive to changes of the surface slope. In terms of the "basic" scheme of Fig.~\ref{fig:dm_setup}(b), if the normal vector direction changes by a small angle, the direction of the reflected "view ray" changes by twice as much, as shown in Fig.~\ref{fig:sensitivity}(b). Moreover, the sensitivity in the measured reflection angle increases as the distance between the surface $S$ and the screen $P$ or the area sensor $D$ grows. Of course, it is not possible to reduce errors arbitrarily by placing $P$ sufficiently far from the object. At some point, wave optics and sensor properties will dominate -- one cannot expect nrad-level results without extra effort. A more detailed analysis~\cite{Faber.2012} places limits on the achievable angular resolution for surface slopes that appear to be on par with those typical to interferometry. For a method using non-coherent light and no precision reference objects, this is a tall order! By contrast, no such scaling law exists for triangulation by fringe projection~\cite{Gayton.2021}.

The same reflection law determines also the main disadvantage of DM -- its poor sensitivity to the absolute surface position. Given a fixed "view ray" of the camera and the respective decoded origin point on the screen, one can find infinitely many surface positions and inclinations consistent with a given observation, as schematically shown in Fig.~\ref{fig:sensitivity}(c). Therefore, surface shape reconstruction in DM typically requires some "regularization"~\cite{Werling.2011} (Sec.~\ref{sec:regular}) in order to constrain the absolute surface position.

Today, different versions of DM are employed in a wide range of applications, such as industrial quality control for household items, mobile phones, automotive parts~\cite{Lippincott.1982, Hofling.2000, Skydan.2007, Armesto.2011, Arnal.2017, Molina.2017, Zhou.2019}, reflective sheet, rod, or tape materials~\cite{Hung.2003, Caulier.2008, Sarosi.2010}, Fresnel lenses~\cite{Kiefel.2016b, Kiefel.2016}, solar concentrator mirrors ~\cite{Arqueros.2003, Fontani.2005, Heimsath.2008b, Scott.2010, Wang.2010, Ulmer.2011, CamposGarcia.2015b, ElYdrissi.2019b}, and even astronomical and synchrotron mirrors of various sizes~\cite{Parks.2011, Schulz.2011, Su.2012, Su.2012b,  Hofbauer.2013, Burge.2014, Olesch.2014c, Su.2014, Oh.2016, Davies.2017}. We have last summarized the state of the art over a decade ago~\cite{Werling.2009b, Balzer.2010d}; driven by applications, the discipline has seen many exciting advances since then. Although brief summaries have appeared on subtopics~\cite{Huang.2018, Xu.2020, Zhang.2021}, we believe that a stock-take of a broader scope is due once again.

Our present tour begins with an exercise in classification and delineation of the field, followed by a brief outline of its history. Sec.~\ref{sec:fundamentals} provides an overview of the theoretical concepts involved and challenges inherent to the interpretation of reflections. In Sec.~\ref{sec:design} we discuss some prominent schemes and variations of practical DM systems. Robust and accurate detection of reflections is of crucial importance to all implementations of DM; Sec.~\ref{sec:processing} is therefore devoted to coding methods and signal processing. Finally, Sec.~\ref{sec:conclusion} concludes the paper.
    
\subsection{What deflectometry is, and what it is not}
\label{sec:whatisdm}

    \subsubsection{Deflectometry vs fringe projection}
    \label{sec:fringe}
        
\begin{figure}
        \centering
        \includegraphics[bb=50 20 1040 940, clip, width=0.9\columnwidth]{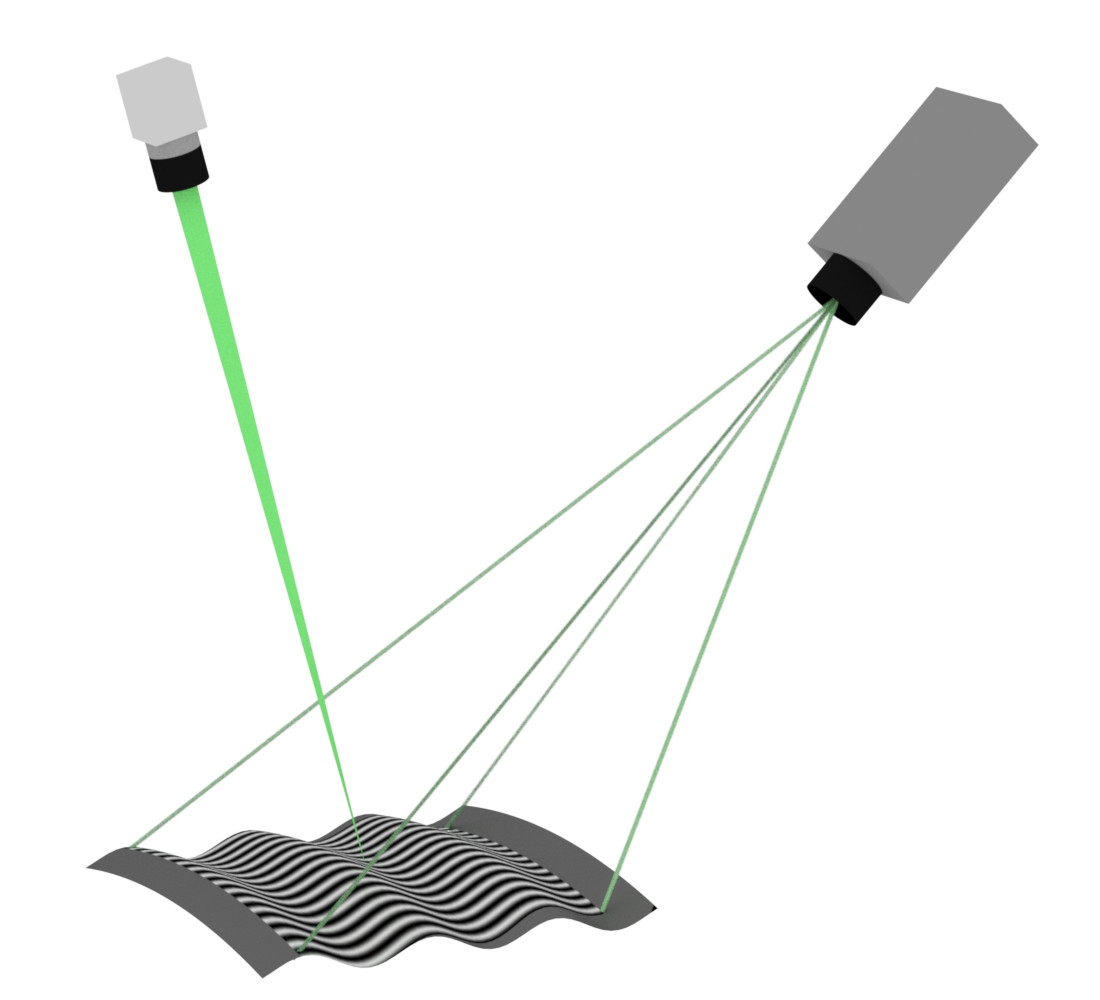}
        \put(-110,170){Projector $B$}
        \put(-180,195){Camera $C$}
        \put(-90,20){Surface $S$}
        \put(-10,0){(a)}
        \vspace{0.1cm}
        \\
        \centering
        \includegraphics[width=0.9\columnwidth]{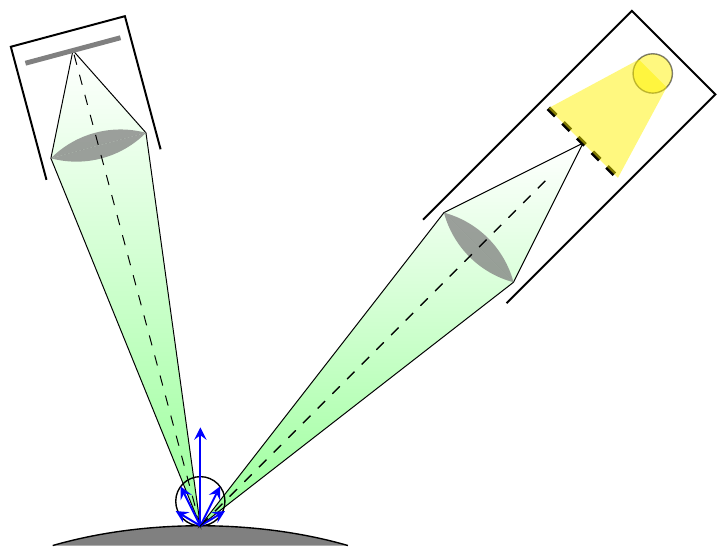}
        \put(-110,145){Projector $B$}
        \put(-180,140){Camera $C$}
        \put(-120,8){Surface $S$}
        \put(-165,60){normal}
        \put(-170,50){vector $\hat{n}$}
        \put(-10,0){(b)}
        \vspace{0.0cm}
        \\
        \centering
        \includegraphics[width=0.9\columnwidth]{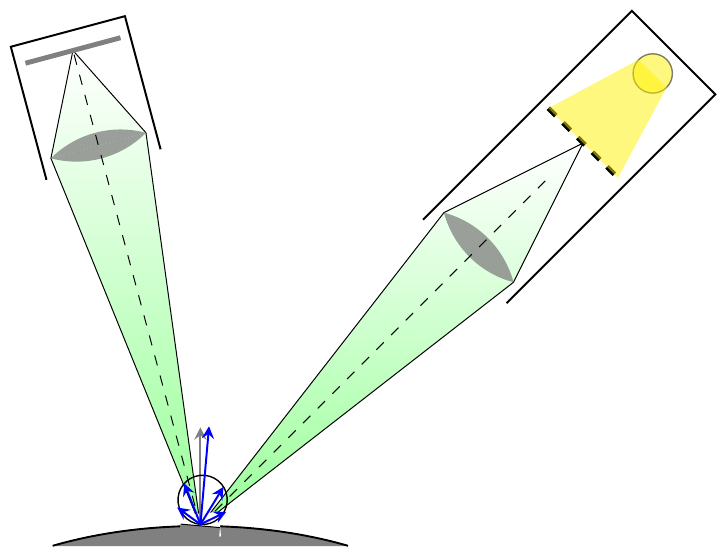}
        \put(-120,8){Deformed surface $S'$}
        \put(-150,40){$\alpha$}
        \put(-10,0){(c)}
        \vspace{-0.1cm}
        \caption{(a) Fringe projection setup: camera $C$ observes a diffusely reflective surface $S$ onto which projector $B$ projects patterns. Object points act as un-directed light emitters; the directions of the incident light and the view rays do not need to satisfy the reflection law, and indeed, should not, so as to avoid specular reflections from $B$. (b) Optical scheme of FP. The registration is sensitive to point positions but not to local slopes, cf. Fig.~\ref{fig:sensitivity}(a). The circle and arrows indicate the scattering lobe for the incident illumination. (c) Tilting a surface element in the same way as in Fig.~\ref{fig:sensitivity}(b): besides a small change in scattered intensity (see \ref{sec:sfs}, the signal on the camera sensor remains unchanged.}
        \label{fig:fproj}
        \vspace{-0.75cm}
    \end{figure}
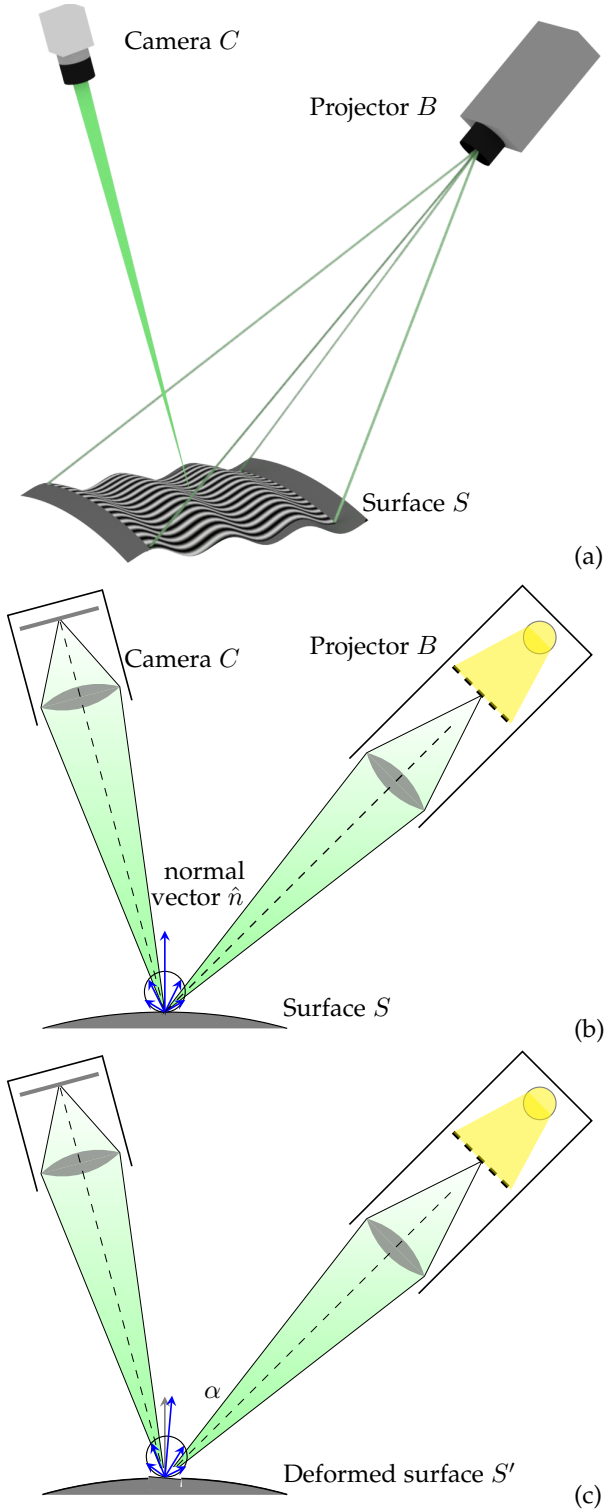

    Novices to DM often expect that a pattern or a reference structure must be projected onto the surface as in the fringe projection (FP) method; this is \underline{not} the case. Using Figs.~\ref{fig:dm_setup}(b) and~\ref{fig:fproj}, the most obvious differences can be summarized as follows: FP applies to diffusely reflective, DM to specular surfaces; in FP, a projector creates a pattern on the surface as a bright (emissive) texture, in DM a bright texture is displayed on a flat screen and is not projected anywhere.
        
    In practice, in any structured illumination setup such as FP one tries to \textit{avoid} capturing the direct specular reflections of the projector beam -- otherwise a bright mirror reflection of the exit lens will appear as a virtual image and overlap or even outshine the useful signal of the fringe patterns. 
        
    Also, the projection and the observation angles in FP are constrained mainly by the object geometry (e.g., shadowing must be avoided if possible), and to a lesser extent by the reflective properties of the surface. Multiple scattering here contributes to random noise or even systematic phase-measurement errors but is permissible.
        
    In contrast to this, in DM the illumination and the observation directions must obey the reflection law. This places much tighter constraints on the setup geometry and introduces a stronger dependence on the object shape. For example, convex surfaces create very small virtual images even of large reference screens, so that only a small portion of the surface can be inspected at once. Multiple reflections from the same surface (sometimes referred to as "inter-reflections") cannot be tolerated at all as they render the data undecodable; we are not aware of even partial solutions to this problem. On the other hand, there exist methods to separate mixed reflections from different surfaces of a transparent object (cf. Sec.~\ref{sec:multi_reflection}).
        
    In order to recover the surface shape in FP, one finds matches between the outgoing ray directions of the projector and the view rays of the camera and establishes the distance to each surface point by triangulation. The method is therefore sensitive to the \textit{absolute surface point positions} in space (zeroth order derivatives of the surface shape). In theory, FP is uniformly sensitive to all spatial frequencies, i.e. the global object shape as well as small-scale surface variations should be recovered equally well. In practice, however, the sensitivity to small defects suffers disproportionately due to the decreasing MTF and the surface roughness.
        
    As discussed above, DM measures \textit{slopes} (first order derivatives of the shape). The sensitivity to higher spatial frequencies is therefore amplified, resulting in excellent (sometimes excessive) sensitivity for small-scale irregularities and at the same time poor sensitivity and stability for low-order surface features. Reconstruction of the surface is more challenging than in FP: one first extracts partial shape derivatives from observations and then integrates them under some boundary conditions (Sec.~\ref{sec:reconstruction}).
        
    As an alternative use of DM data, one may differentiate them and recover surface \textit{curvatures} (combinations of second order shape derivatives; typically one uses Gaussian or mean curvatures). Unlike point positions and slopes, the latter are intrinsic local characteristics of the surface~\cite{Weingartner.2001c, Pak.2017} that are independent of its embedding in 3D space. As such, curvature maps are useful observables for various quality inspection tasks. Derivation of curvatures is less error-prone than shape integration and does not require accurate prior knowledge of the distance to the object. Differentiation amplifies noise~\cite{Komander.2019, Komander.2019b} but (unlike integration) does not spread correlated errors over the surface. 
        
    In order to decide whether the sensitivity profile of DM suits a given application, in addition to the above considerations one has to analyze task-specific objectives of the inspection and the statistics of surface features for typical samples. Respective power spectra for certain types of technical surfaces can be found in the literature~\cite{Falconi.1964, Wagner.2003b, Su.2015, Coniglio.2021, Choi.2021}.
        
    \subsubsection{Scanning deflectometric techniques}
    \label{sec:scanning}
        
    In some applications, the basic setup of Fig.~\ref{fig:dm_setup}(b) is impractical or even impossible to implement; the reasons can be quite diverse, as are workarounds developed to take advantage of the high sensitivity of DM anyway.
        
    For surfaces too complex to measure with the "inverse" setup, e.g. the so-called "wild aspherics" or free-form shapes, one can use point-scanning techniques. This approach is also known as "experimental ray tracing"~\cite{Hausler.1988, Binkele.2021}: a single, narrow, well-collimated (typically laser) beam probes a reflective surface in a scanning pattern and produces a grid of deflection measurements using the "direct" scheme of Fig.~\ref{fig:dm_setup}(a). The process is slow but can achieve high performance in practice -- uncertainties in the surface shape of tens to hundreds of nm have been reported~\cite{Ceyhan.2013}.
        
    Low-uncertainty flatness tests are usually accomplished with interferometry. Remarkably, some varieties of DM are able to improve on its performance and reach sub-nm uncertainty levels~\cite{Illemann.2002, Pedreira.2019, Ehret.2021}. Slope maps are recorded here with a scanning autocollimator, and the surface reconstruction is based on several 1D scans. The advantage over interferometers is that large objects (of sizes of order 1 m) can be measured without an extremely large and well-calibrated aperture and/or that the added complication of sub-aperture stitching can be avoided.
                
    Yet another scanning technique enables the inspection of nearly-flat objects in a linear motion (e.g., on a conveyor belt~\cite{Wedowski.2012b, Wedowski.2012c, Hugel.15.02.2016, Meguenani.2019, Penk.2020}). A laser line illuminates the surface and reflects towards a flat diffusely reflective or transmissive projection screen while a camera observes it. From the recorded images, one extracts the (distorted) curve shape. In case the deviations of the surface from a plane are small, this suffices to constrain one of the two slope components, which is often good enough for industrial use.
        
\begin{figure}
        \includegraphics[bb=0 40 510 355, clip, width=0.82\columnwidth, height=5.3cm]{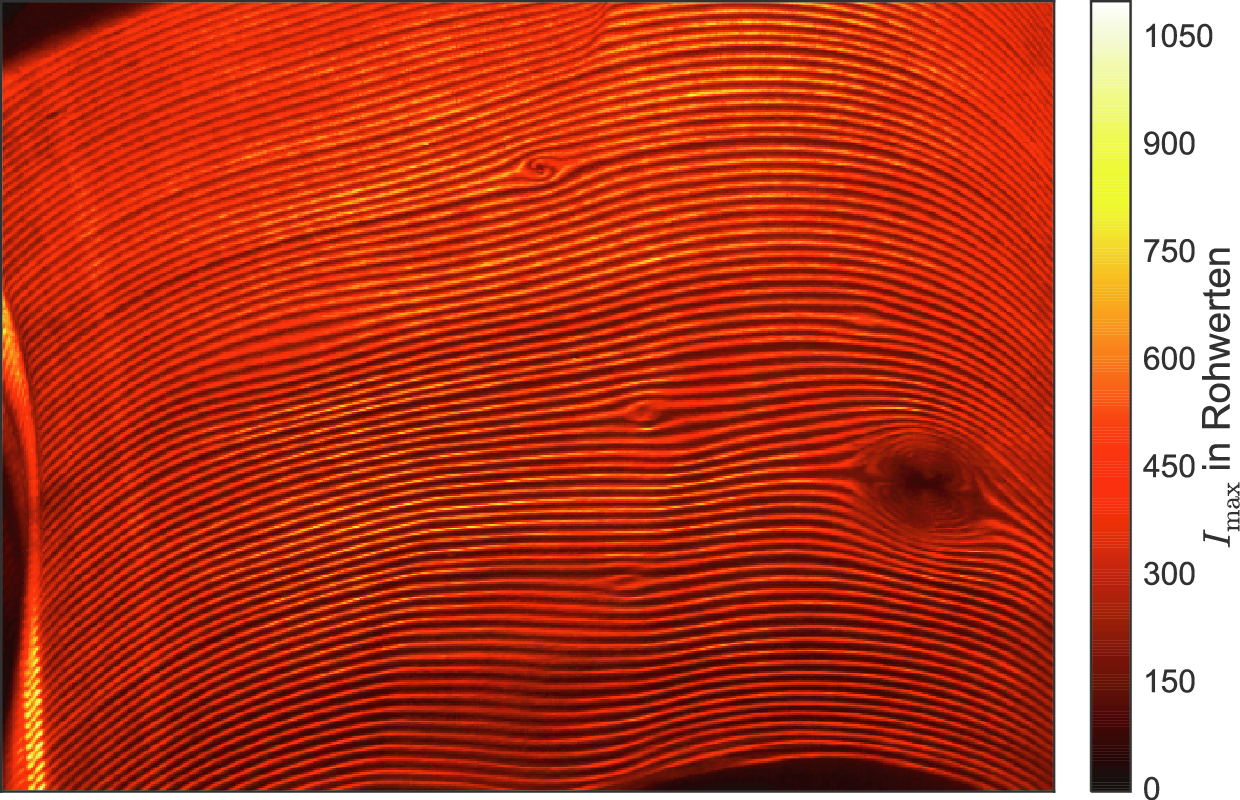}
        \put(5, 0){(a)}
        \\
        \includegraphics[bb=10 45 495 400, clip, width=0.82\columnwidth, height=5.5cm]{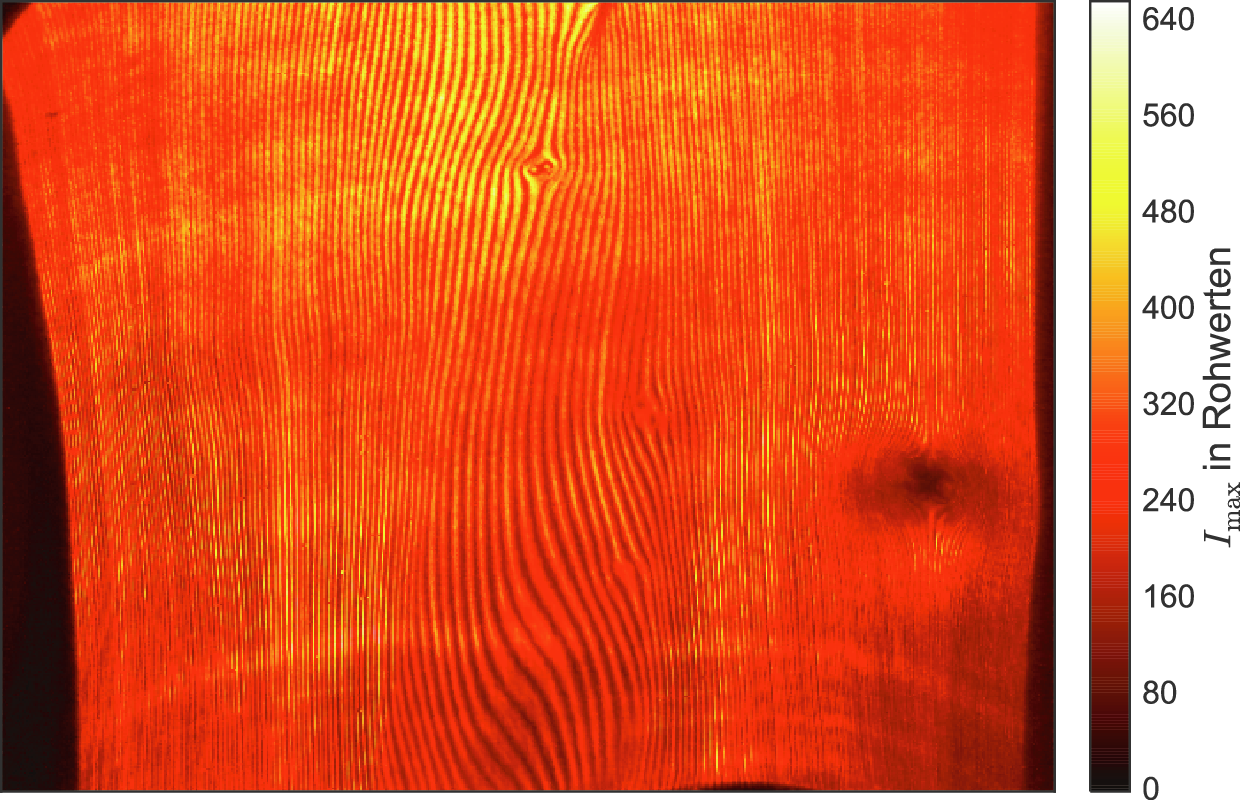}
        \put(5, 0){(b)}
        \\
        \includegraphics[bb=40 0 945 605, clip, width=0.95\columnwidth]{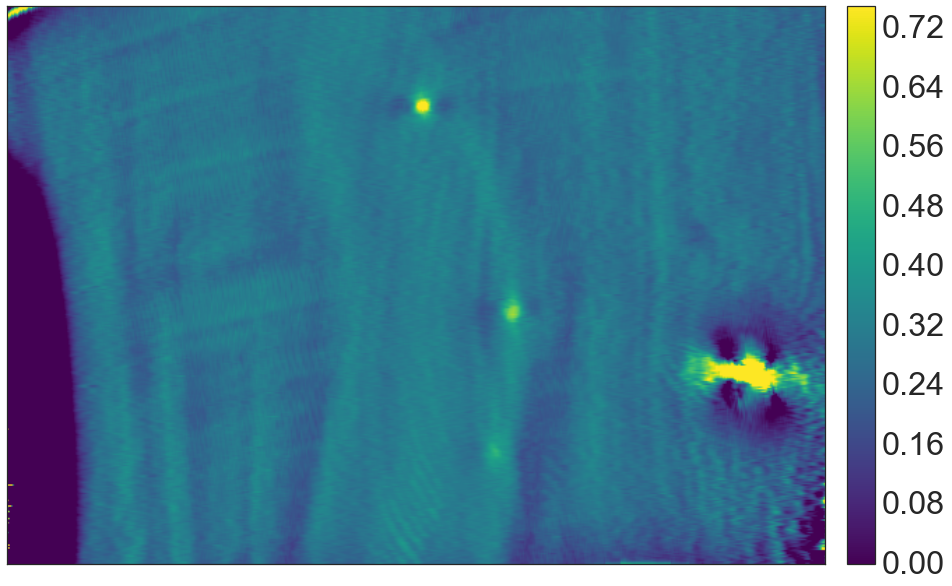}
        \put(5, 0){(c)}
        \caption{Sample data from 2D scanning IRDM inspection. (a) and (b): composite images from vertical and horizontal line scans, respectively. Brighter colors correspond to higher recorded temperatures (relative units). (c): magnitudes of combined slope gradients after decoding, in relative units. Images: S. H{\"o}fer.}
        \label{fig:ir_scan}
        \vspace{-0.6cm}
    \end{figure}

    For wavelengths outside the visible range, rapidly switchable display screens are not easily available. Scanning techniques again may be the solution; however, here one typically keeps the object at rest and moves the reference structure. For example, in thermal-infrared deflectometry (Sec.~\ref{sec:irdm}) one can use specialized image-generating devices (see the references in~\cite{Hofer.2017c}). More affordable solutions, however, rely on tension-loaded heated wires or other static emissive structures~\cite{Graves.2019b} that have a high contrast in thermal IR and a simple geometry. In a typical implementation, such a structure makes two scanning passes in two orthogonal directions~\cite{Su.2013d, Su.2014b, Hofer.2016b, Hofer.2016c}, sweeping the surface of some "virtual screen", as shown in Fig.~\ref{fig:ir_scan}. Point positions on this "virtual screen" are found using pattern recognition, and the subsequent processing proceeds as in "standard" DM.
        
    A variation of this technique that uses only one scanning direction but records both spatial derivatives is also known for DM in the ultraviolet range, with the purpose of suppressing double reflections~\cite{Sprenger.2010} (Sec.~\ref{sec:uvdm}).
        
    \subsubsection{Deflectometry in transmission}
    \label{sec:in_transmission}
        
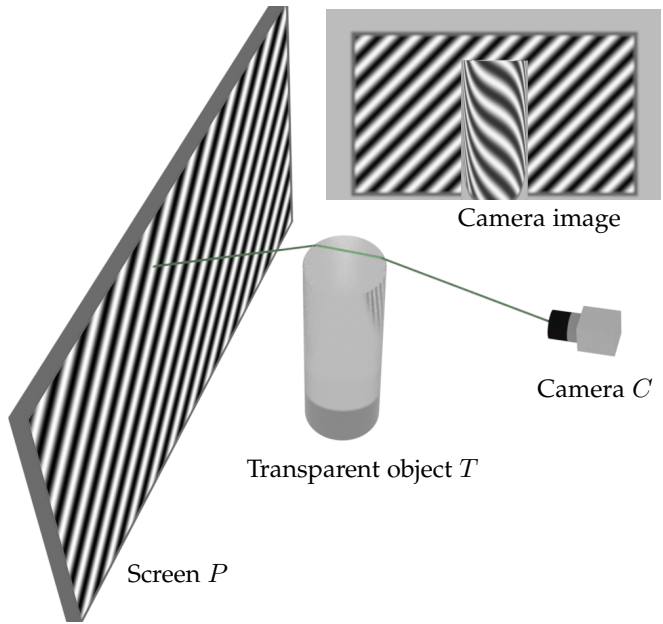
\begin{figure}
        \centering
        \includegraphics[bb=20 30 990 970, clip, width=0.95\columnwidth]{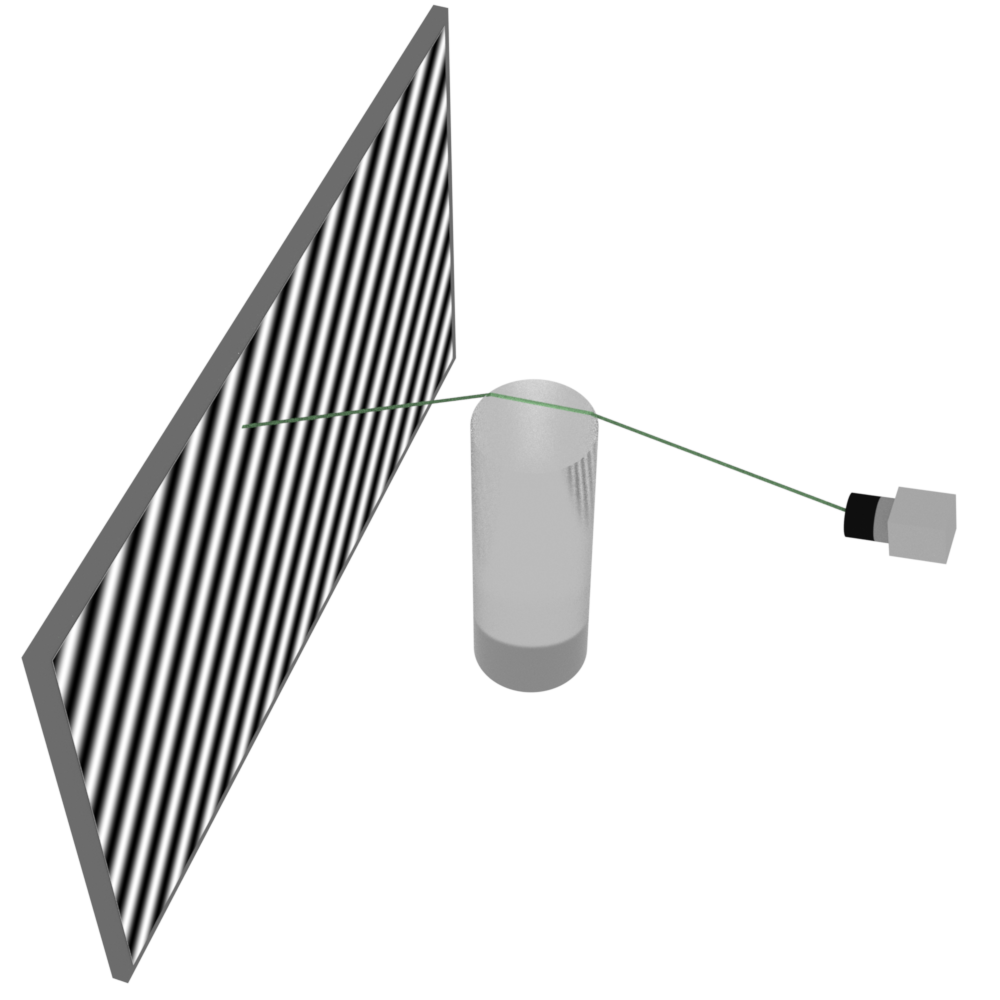}
        \put(-120,160){\includegraphics[bb=20 0 1250 700, clip, width=0.5\columnwidth]{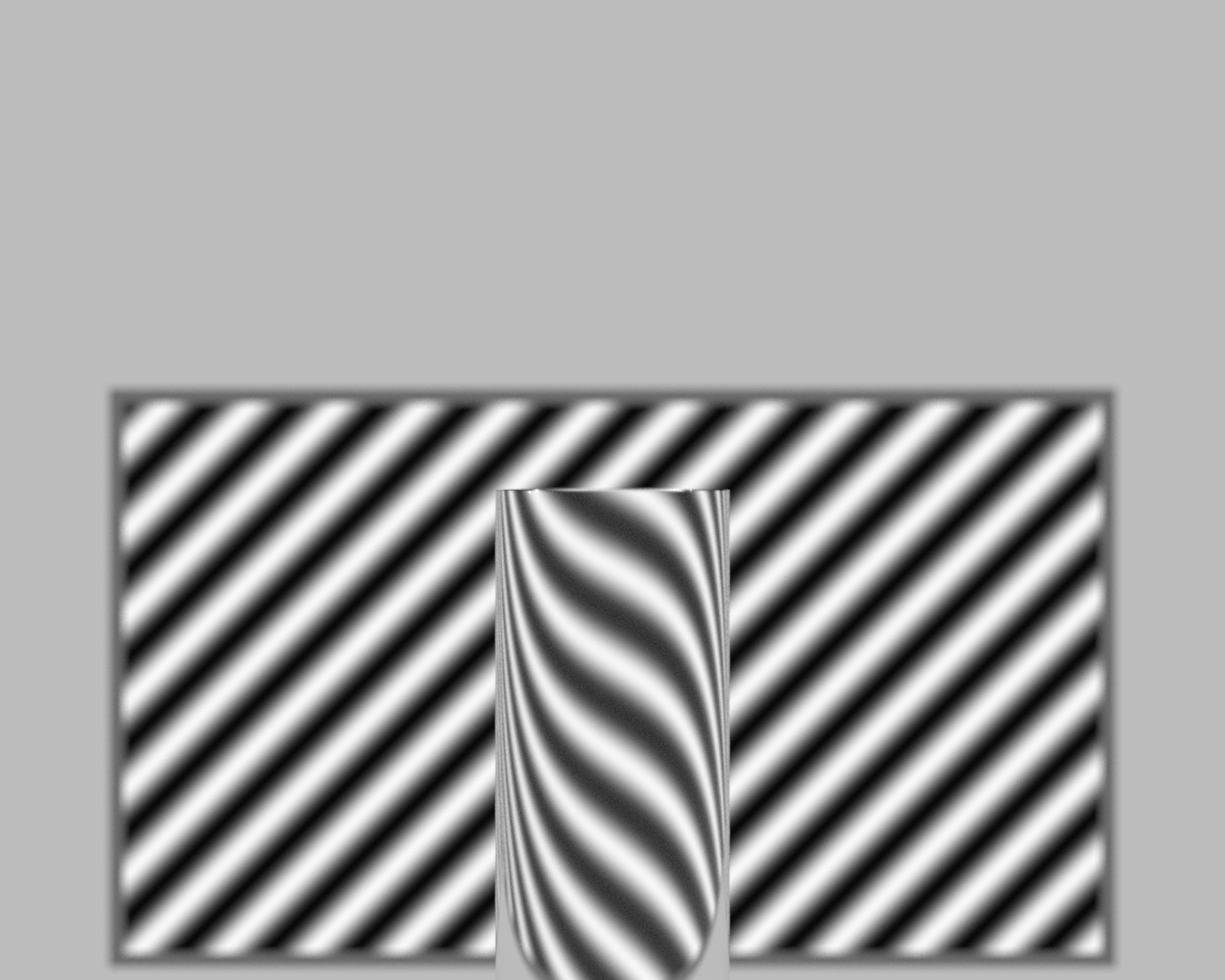}}
        \put(-70,150){Camera image}
        \put(-40,85){Camera $C$}
        \put(-195,15){Screen $P$}
        \put(-150,55){Transparent object $T$}
        \caption{DM in transmission: a camera observes a pattern through a transparent studied object. View rays deflect due to reflection and refraction on the outer surfaces and on the inhomogeneities inside the object. Inset picture shows the view of the camera with the fringe distortions as measurand.}
        \label{fig:transmission}
        \vspace{-0.6cm}
    \end{figure}

    With some minor modifications, DM can be applied to transparent objects as in Fig.~\ref{fig:transmission}. The same pattern sequences and decoding algorithms as in the reflective case can be used to obtain the deflection maps. The latter, however, now characterize the refraction of light by the outer surfaces, and possibly additional internal deflections due to discontinuities and gradients of the refractive index inside the studied object. Depending on the setup, external and internal reflections may also contribute to the signal. Data processing in this case significantly differs from solely reflection-based shape reconstruction: instead of a single surface, one typically has to recover the entry and exit surface shapes and at the same time map the volumetric distributions of the refractive index. The theory and implementation of such schemes are non-trivial and deserve a separate review; therefore, we exclude them from the subsequent discussion and in the rest of this section only briefly mention some notable developments.
        
    A number of schemes for DM in transmission have been proposed in the metrology~\cite{Kafri.1985, Massig.1999, Canabal.2002, Vargas.2010, Atcheson.2012, Meriaudeau.2012, Fischer.2016, Li.2018, Wang.2018c, Binkele.2019, Wang.2020b, Wang.2021c} as well as in the computer vision~\cite{Trifonov.2006, Yamazaki.2007, Kutulakos.2008} communities, often under the title of "optical deflectometric tomography"~\cite{Gonzalez.2013, Sudhakar.2015}. Various proposed implementations employ laser scanning, active and static patterns, distant detection (Fourier regime), etc. However, most schemes assume very small deflection angles (or, equivalently, small variations of the refractive index) and therefore apply only to simple object shapes and/or require that the studied object be immersed in an index-matching fluid. Alternative (large-deflection) approaches often simplify the problem and e.g. assume that the refractive index inside the object is constant and only recover its shape~\cite{Petz.2009} or focus on selected (application-specific) optical parameters of samples (such as "refractive power")~\cite{Knauer.2008, Vargas.2010}. The traditional schlieren technique, a form of deflectometry~\cite{Toepler.1864, Greenberg.1995}, has also been studied in combination with moir{\'e}~\cite{LEsperance.2017} and phase-shifting methods~\cite{Joannes.2003, Beghuin.2009, Antoine.2019b} in order to characterize optical components in direct transmission or tomographically~\cite{Foumouo.2010, Gonzalez.2011b}.
        
    \subsubsection{Shape from shading}
    \label{sec:sfs}
        
\begin{figure}
        \centering
        \includegraphics[bb=0 10 458 385, clip, width=0.49\columnwidth]{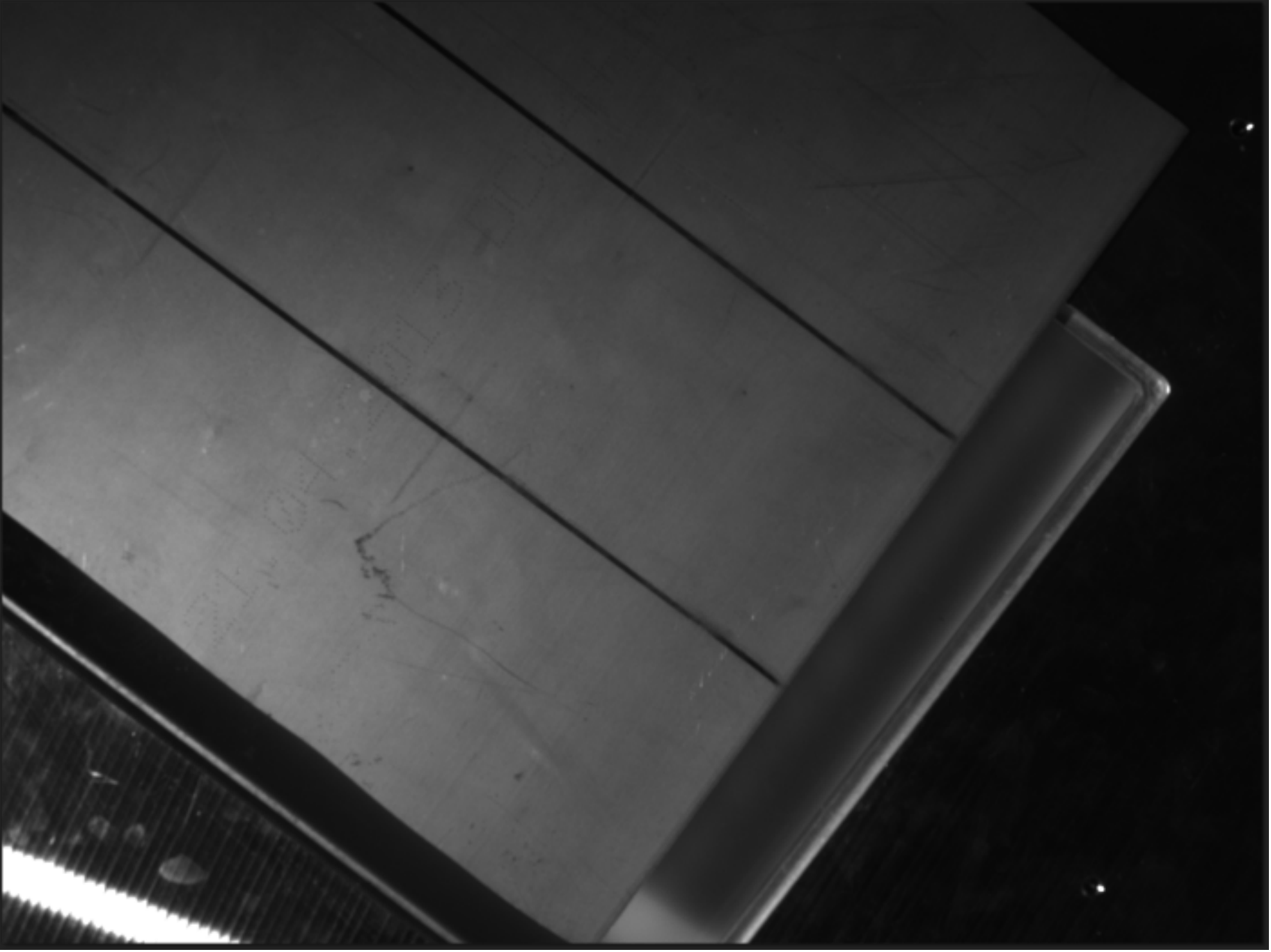}
        \put(-20, 10){\fcolorbox{black}{white}{(a)}}
        \hfill
        \includegraphics[bb=0 10 458 385, clip, width=0.49\columnwidth]{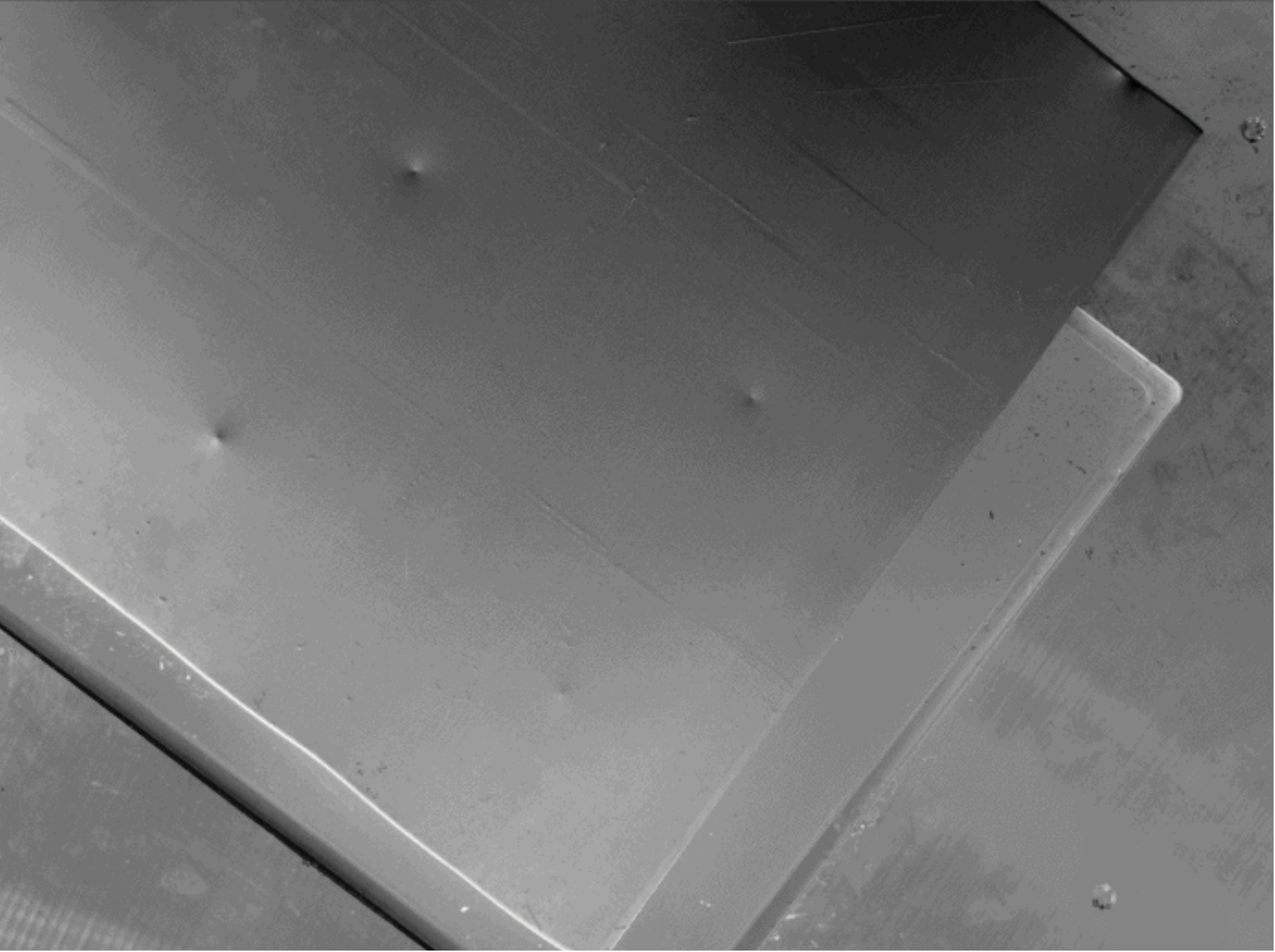}
        \put(-20, 10){\fcolorbox{black}{white}{(b)}}
        \put(-120, 10){
            \begin{tikzpicture}
                \draw [rotate=50, -stealth, line width=0.7mm, white](0,0) -- (1,0);
            \end{tikzpicture}
            }
        \\
        \includegraphics[bb=0 10 458 385, clip, width=0.49\columnwidth]{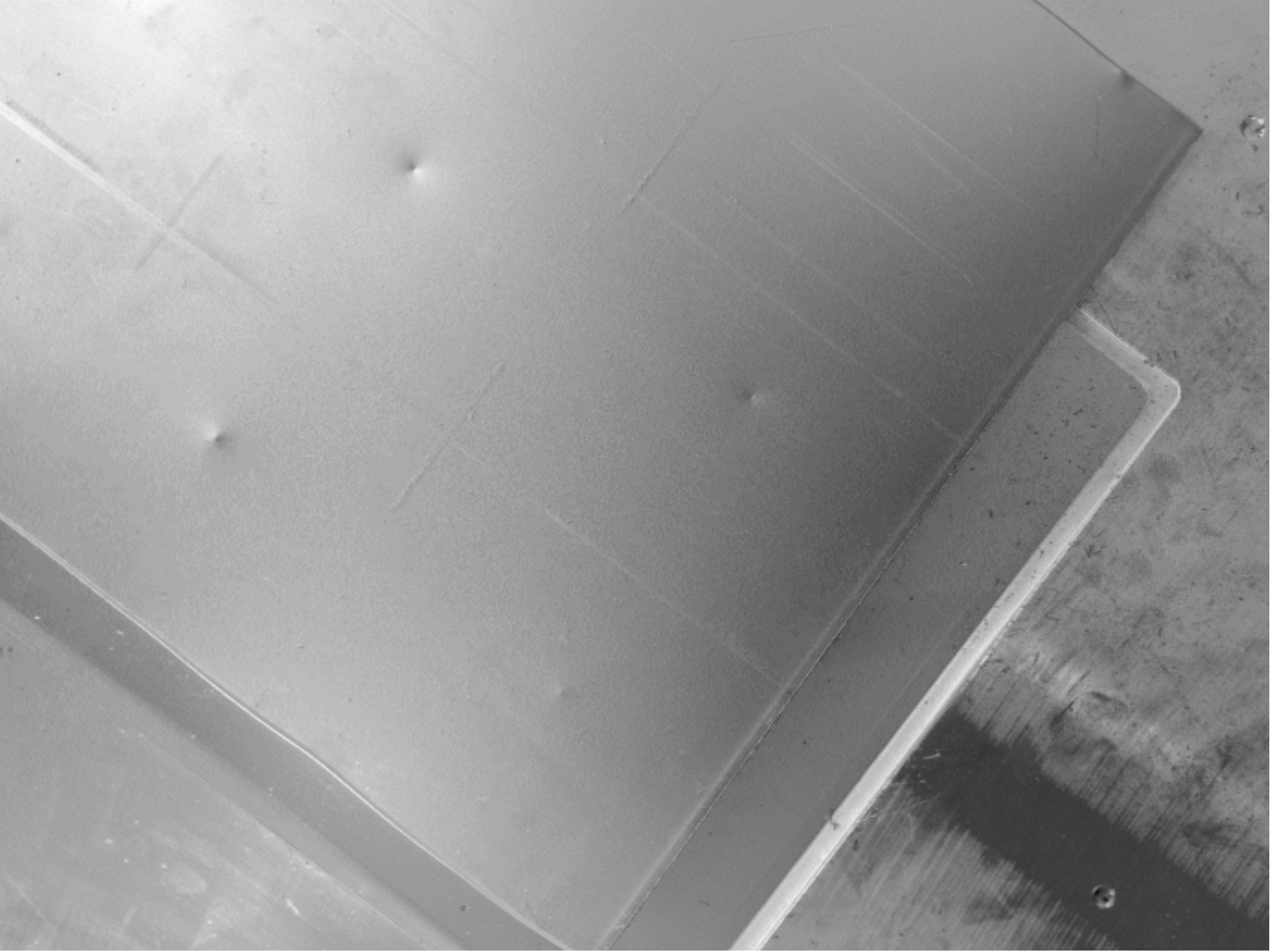}
        \put(-20, 10){\fcolorbox{black}{white}{(c)}}
        \put(-120, 75){
            \begin{tikzpicture}
                 \draw [rotate=-40, -stealth, line width=0.7mm, black](0,0) -- (1,0);
            \end{tikzpicture}
            }
        \hfill
        \includegraphics[bb=0 0 210 180, clip, width=0.49\columnwidth]{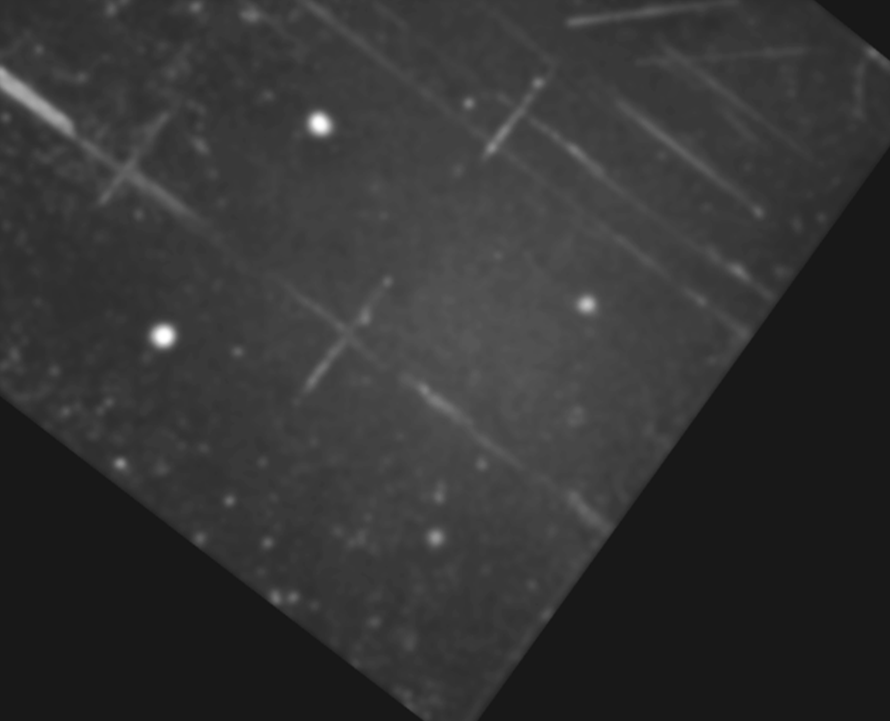}
        \put(-20, 10){\fcolorbox{black}{white}{(d)}}
        \caption{SFS measurement of a metal sheet with sample dents and bumps. (a) Appearance of the object in undirected illumination. (b), (c): slope maps obtained under two different orthogonal directions of incident illumination, indicated by arrows. The "relief" impression is not coincidental but reveals the principle and gives an idea of the sensitivity. (d): magnitudes of the detected slope gradients. Images: S. Werling / M. Heizmann.}
        \label{fig:sfs}
        \vspace{-0.6cm}
    \end{figure}
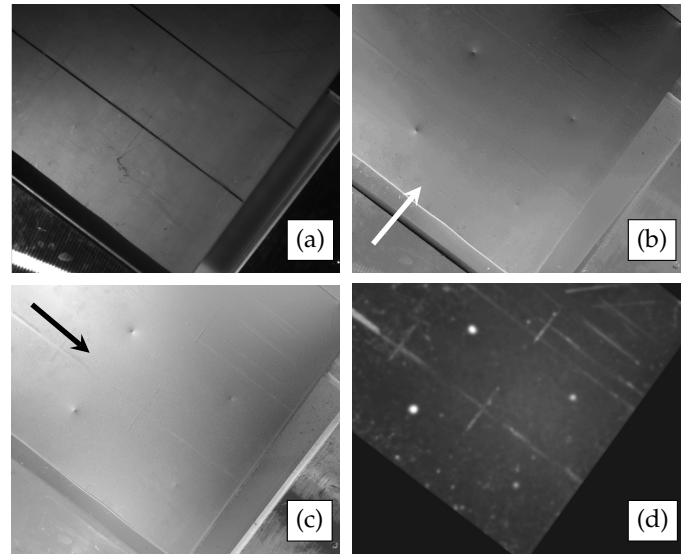

    Let us assume that the studied surface reflects diffusely according to some simple (e.g. Lambertian) reflectance model (cf. Fig.~\ref{fig:reflection_components}). Under illumination at some fixed angle, the intensity of the reflected light along a given observation direction is sensitive to the surface slope. This combination of diffuse reflective geometry measurement with radiometry is known as "shape from shading" (SFS)~\cite{Zheng.1991, Szeliski.1991, Prados.2006b, Balzer.2006, Lellmann.2008, Beyerer.2016}. Typically (but not necessarily) objects are illuminated at a shallow angle and observed from a nearly normal direction in order to increase sensitivity. With proper radiometric calibration, SFS can measure very small surface slopes and locate minor defects. SFS can be considered a subset of the "photometric stereo" methods~\cite{Woodham.1980, Ikeuchi.1981, Woodham.1989, Stephan.2016b} that use active illumination and multiple exposures. Similarly to DM, it is best suited for the detection and estimation of local \textit{variations} of slope. An example SFS measurement is shown in Fig.~\ref{fig:sfs}.
        
    \subsubsection{Shape from specular flow}
    \label{sec:sfsf}
        
    Arguably the most "natural" kind of DM, the shape-from-specular flow (SFSF) approach in computer vision has been proposed by Roth and Black\cite{Roth.2006} based on differential analysis by Blake and B{\"u}lthoff\cite{Blake.1991}. Unlike "canonical" DM, SFSF aims to recover shapes of mirrors in motion by observing the evolution of reflections {\itshape without control over (or prior knowledge of) the reflected environment}. We intuitively use SFSF when e.g. we look at a shiny car on the street -- slightly moving the head helps us notice surface defects even if we do not know the geometry of the surrounding buildings or trees. Similarly, in SFSF one slightly changes the setup geometry (e.g., adjusts the camera position or rotates the object) while letting the camera record successive frames. As in DM, the camera here observes a specular object that reflects some uncalibrated environment. From its raw frames, one then derives an {\itshape optical flow} (OF) field: a dense map of apparent 2D displacements of prominent image elements (edges, corners) or pixel values between the subsequent camera images. This field then serves as input for surface shape reconstruction (often under the assumption of an infinitely distant environment) or is used in qualitative analysis.
        
    While virtually unknown in the metrology community, OF is the foundation of numerous applications in computer vision. Its estimation is an important research field and multiple algorithms are implemented in popular libraries~\cite{Bradski.2000}. If the OF corresponds to a moving reflection in a curved mirror, it is known as specular flow (SF). Conceptually, OF and SF are the same quantity; but the motion fields of specularities often feature large irregularities (distortions and discontinuities). This impacts various statistics and necessitates dedicated estimation algorithms~\cite{Canas.2009, Adato.2010b, Adato.2011c}. Despite its inherently heuristic nature, under certain conditions OF or SF can be estimated with sub-pixel uncertainty which potentially may enable novel metrological methods.
        
    In the original formulation SFSF was only used to estimate parameters of primitive shapes. Later, the problem was re-formulated in terms of global variational reconstruction~\cite{Solem.2004, Lellmann.2008}. Adato and others ~\cite{Adato.2007, Canas.2009, Adato.2010, Vasilyev.2011} suggested a general reconstruction method based on a system of linear partial differential equations. In their setup, a telecentric camera is fixed with respect to the object and the distant textured environment undergoes a global rigid rotation. Later, SFSF equations and reconstruction methods have been extended~\cite{Pak.2016d, Pak.2017} to a theoretically more demanding but also more realistic scenario where a perspective camera linearly moves with respect to a static object and a static environment. These developments appear promising; unfortunately, a solid experimental validation beyond a basic proof of principle is still missing.
        
    One notable statement of~\cite{Pak.2017} is that the SF is fundamentally sensitive to the curvatures of the mirror surface (second order shape derivatives) and has thus even better sensitivity to high spatial frequencies than "canonical" DM. An interesting direction of future research could be a combination of DM and SFSF in a single inspection setup in order to mimic the human assessment as in Fig.~\ref{fig:examples}(a).
        
    \subsubsection{Fourier and telecentric techniques}
    \label{sec:telecentric}
        
    Consider a special version of the setup in Fig.~\ref{fig:dm_setup}(a) where the object is small compared to the distance $d$ between object and detector, and is fully illuminated by a collimated light source. In this "Fourier regime"\footnote{Not to be confused with Fourier decoding mentioned in Sec.~\ref{sec:phase_measuring}.} (also known as "direction-coded deflectometry"~\cite{Sener.2004, Sener.2009}) the geometrical equations simplify ($d$ drops out) and the detected signal can be directly converted to deflection angles, from which one can easily recover the surface shape. This simplification happens at the cost of more expensive telecentric optics, whose quality (angular uncertainty, or "non-parallelism" of rays) then directly contributes to the residual measurement errors.
        
    In a related technique, the intensity of a uniform parallel beam of light is re-distributed by small irregularities of the reflective surface, which causes a spatial modulation of irradiance incident on a distant radiometric sensor. This "Makyoh imaging" scheme owes its name to the ancient art of making "magic mirrors": a master would emboss secret symbols on the back side of a polished metal mirror. Invisible in direct inspection, the symbols can be revealed by reflecting sunlight onto a distant wall.
        
    The theory behind the method is well studied~\cite{Saines.1999, Berry.2005, Riesz.2011}; its sensitivity is sufficient to e.g. inspect semiconductor wafers for irregularities~\cite{Kugimiya.1988, Blaustein.1989, Hahn.1990, Szabo.1995, Finck.2009, Tobisch.2012, Hologenix.2020}.

    \subsubsection{Qualitative and semi-qualitative methods}
    \label{sec:qualitative}
        
    Typical uses of DM for qualitative surface inspection will be outlined in Sec.~\ref{sec:defect_detection}; here we mention some alternative ideas that have caught our attention. They mostly originate from the computer vision community and typically deal with incomplete data and/or uncalibrated configurations.

    An interesting DM setup for industrial QA has been reported in~\cite{Tornero.2012b} where a camera that is fixed with respect to the object (a car body) moves along with it through a light tunnel. The captured frames are accumulated into a synthetic "integral image", from which one can easily identify defects as local extrema of pixel values (too dark or too bright spots) according to a pre-defined mask.
        
    Another work~\cite{Godard.2015} reports the qualitative shape reconstruction of small reflective objects based on multi-stereo views and panoramic environment maps. Unlike the exact DM reconstruction, normal vectors here are matched probabilistically based on local color distributions in the environment. Similarly,~\cite{Jaquet.2013} demonstrates the reconstruction of nearly-flat reflective surfaces under the assumption that the reflected scene contains only straight lines that appear upon reflection as curves in the camera images (a typical scenario is a shop window in a city that reflects the surrounding office buildings). Using even less prior information, local curvatures of a smooth mirror can be deduced from auto-correlation patterns in the images captured within a sufficiently richly-textured environment~\cite{Tappen.2011}. Finally, reflective shapes may be partially recovered based on certain invariants identified in the images captured in an unknown environment~\cite{Sankaranarayanan.2010}.
        
    More recently, artificial neural networks (ANNs) have been employed in order to identify defects in DM data (see, e.g.,~\cite{MaestroWatson.2018, Zhang.2019, Zhou.2020, Guan.2022}), sometimes with pre-processing by classical image processing~\cite{Qi.2020}. As with many other applications of ANNs, it remains unclear to which extent the models trained in a given constellation with certain objects can be transferred to different problems. Presently we leave these methods out of the discussion until their "generalization power"~\cite{Zuo.2022} is understood better.

    \subsection{Historical overview}
    \label{sec:history}
    
    The "magic mirrors" mentioned in Sec.~\ref{sec:telecentric} may well be the first documented utilization of the high sensitivity of specularities to slope variations of a mirror, discovered centuries or even millennia ago~\cite{Saines.1999}. The opposite case (manufacturing of mirrors that do \textit{not} create patterns in the reflected wavefront) has been documented in the past two centuries as the wire test~\cite{MalacaraHernandez.2006, JuarezReyes.2018} and the Foucault test~\cite{Foucault.1858}. The goal of these methods was to provide sensitive visual cues for defects or aberrations in the context of manual fabrication of optics; and although the (quantitative) theory of reflection was well understood, it was mainly used to create qualitative indication tools.
    
    In parallel, schlieren techniques (based chiefly on the redistribution of intensity) have been developed for the inspection of optics in transmission~\cite{Toepler.1864, Kafri.1980, Marguerre.1985, Settles.2001, Settles.2017}.
        
    Testing apertures evolved over time: from the initial pinholes, slits or wires, the principle was extended to one- and two-dimensional grids~\cite{Hartmann.1907, Ronchi.1927, MalacaraHernandez.2015}. Later, these works have paved the way for quantitative evaluations~\cite{Rayces.1964, SalasPeimbert.2005}. There exists a considerable freedom to choose the shapes of static reference patterns~\cite{CarmonaParedes.2007}: in addition to generic planes~\cite{DiazUribe.2000b}, also cylinders~\cite{DiazUribe.2000, CamposGarcia.2004}, cones~\cite{CamposGarcia.2015, CamposGarcia.2019b}, boxes~\cite{CamposGarcia.2011}, and custom geometries~\cite{Perard.2001} have been used. One may assemble points or lines into 1D or 2D arrays; otherwise, one may use rectangular, Gaussian or sinusoidal intensity profiles. Patterns can be printed or displayed in Cartesian, polar or spiral arrangements (e.g.~\cite{Klass.1980, Lippincott.1982, Massig.2001, DiazUribe.2009, Li.2014c, Kludt.2018b, Riesz.2018, Carvalho.2021, Fontani.2022}) or in fact with any pre-distortion that is transformed into a regular pattern if the reflective surface complies with certain specifications~\cite{Perard.1995, Werling.2007c, Liang.2016, Zhang.2017b} (see also \ref{sec:inverse_patterns}).

    The moir{\'e} approach already used in the Ronchi test has evolved into moir{\'e} deflectometry~\cite{Kafri.1980, Kafri.1981, Ritter.1982, Karny.1982, Kafri.1985, Servin.1990}; its extension to two dimensions is better known as raster reflection and has been used in tests for mechanical responses to loads~\cite{Ligtenberg.1952, Rieder.1965, Ritter.1983, Ritter.1991, Massig.2001}, surface defects~\cite{Lippincott.1982, Sanderson.1988}, glass windows~\cite{Skydan.2007, Chambard.2009, Xu.2010, Aprojanz.2019}, and phase objects~\cite{Massig.1999, Beghuin.2009}.
        
    The final stepstone for convenient low-uncertainty evaluation was the introduction of the phase-shifting technique, created for interferometry in the 1970s and immediately adopted for fringe projection after sufficient digital storage and processing means became available~\cite{Takeda.1982, Halioua.1983, Srinivasan.1984}. Surprisingly, phase shifting has spread to deflectometry only around the turn of the millennium~\cite{Pfeifer.1995, Hofling.2000, Perard.2001, Horneber.2001, Petz.2001,  Surrel.2004, Bothe.2004f, Knauer.2004c, MorenoOliva.2008} -- one reason may be that convenient flat-screen monitors for displaying modulated patterns were starting to become available at that time. As a result, moir{\'e} techniques are now largely obsolete in DM, and can in hindsight be interpreted as a complicated way (but necessary at the time) to reduce uncertainties.

    Some work has also been dedicated to microscopic applications of deflectometry~\cite{Krasinski.1985, Bitte.2002, Bothe.2007d, Hausler.2008, Huang.2013e}, but it appears that these have not displaced the sensitive and semi-quantitative methods that have been previously in use.
        
    In the past century, the majority of developments have been contributed by groups in Germany (e.g., those from Braunschweig, Bremen, Erlangen and Karlsruhe); some twenty years ago, several special-interest groups have appeared e.g. in Beer-Sheva (IL) and Mexico City. In the past decade, the field has been taken up by groups and schools in e.g. Nivelles (BE); Arrasate, Barcelona, and València (ES); Huddersfield (UK); Singapore; Chengdu, Shanghai, and Tianjin (CN); Charlotte and Tucson (USA) as well as numerous other schools and labs with maybe fewer researchers but no lesser results. The output of the community in a wider sense also includes an astounding number of stand-alone theses at all levels, which demonstrates how quickly useful data can be obtained from deflectometry.
    

\section{Fundamentals}
\label{sec:fundamentals}

In this section we discuss the physical and mathematical foundations of DM as a metrological method for measuring the shapes of specular surfaces. Unless mentioned otherwise, the notation and conclusions refer to the basic setup in Fig.~\ref{fig:dm_setup}(b); however, they can be easily adapted for most alternative implementations and variations of DM.

\subsection{Measurement goal specification}
\label{sec:aesthetics}
    
    The most essential characteristic of a metrological method is the target uncertainty of measurements. For DM, one has to specify the required sensitivity to shape deviations and defects in terms of unambiguous objective metrics. For some technical surfaces, such information is readily available: for instance, the maximum allowed divergence of the reflected rays for a telescope mirror determines the scale of tolerable deviations from the design. One convenient formulation of tolerances for such precision optics (on-axis, with round or hexagonal aperture), for instance, is in terms of Zernike coefficients~\cite{Li.2014f}. In certain cases such polynomial-based models may even offer the possibility to use DM measurands directly, avoiding the integration of the surface~\cite{Burge.2010b, Dominguez.2012}.
    
    However, in a quite common case when DM is used to assess surfaces that have a purely aesthetic function (e.g., car bodies), target criteria are much harder to quantify. Of course, one can design a sensor that is more sensitive than any human, but this will only increase the rate of false positives in quality control while providing no apparent benefits (why measure a car to nanometers?). When performing manual inspection, one often has to decide if a surface is "smooth" or "free from defects" -- but what does that mean? One possible approach is to (painstakingly) describe various defects: their nature, primary sizes, and tolerable "severities" for different classes of surfaces. However, this solution is not perfect: the perception of smoothness depends on the wavelength of light and the surface roughness, while the visibility of defects differs from person to person~\cite{Kessler.1997} (and buyers will judge any perceptible defect as serious in order to lower the selling price). Guided by similar considerations, multiple attempts have been made e.g. in the automotive field to relate the visibility of surface deformations as reported by customers to measurable parameters such as defect sizes or curvatures~\cite{Kessler.1997, Hsakou.2006, Andersson.2009, Fernholz.2013, Aprojanz.2019}. Such works, however, typically analyze a very limited number of cases and make little or no effort to provide a general theoretical justification.

\begin{figure}
        \centering
        \includegraphics[bb=5 0 270 235, clip, width=0.95\columnwidth]{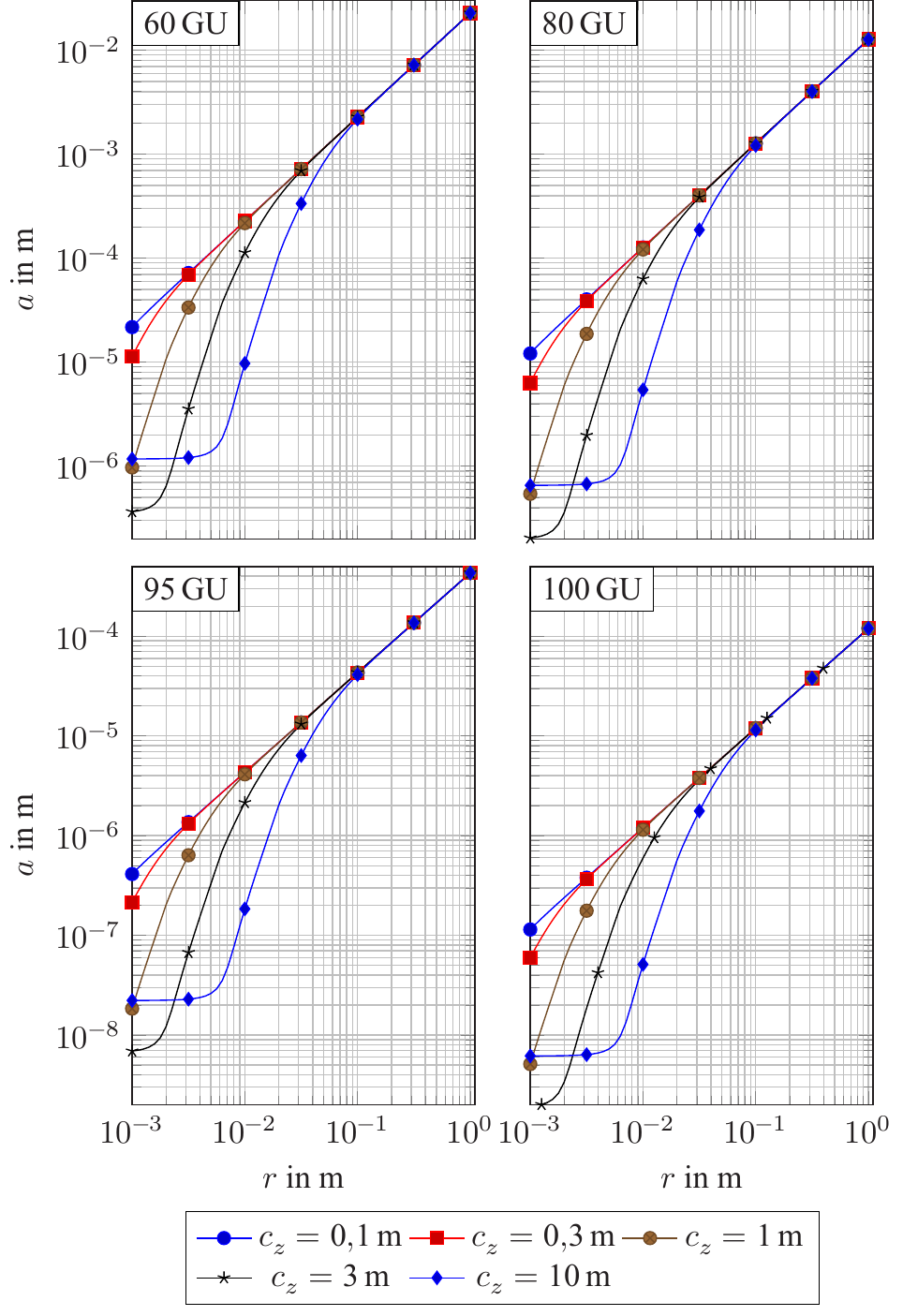}
        \caption{Theoretical lower bounds on the sizes of an isolated shape defect that is "barely visible" on top a planar specular surface. The plots correspond to the highest achievable acuity of human vision. The defect is modeled as a symmetric Gaussian shape with a characteristic radius $r$ and height $a$; the observation point is displaced from the defect center by 1 m in the lateral and by $c_z$ in the normal direction with respect to the surface. The two panels correspond to reflectivities of 95 and 100 gloss units (GU). Using simple scaling rules, these curves may be adjusted for any alternative observation geometry. Image adopted from~\cite{Ziebarth.2019}.}
        \label{fig:ziebarth2019}
    \end{figure}

    A recent work~\cite{Ziebarth.2019} has proposed universal theoretical constraints on the visibility of surface defects on specular surfaces. The approach uses only a few assumptions and utilizes the concept of specular flow (Sec.~\ref{sec:sfsf}) and its relation to human perception~\cite{Blake.1990, Blake.1991, Waldon.1993, Roth.2006}; in principle, it can be adapted to arbitrary DM measurement scenarios. As of writing, though, these results still need a better validation and an extension to important practical cases (such as partially specular surfaces). Some results of~\cite{Ziebarth.2019} are shown in Fig.~\ref{fig:ziebarth2019}; one can see that people indeed are quite sensitive to small deformations of mirrors and that laterally smaller defects are in general easier to detect.

    \subsection{Phase-shifted cosine patterns}
    \label{sec:cos_patterns}

    Encoding of screen positions via pattern sequences is a vast topic; some methods relevant for DM are discussed in detail in Sec.~\ref{sec:coding}. However, to facilitate the discussion in this section we very briefly introduce the most commonly used technique based on phase-shifted cosine patterns.
    
\begin{figure}[t]
        \centering
        \includegraphics[width=0.24\columnwidth]{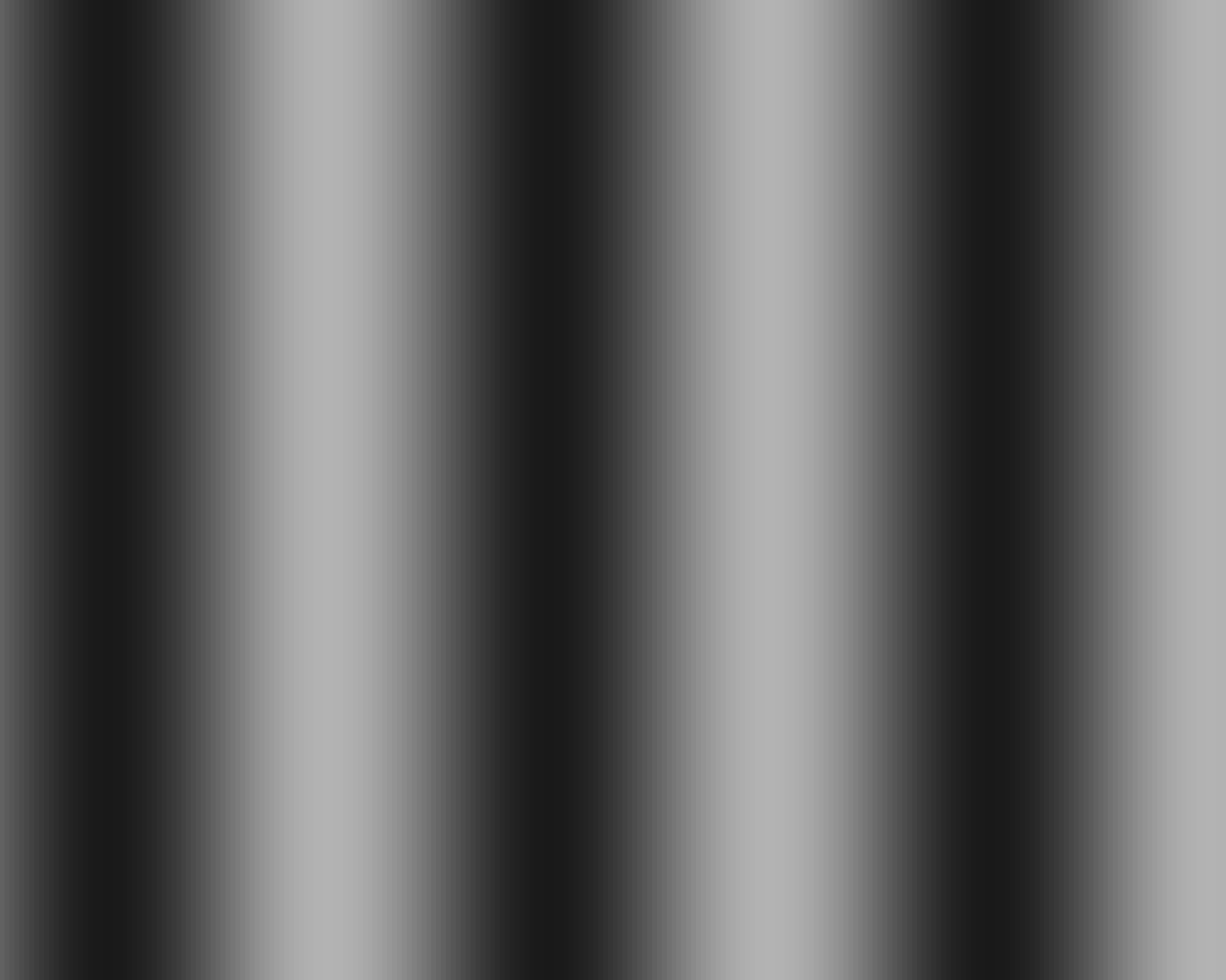}
        \put(-20,10){\colorbox{white}{(a)}}
        \hfill
        \includegraphics[width=0.24\columnwidth]{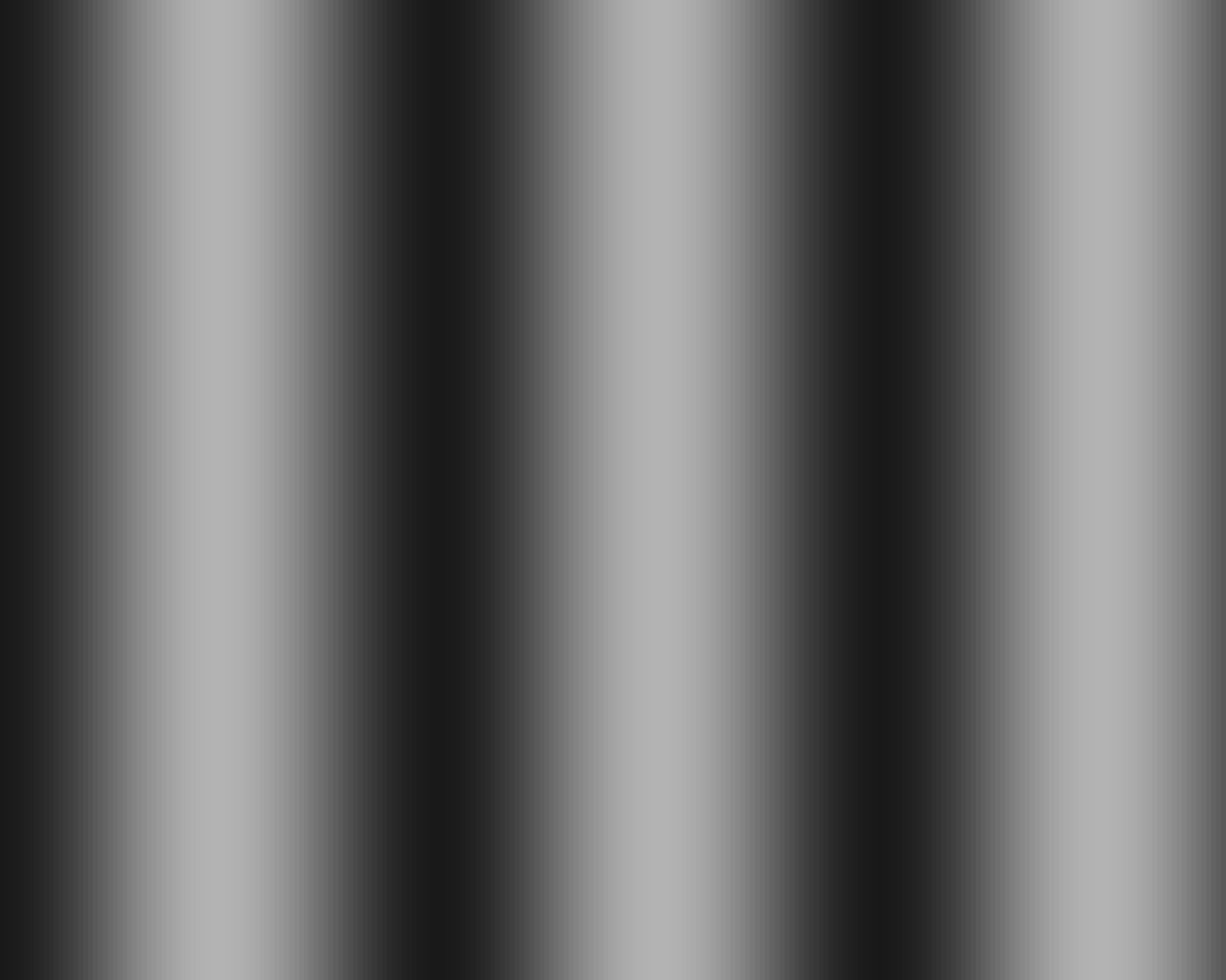}
        \put(-20,10){\colorbox{white}{(b)}}
        \hfill
        \includegraphics[width=0.24\columnwidth]{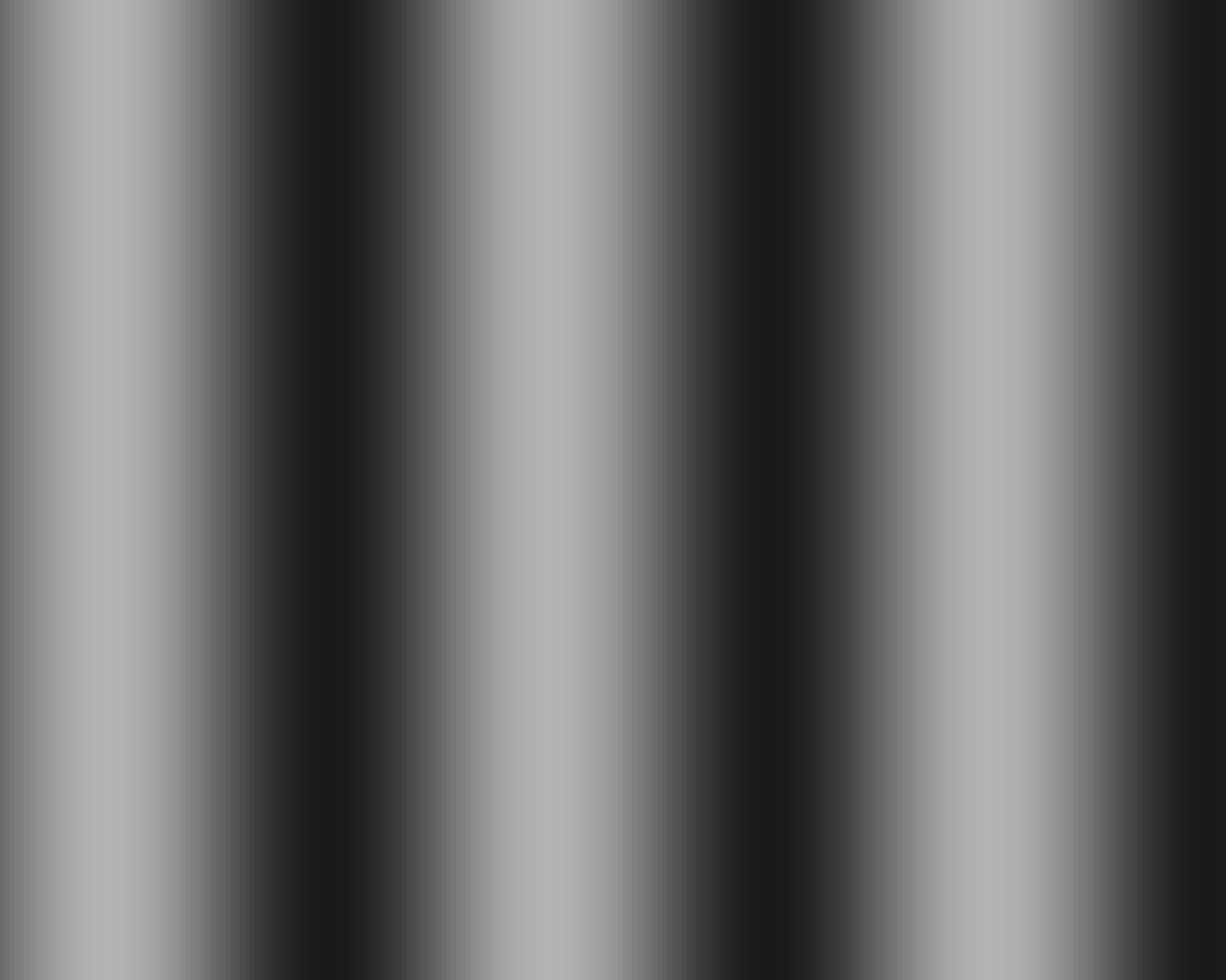}
        \put(-20,10){\colorbox{white}{(c)}}
        \hfill
        \includegraphics[width=0.24\columnwidth]{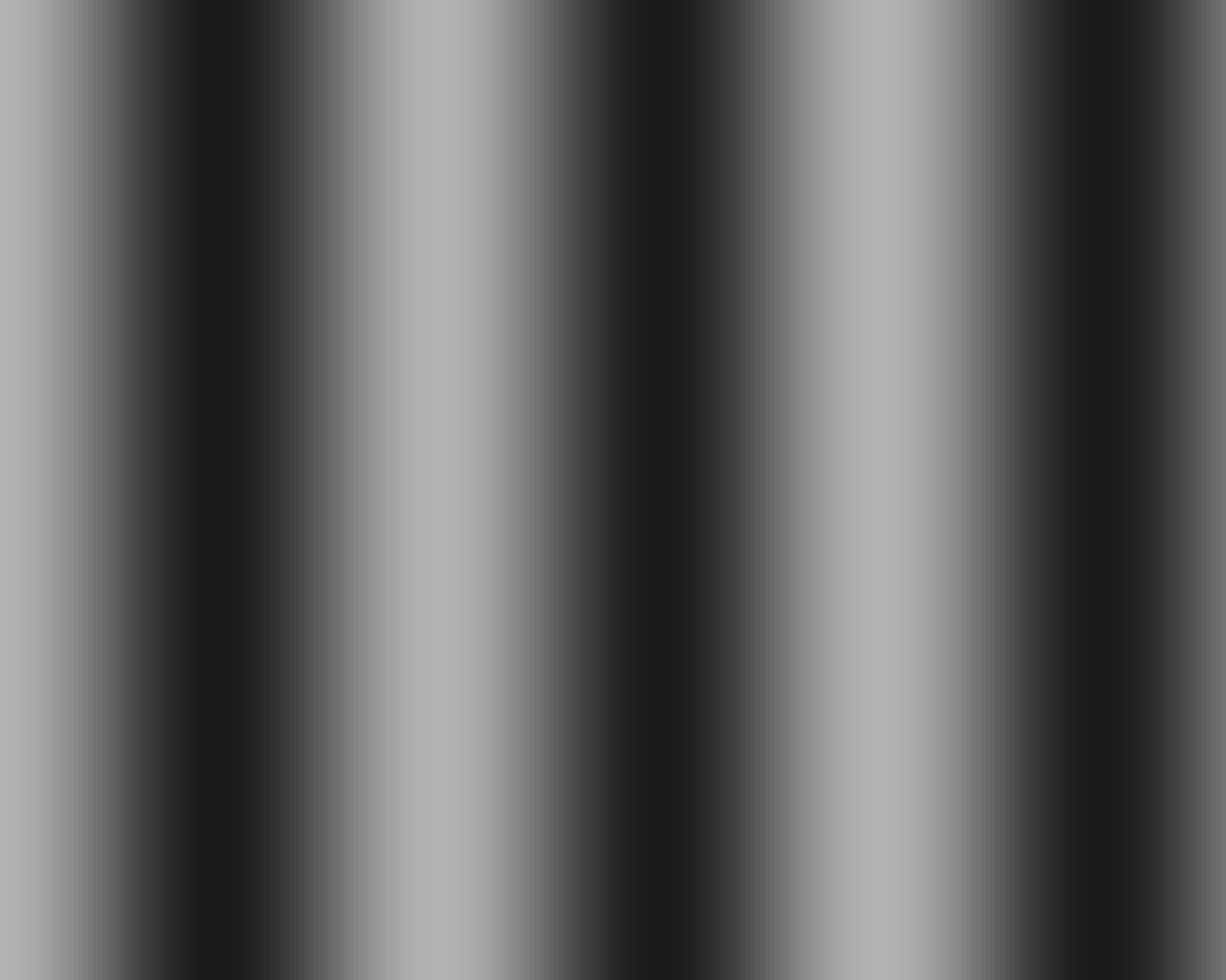}
        \put(-20,10){\colorbox{white}{(d)}}
        \\ \vspace{0.1cm}
        \includegraphics[width=0.24\columnwidth]{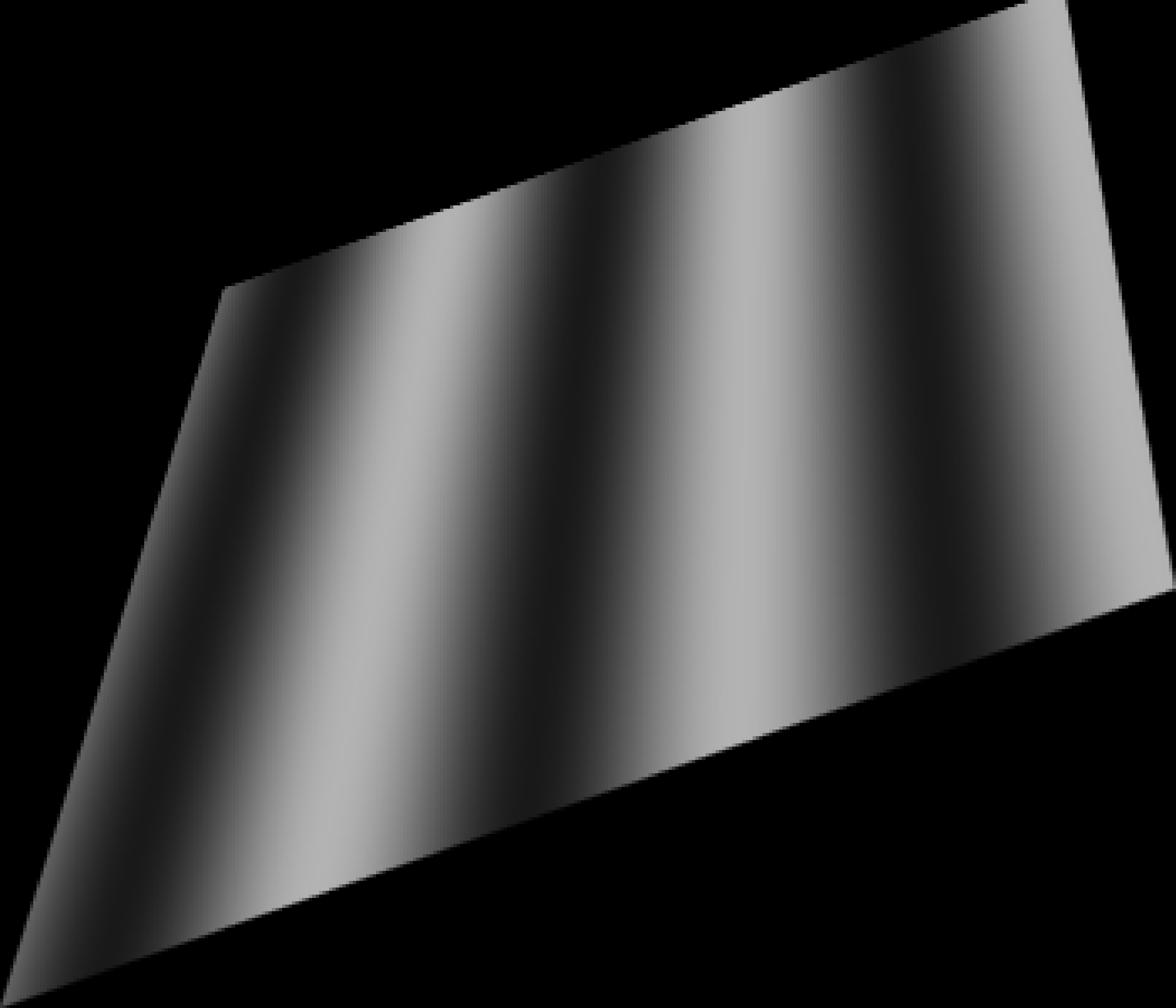}
        \put(-20,10){\colorbox{white}{(e)}}
        \hfill
        \includegraphics[width=0.24\columnwidth]{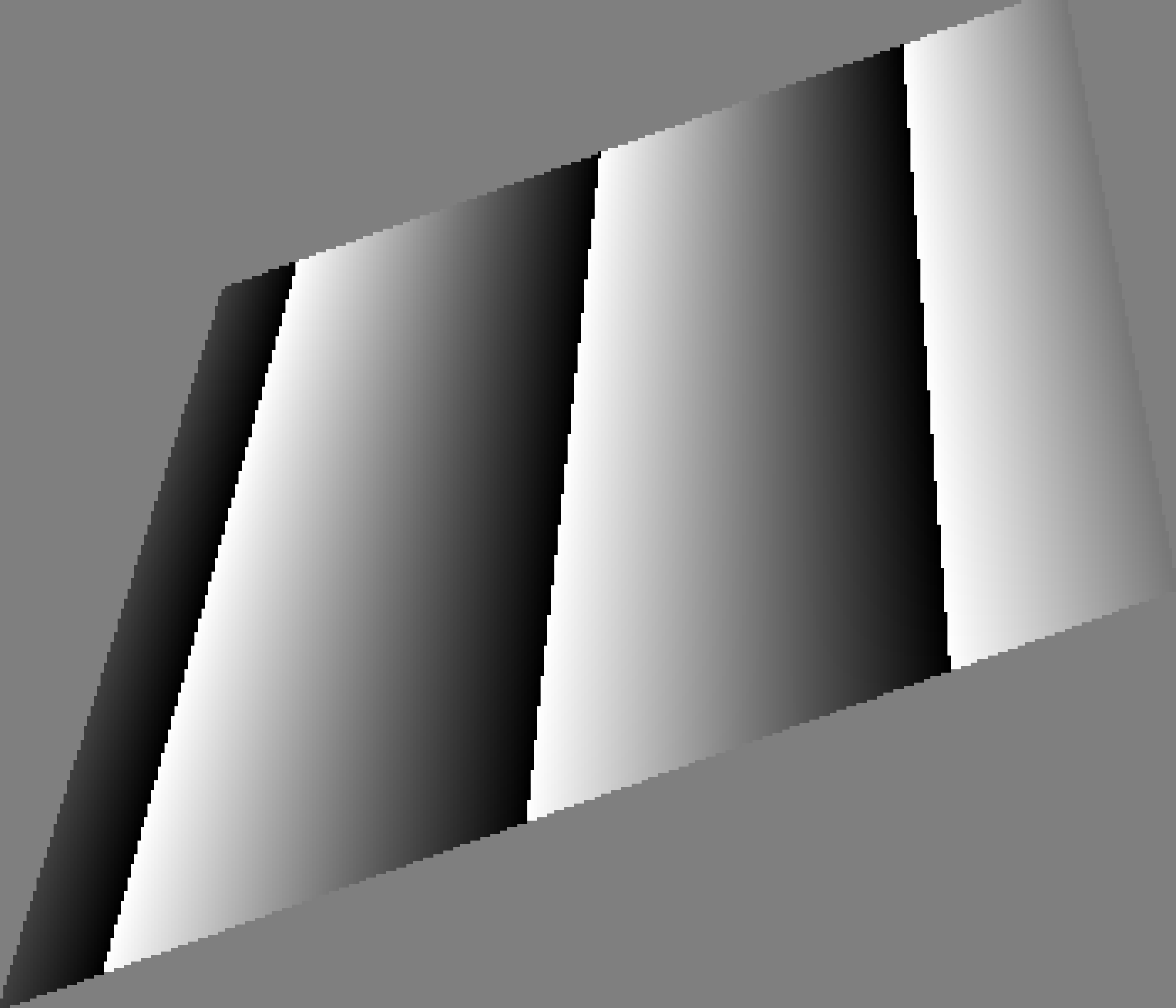}
        \put(-20,10){\colorbox{white}{(f)}}
        \hfill
        \includegraphics[width=0.24\columnwidth]{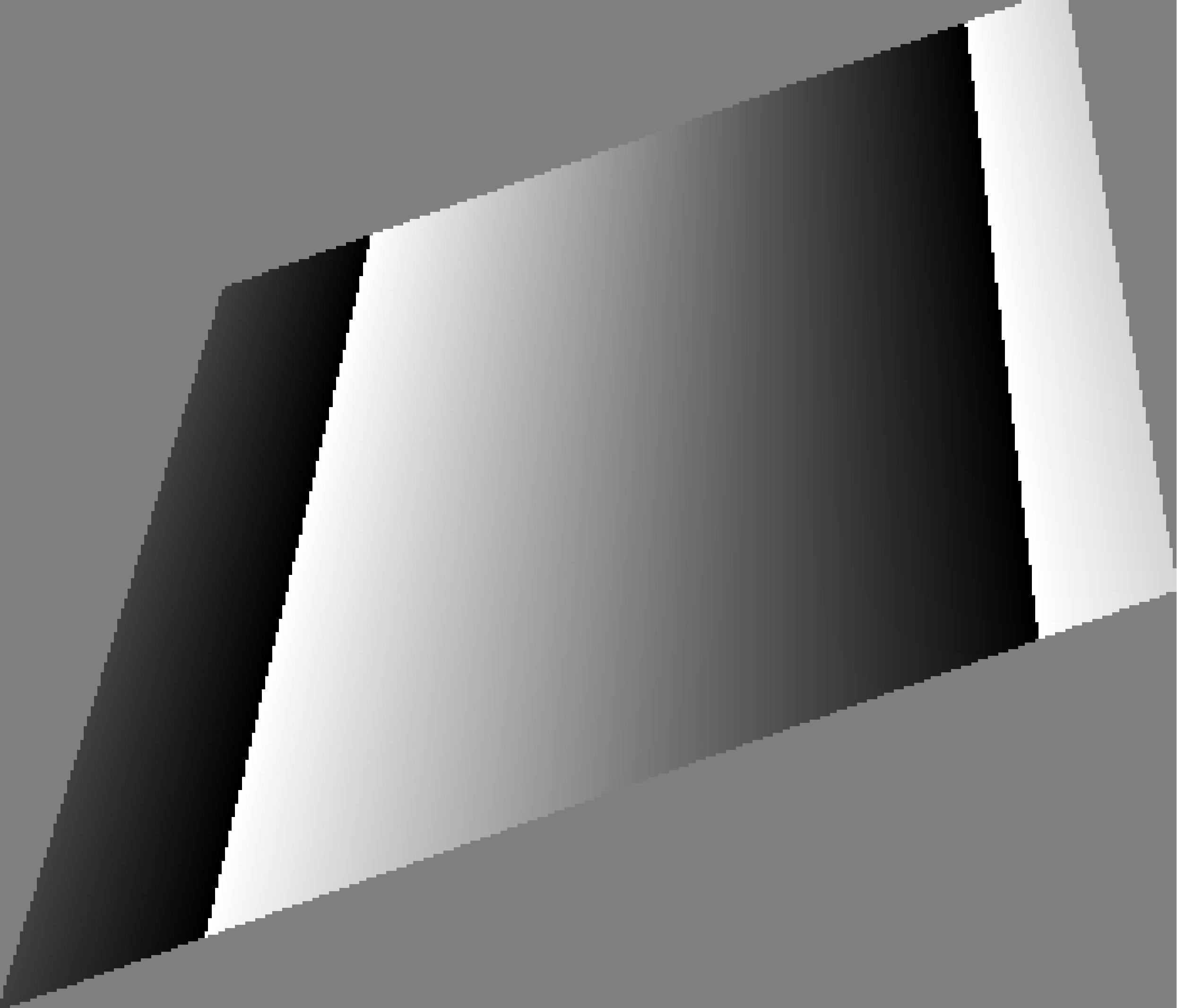}
        \put(-20,10){\colorbox{white}{(g)}}
        \hfill
        \includegraphics[width=0.24\columnwidth]{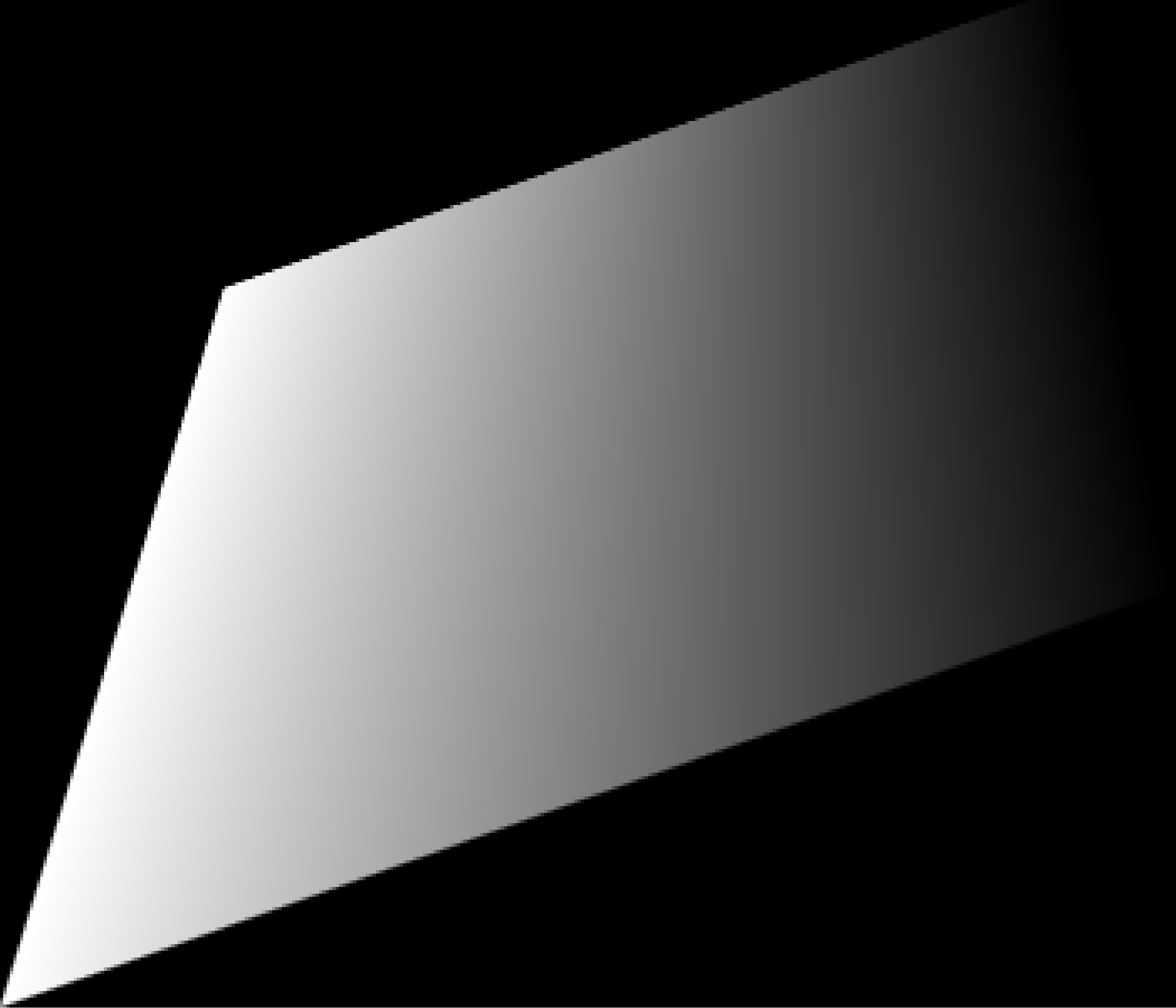}
        \put(-20,10){\colorbox{white}{(h)}}
        \\
        \caption{Encoding with phase-shifted cosine patterns. (a-d): patterns (Eq.~\ref{eqn:encode}) for $N = 4$. (e): a sample (synthetic) camera image corresponding to a reflection of the pattern (a). (f): respective map of $\hat{x}$ decoded according to Eq.~\ref{eqn:decode} (with $m = 0$). (g): map of decoded $\hat{x}$ for a different modulation period $L$. (h): final unambiguously decoded $x$-coordinates. Gray pixel values in (f-h) have been re-scaled for better visibility.}
        \label{fig:patterns}
        \vspace{-0.6cm}
    \end{figure}

    Consider a sequence of $N$ grayscale patterns where the pixel value in the $k$-th pattern at some position $(x, y)$ is
    \begin{align}
        g_k^{\rm screen}(x, y) &= A + B \cos\left(\frac{2\pi x}{L} + \psi_k\right), ~~\psi_k = \frac{2\pi k}{N},
        \label{eqn:encode}
    \end{align}
    where $A$, $B$, and $L$ define the mean brightness, modulation amplitude, and the period, or wavelength (inverse of spatial frequency) of the pattern, respectively. An example of such a sequence is shown in Figs.~\ref{fig:patterns}(a-d).
    
    The patterns are displayed on a flat screen and are eventually observed by a camera as e.g. in the scheme of Fig.~\ref{fig:dm_setup}(b) so that the screen pixel $(x, y)$ maps to a camera pixel $(u, v)$. In the simplest case, the recorded pixel value is
    \begin{align}
        g_k^{\rm camera}(u, v) &= C(u, v) + D(u, v) ~g_k^{\rm screen}(x, y),
        \label{eqn:transfer}
    \end{align}
    where $C$ and $D$ parameterize the transfer function. An example camera image is shown in Fig.~\ref{fig:patterns}(e). It is easy to show that $x$ can be recovered from the sequence of $N$ observations up to a multiple of $L$ as follows:
    \begin{align}
        \hat{x}(u, v) &= \frac{L}{2\pi} \tan^{-1}\left(- \frac{a(u, v)}{b(u, v)}\right) + m L, ~~m\in Z, \nonumber \\
        a(u, v) &= \sum_{k} g_k^{\rm camera}(u, v) \sin\psi_k, ~~\mbox{and} \nonumber \\ 
        b(u, v) &= \sum_{k} g_k^{\rm camera}(u, v) \cos\psi_k. 
        \label{eqn:decode}
    \end{align}
    It is also known that this "DFT" (discrete Fourier transform)-like formula~\cite{Surrel.1996} is the optimal solution in the least-squares sense~\cite{Greivenkamp.1984} when the random noise is independent of the fringe phase~\cite{Surrel.1997}. If the period $L$ is larger than the screen size, the decoding is unambiguous. In practice, though, large $L$ leads to large decoding errors (the reasons are given in \ref{sec:phase_measuring}). Therefore, one usually uses several coding sequences with different periods $L_1$, $L_2$, ... and then recovers the proper coordinates from the respective maps $\hat{x}_1$, $\hat{x}_2$, ... . For example, the sequence of observations as in Fig.~\ref{fig:patterns}(e) results in the decoded $\hat{x}$ values as in Fig.~\ref{fig:patterns}(f) (we set $m = 0$), and another sequence generates a map as in Fig.~\ref{fig:patterns}(g). Together they lead to the final unwrapped coordinates as in Fig.~\ref{fig:patterns}(h). This "disambiguation" is very similar to phase unwrapping in multi-wavelength interferometry.
    
\begin{figure}
        \centering
        \includegraphics[bb=10 10 717 565, clip, width=0.90\columnwidth]{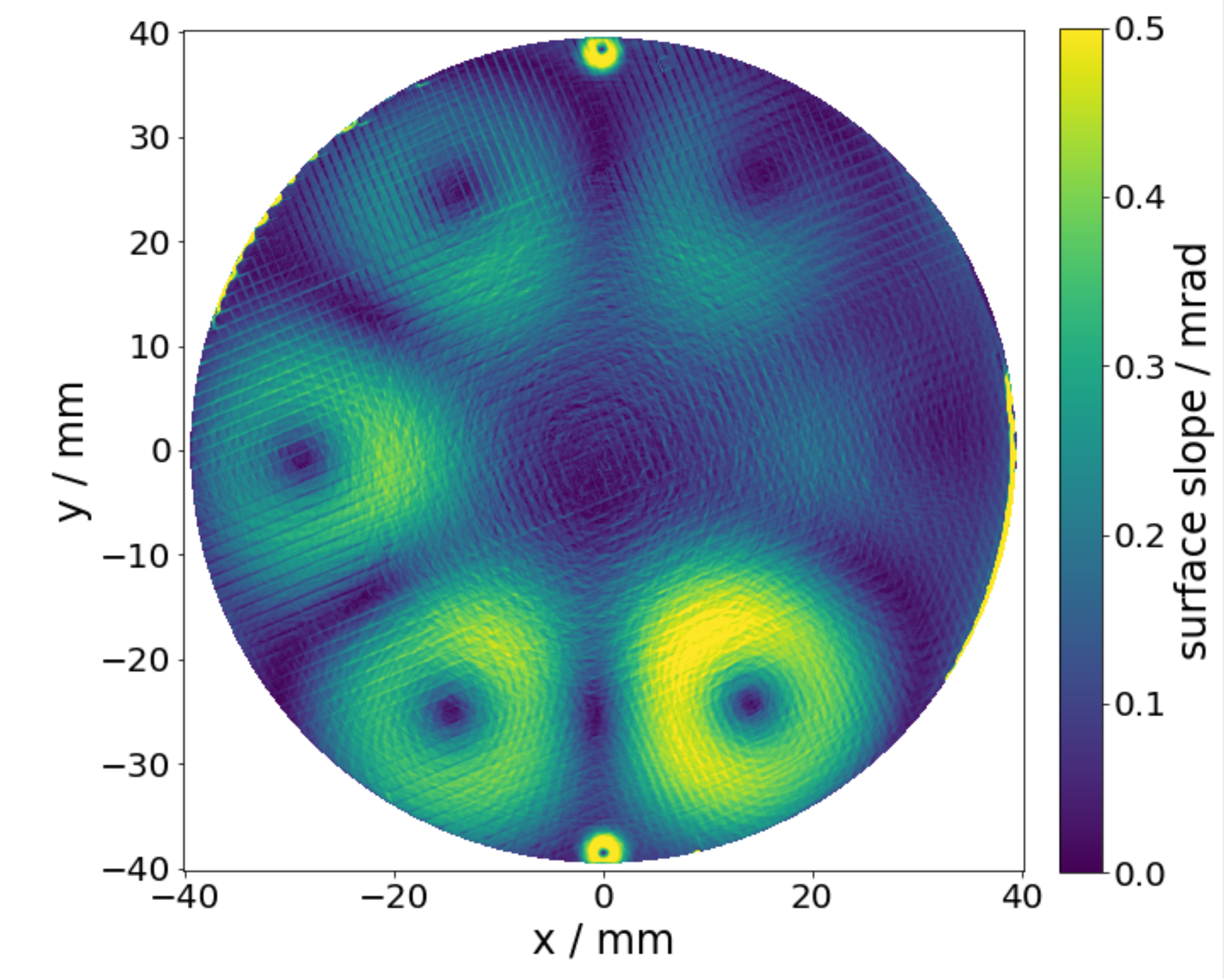}
        \put(0,10){(a)}
        \vspace{-0.0cm}
        \\
        \includegraphics[bb=10 32 680 548, clip, width=0.92\columnwidth]{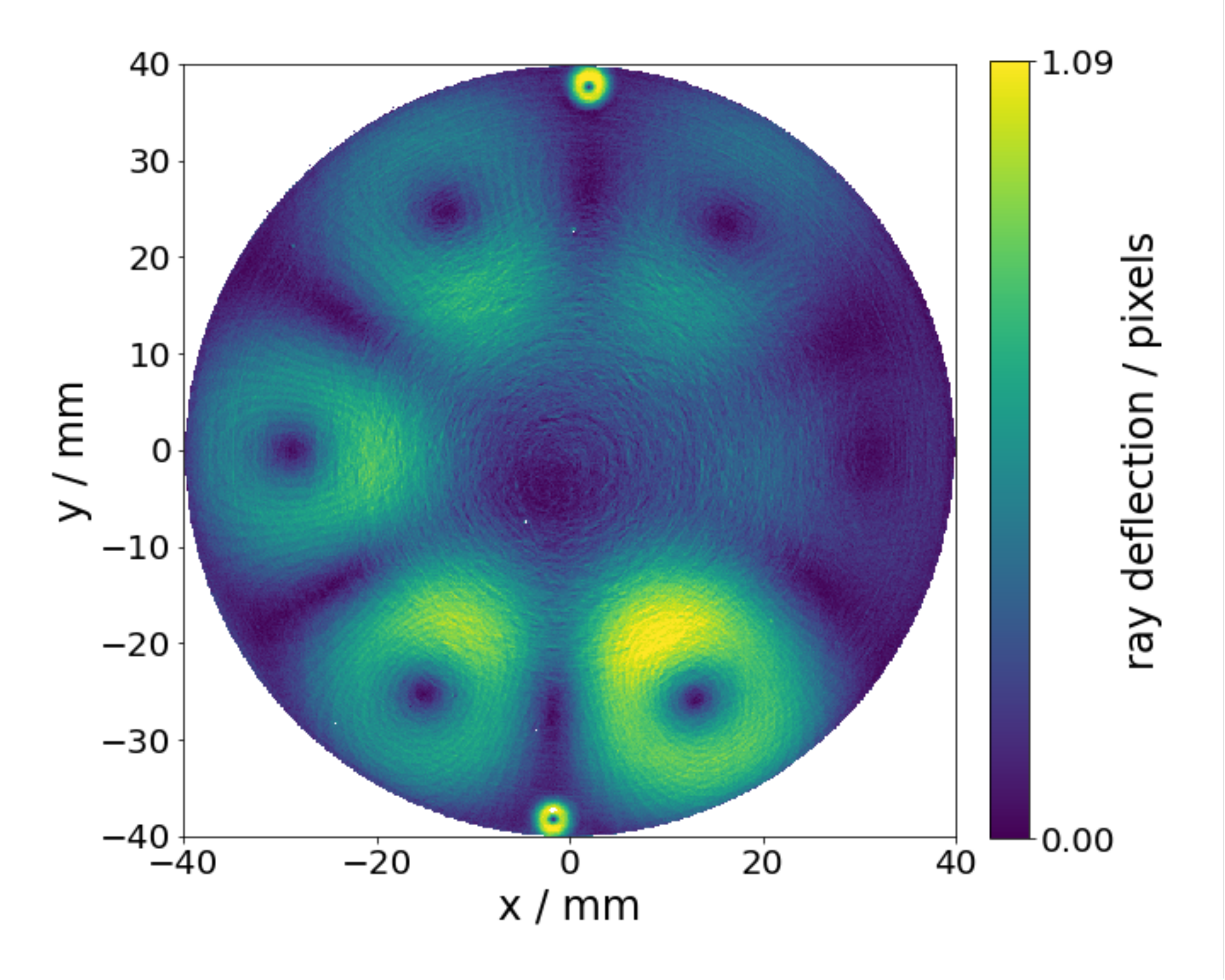}
        \put(0,10){(b)}
        \vspace{-0.0cm}
        \\
        \includegraphics[bb=10 32 680 548, clip, width=0.92\columnwidth]{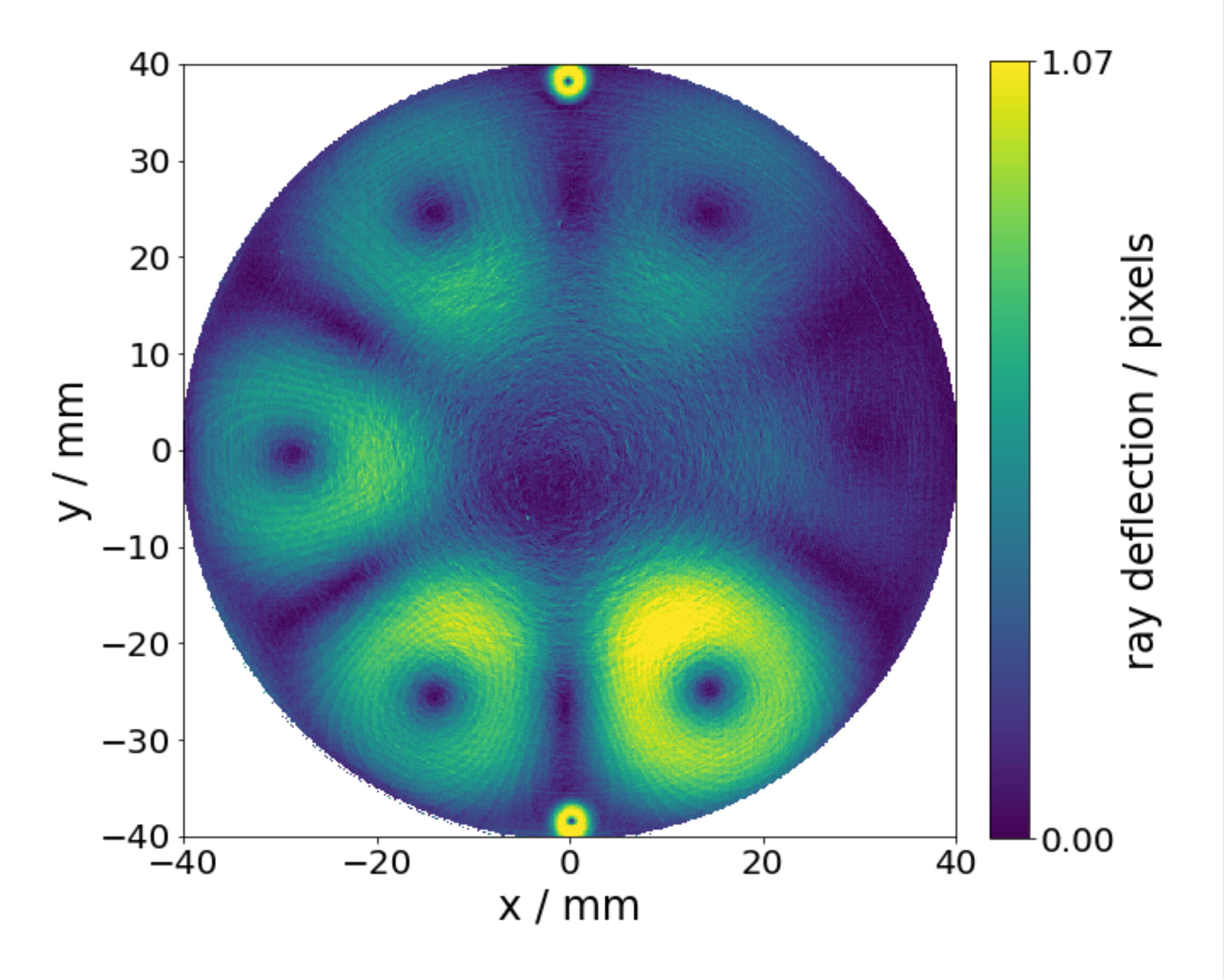}
        \put(0,10){(c)}
        \vspace{-0.1cm}
        \caption{Sensitivity demonstration of phase-measuring DM with a precision-turned slope standard ($\approx$100 GU), featuring six Gaussian peaks of different heights, designed to be at the limit of human detection capability at a distance of several meters and with suitable reflected patterns. The absolute scales are matched within about 2\%. (a) Slopes inferred from a surface map, stitched together from several thousand measurements with a white-light interferometric microscope ~\cite{Ziebarth.2019}: values under 0.1 mrad are detectable. (b) A DM measurement of the same object, the screen to object distance is $\approx$300 mm. (c) A DM measurement in a different instrument with a screen-to-object distance of $\approx$590 mm but larger screen pixels, leading to similar deflections in pixel units. (b) and (c) are uncalibrated, thus the unit is the measured ray deflection in pixels. The small features at the very top and bottom are alignment fiducials.}
        \label{fig:sensitivity_demo}
    \end{figure}

    In a similar fashion one may encode and decode the $y$-coordinates and thus uniquely recover the screen positions corresponding to the modulated camera pixels; in other words, each camera pixel "knows" the screen pixel that it is looking at (in DM, the respective optical path includes the reflection from the surface). As Fig.~\ref{fig:sensitivity_demo} demonstrates, the associated deflections can be measured with an astounding SNR: with a set-up of less than 1 m$^3$ in size and no special precautions during data acquisition, one can easily detect slopes of order 0.1 mrad and outperform human vision.
    
    \subsection{Basic deflectometric setup geometry}
    \label{sec:geometry}
    
\begin{figure}[b]
        \vspace{-0.2cm}
        \centering
        \includegraphics[bb=80 70 540 375, clip, width=0.95\columnwidth]{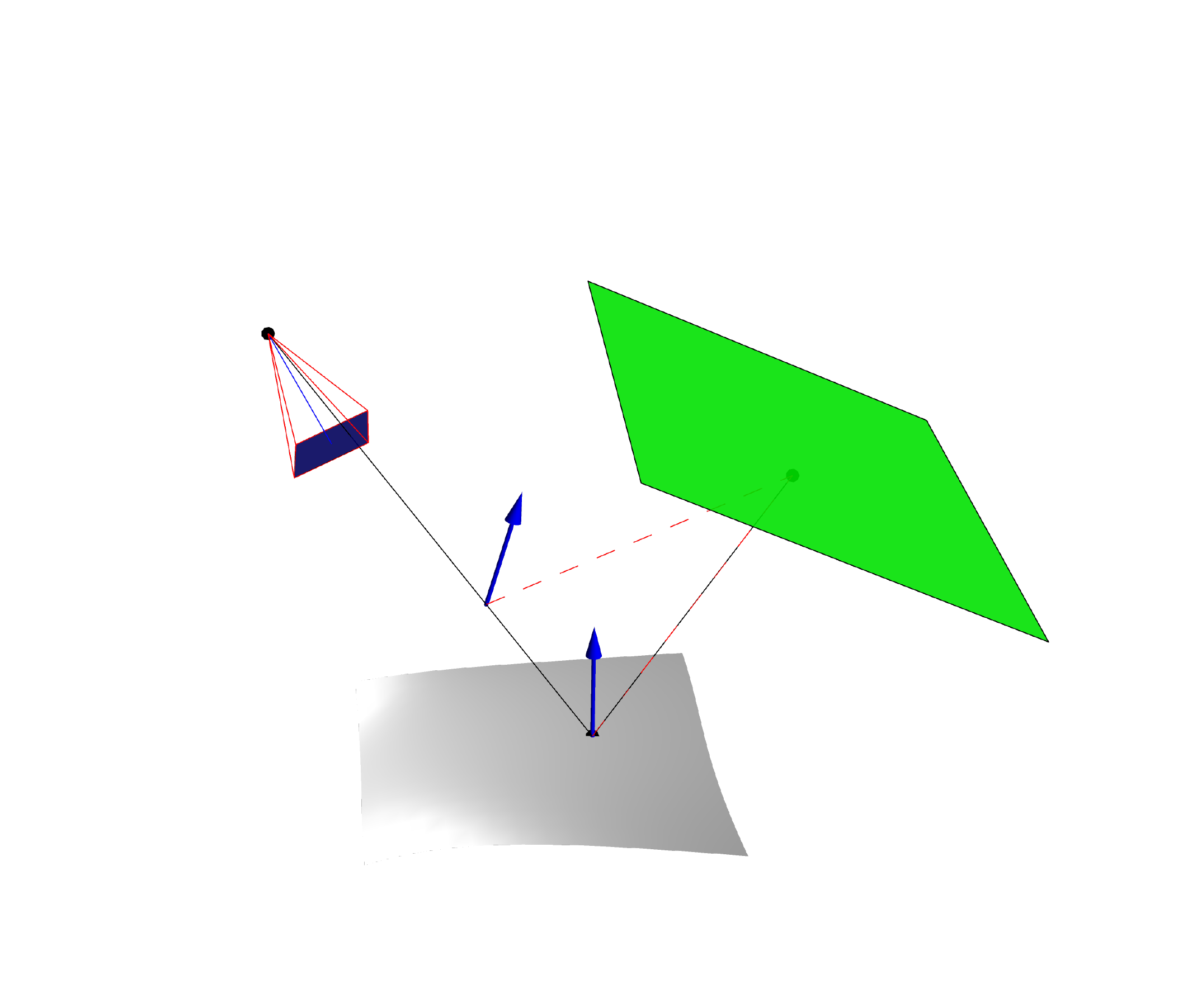}
        \put(-195,138){Camera $C$}
        \put(-217,150){$\vec{o}$}
        \put(-40,130){Screen $P$}
        \put(-75,115){$\vec{m}$}
        \put(-85,20){Surface $S$}
        \put(-240,93){Image plane}
        \put(-127,25){$\vec{p}$}
        \put(-135,69){$\hat{n}(\vec{p})$}
        \put(-165,63){$\vec{p}^{~\prime}$}
        \put(-158,105){$\hat{n}(\vec{p}^{~\prime})$}
        \caption{Geometry of a DM observation corresponding to the basic setup of Fig.~\ref{fig:dm_setup}(b); the description is in the text.}
        \label{fig:geom}
    \end{figure}
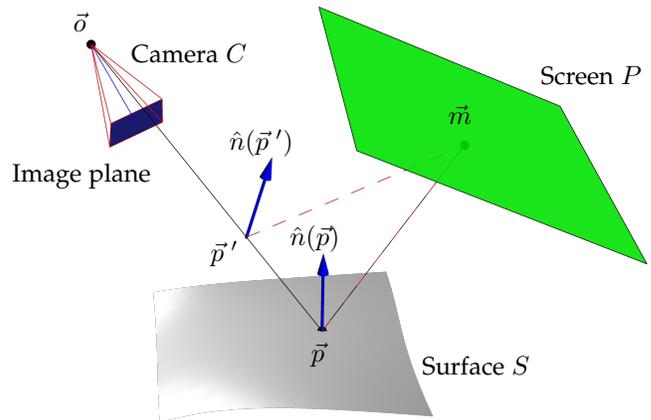

    Consider a DM setup as schematically shown in Fig.~\ref{fig:geom}. A camera $C$ observes a specular surface $S$, in which a flat screen $P$ is reflected. We assume a pinhole camera with the projection center located at point $\vec{o}$ and for simplicity indicate its image plane in front of $\vec{o}$. The processing of camera frames as outlined in Sec.~\ref{sec:cos_patterns} delivers the correspondences between the screen and the camera pixels. In particular, a screen pixel at point $\vec{m}$ emits light that reflects from the mirror at point $\vec{p}$ and arrives at the camera; equivalently, a "view ray" is emitted by the camera, hits the mirror at $\vec{p}$, and upon reflection ends up at $\vec{m}$. If we denote the local unit normal vector to $S$ as $\hat{n}(\vec{p})$, the reflection law states
    \begin{align}
        \hat{n}(\vec{p}) &= \frac{\vec{n}(\vec{p})}{\|\vec{n}(\vec{p})\|}
        ~~\mbox{with}~~
        \vec{n}(\vec{p}) = \frac{\vec{o} - \vec{p}}{\|\vec{o} - \vec{p}\|} + \frac{\vec{m} - \vec{p}}{\|\vec{m} - \vec{p}\|}.
        \label{eqn:reflaw}
    \end{align}
    Eq.~\ref{eqn:reflaw} in fact defines a unit vector not only at $\vec{p}$ on the surface $S$: if a mirror element is placed at some other point $\vec{p}^{~\prime}$ on the view ray and is orthogonal to $\hat{n}(\vec{p}^{~\prime})$, then the recorded data for the given camera pixel will remain unchanged. Therefore, a (dense) recorded dataset for all camera pixels induces via Eq.~\ref{eqn:reflaw} a volumetric {\itshape normal field} in the view frustum of the camera. The surface reconstruction problem then is equivalent to the {\itshape integration} of this field, or finding a surface in space that is at all points orthogonal to it. Whether such a surface is unique, and if so, how to find it, are two non-trivial questions that will be discussed later.
    
    \subsection{Component models and calibration}
    \label{sec:calibration}
    
    A measurement as in Fig.~\ref{fig:geom} requires an accurate characterization of the camera and the screen; in addition, one needs to establish their relative positions and orientations in space. (In some cases, one also needs to know the location of the object.) The quality (uncertainty) of the respective models and parameters should be consistent throughout the setup and adequate to the requirements of the task at hand. In what follows we briefly mention several tools and techniques that have proven useful in DM applications.

        \subsubsection{Camera models}
        \label{sec:camera_models}
        
        In terms of Fig.~\ref{fig:geom}, a camera model defines a mapping from image pixels to the view rays in the camera's own reference frame. Once the respective (intrinsic) parameters are fixed, one only needs six extrinsic parameters (the 3D camera position and three Euler's rotation angles) to completely determine all the view rays needed for the reconstruction. Camera calibration (estimation of intrinsic and extrinsic parameters) is a fundamental problem and we cannot possibly discuss it in depth here; some techniques often employed in the context of DM are as follows:
        \begin{itemize}
            \item A central-projection camera with low-order polynomial distortions is the dominant model in computer vision (perhaps in part due to support by popular libraries~\cite{Bradski.2000}). Until about 2005, this model was frequently used in DM~\cite{Gruen.2001, Salvi.2002, Remondino.2006}; the calibration relied on static checkerboard targets~\cite{Tsai.1987, Zhang.2000b} that yielded a sparse array of detected points. In more demanding applications, it has now been replaced by generic camera calibration techniques.
                
            \item An undistorted central projection model (with an actual pinhole aperture) is used in extremely demanding applications where it simplifies calculations and improves reconstruction results~\cite{Tang.2009, Su.2012b, Su.2013c}. However, the position of the pinhole with respect to the sensor must be controlled very carefully, and measurements can take many minutes due to the very small aperture. By contrast, the calibration is trivial and can be implemented in many ways~\cite{Huang.2015c}.
            
            \item Various generic camera models aim to accurately reproduce the imaging geometry of real optics; in particular, one can allow and account for non-central projection of complex multi-lens objectives and define slightly different origin points for each view ray. Another important use-case is non-standard optics (omnidirectional, telecentric, catadioptric, etc.). In precision DM, generic and model-free approaches have been shown to reduce measurement errors even for paraxial imaging, so the topic has attracted a lot of attention~\cite{Sturm.2004, Wang.2005, Grossberg.2005, Kannala.2006, Barreto.2008, Ramalingam.2010, Luber.2010, Bothe.2010b, Rosebrock.2012, Xiang.2013, Pak.2016, Pak.2016b, Prinzler.2018, Schops.2020, Uhlig.2021, Gauchan.2021}.
        \end{itemize}
        
        Any of these calibrations will benefit from better-quality input data, and printed checkerboards are not the best solution in existence (generic models, for instance, cannot rely on such sparse and noisy datasets). Ideally, calibration would require a dense collection of reference points~\cite{Sagawa.2005, Schmalz.2011, Huang.2013f, Ma.2014, Xu.2017}. Coded pattern sequences, and particularly the phase shifting technique (Sec.~\ref{sec:coding}), nicely fit the bill here and are often used in practice.
        
        \subsubsection{Coding screen models}
        \label{sec:screen_models}
                            
        The reference screen (most often, a flat LC display) is typically modeled as a plane with pixels arranged as a rectangular grid. Scale calibration is possible even by a direct manual measurement of active area sizes - with more than 1000 pixels in each screen direction, this yields \si{\micro\metre} uncertainty for the pixel sizes even in absence of appropriate manufacturer specifications. In more demanding cases, the details of a real LCD matrix have to be taken into account. In particular, the refraction in the cover glass may be modeled explicitly~\cite{Huang.2015c, MaestroWatson.2017, Li.2018, Bartsch.2018, Petz.2019, Petz.2020} or accounted for by some low-order deformation terms~\cite{Petz.2004, Petz.2005c, Reh.2014, Bartsch.2019}. The latter may also describe the deformation of large screens due to gravity. The estimation of respective parameters requires a dedicated measurement or may be implemented in parallel with the surface reconstruction as part of a global optimization.
            
        Other effects that may contribute to the decoding error include "jitter" in the detected pixel positions due to random scattering through the rough cover glass, and moiré-type interference between the pixel grids of the screen and the camera (both have a fill-factor below unity). We are not aware of any attempts to model and correct these effects; instead, one typically adjusts the setup in order to suppress them.
        
        When a measurement task necessitates reference pattern sizes larger than the available LC displays, one possible solution is a projection screen illuminated by a digital projector, as shown in Fig.~\ref{fig:geom}(c). Such systems may in many possible ways deviate from a flat or otherwise regular grid of pixels; the efforts to characterize them strongly depend on the target uncertainty goals~\cite{Horbach.2009, Hornung.2014}.
        
        \subsubsection{Relative component positions and orientations}
        \label{sec:component_positions}
        
        When setting up a DM measurement, one points the camera at the object and then rotates and moves setup pieces until the camera can see the reflected screen. Optics may also need to be adjusted; for instance, a short-focus lens observes a larger surface patch from a shorter distance (i.e., the setup is compact). However, if the surface is convex, it magnifies the (already large) spread of reflected view rays, requiring the coding screen to cover an even larger solid angle around the object; in this situation, one may prefer a longer-focus lens placed further from the object, as this narrows down the spread of rays again. Simple measurements can be aligned by hand; efficient multi-camera and/or multi-screen arrangements for complex surfaces need to be found via simulation-based optimization, which is a topic of active research (Sec.~\ref{sec:planning}). This complexity is due to an (unknown) object being part of the optical scheme: unlike e.g. a laser triangulation sensor, one cannot fine-tune a universal DM setup once and for all.
            
        Once a viable configuration is fixed, it has to be characterized, which in the simplest case means finding the 3D camera position relative to the screen (six extrinsic parameters). Some DM applications tolerate a relatively low-quality setup calibration, which can be performed with the simplest means. This is often the case when measuring small differences with respect to a reference object (e.g., surface deformations~\cite{Li.2014f, Kim.2021, Quach.2022}), or even shearing measurements followed by a reconstruction of a quasi-flat surface~\cite{E.2016}. However, in general the calibration uncertainty impacts the surface reconstruction quality in a non-trivial way.
            
        Finding a camera position in space is again a well-known problem in computer vision. If the camera has a direct view of the screen, the task reduces to a straightforward bundle adjustment based on camera and screen parameters and a dataset collected with the same coding technique that is used in DM. Unfortunately, in most cases this option is excluded. Many proposed solutions to this problem therefore require an additional reflective object that optically couples the camera to the screen~\cite{Knauer.2002, Bonfort.2006b, Rodrigues.2010, Hofer.2010, Werling.2011}.
        
        This new object, in turn, introduces new uncertainties, and strategies to cope with them may be quite diverse. For instance, a calibration object may have an arbitrary geometry or be a precision mirror with a simple (flat or spherical) shape~\cite{Sigrist.2015, Han.2019, Niu.2020}, it may or may not carry visual markers~\cite{Knauer.2004c, Petz.2005c, Rose.2009, Xiao.2012b}, etc. Calibration procedures may involve accurately moving and tilting this object~\cite{Zhou.2016b}, referencing the measurements against a precision mirror in nominally identical position~\cite{E.2017}, recording the setup with multiple cameras in order to reduce ambiguity~\cite{Ren.2015b, Li.2018b}, or generating a synthetic reference pattern to guide the alignment~\cite{Kang.2021}.
            
        An alternative approach that potentially may reduce the need for a precision calibration object (or eliminate it entirely) is to include these six parameters in the "global optimization loop" so that the object reconstruction and the setup calibration happen in parallel within some "holistic" or probabilistic framework. As the cost function to minimize, one often chooses ray re-projection errors or similar metrics~\cite{Olesch.2010, Olesch.2011, Rapp.2012, Faber.2012, Ren.2015b, Xu.2018c, Allgeier.2020}. As a price, however, one may need to record several object poses per measurement.
        
        In summary, the broad spectrum of ideas outlined above indicates that the questions related to achieving and maintaining a stable calibration in a deflectometric setup (as well as understanding the effects of calibration errors on the surface reconstruction outcomes) are far from settled and will keep the community occupied for years to come. 

    \subsection{Surface reconstruction}
    \label{sec:reconstruction}
    
    The task of reconstructing a function from its gradients predates DM in its current form: the well-known Hartmann test is also a gradient technique whose most demanding application is adaptive optics for telescopes, and has recently also been utilized to monitor deformations while coating mirrors~\cite{Arnoult.2021}. Another major driver for wavefront reconstruction methods is shearing interferometry. As mentioned above, the integration aims to find a surface consistent at all points with the volumetric normal field $\hat{n}(\vec{p})$ of Eq.~\ref{eqn:reflaw} induced by a measurement.

    The slope reconstruction is based on the knowledge of spatial relationships between the reference structure, the tested object, and the camera. Most frequently (but not necessarily) the reference structure is a periodical pattern, and in order to find a unique mapping between the reference and the sensor coordinates one has to use a tandem of encoding and decoding (e.g. as in Sec.~\ref{sec:cos_patterns}). There is no shortage of summaries on absolute position coding~\cite{Salvi.2004, Hofer.2013b, Zhang.2018, Falaggis.2018, Gupta.2018}; we mention some possible procedures in Sec.~\ref{sec:coding}.

        \subsubsection{Integrability condition and direct reconstruction}
        \label{sec:integrability}
        
        If we parameterize the surface shape e.g. as a function of camera pixel coordinates $(u, v)$ and impose the consistency condition on its second derivatives, then the surface position ambiguity of Fig.~\ref{fig:sensitivity}(c) disappears and one can explicitly find the surface point $\vec{p}(u, v)$ on the respective view ray as a function of $\vec{m}(u, v)$, $\partial_u \vec{m}(u, v)$, and $\partial_v \vec{m}(u, v)$~\cite{Liu.2013d, Pak.2014}.
        
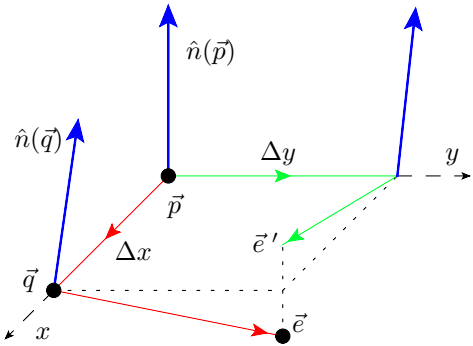
\begin{figure}
            \centering
            \includegraphics[width=0.7\columnwidth]{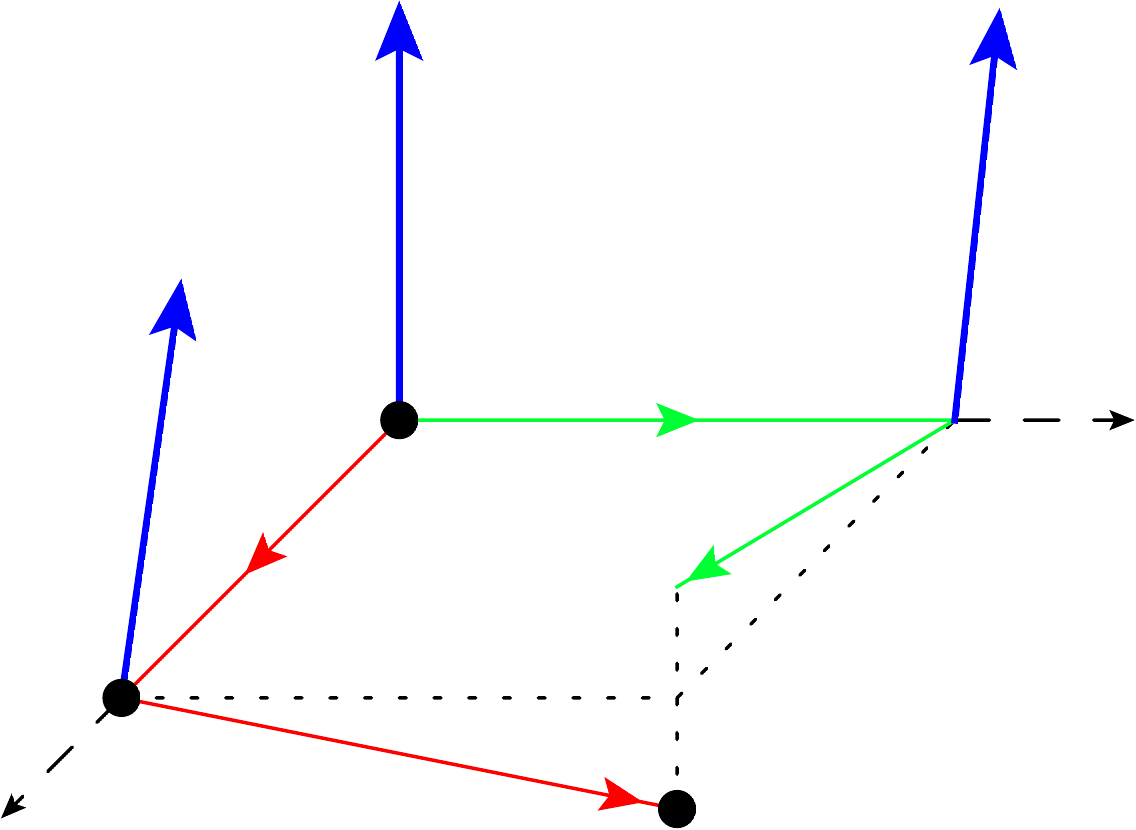}
            \put(-165,2){$x$}
            \put(-10,70){$y$}
            \put(-108,108){$\hat{n}(\vec{p})$}
            \put(-173,75){$\hat{n}(\vec{q})$}
            \put(-115,50){$\vec{p}$}
            \put(-170,22){$\vec{q}$}
            \put(-68,5){$\vec{e}$}
            \put(-83,35){$\vec{e}^{~\prime}$}
            \put(-135,32){$\Delta x$}
            \put(-80,70){$\Delta y$}
            \caption{Local integrability of the normal field. Let us choose coordinates where the induced normal vector at point $\vec{p} = (0, 0, 0)^T$ is $\hat{n}(\vec{p}) = (0, 0, 1)^T$. If we move first along the x-, and then y-directions by infinitesimal steps $\Delta x$ and $\Delta y$ while remaining orthogonal to the normal field (red path), we end up at the point $\vec{e} = \left(\Delta x, \Delta y, \Delta x \Delta y \left(\hat{u}^T_y \frac{\partial \hat{n}}{\partial x}\right)\right)^T$, where $\hat{u}_y = (0, 1, 0)^T$. Alternatively, if we follow the green path, we end up at $\vec{e}^{~\prime} = \left(\Delta x, \Delta y, \Delta x \Delta y \left(\hat{u}^T_x \frac{\partial \hat{n}}{\partial y}\right)\right)^T$ with $\hat{u}_x = (1, 0, 0)^T$. If the normal field is integrable at $\vec{p}$, these points must coincide. Eq.~\ref{eqn:integrability} generalizes this argument.}
            \label{fig:no_long_rot_comp}
            \vspace{-0.5cm}
        \end{figure}

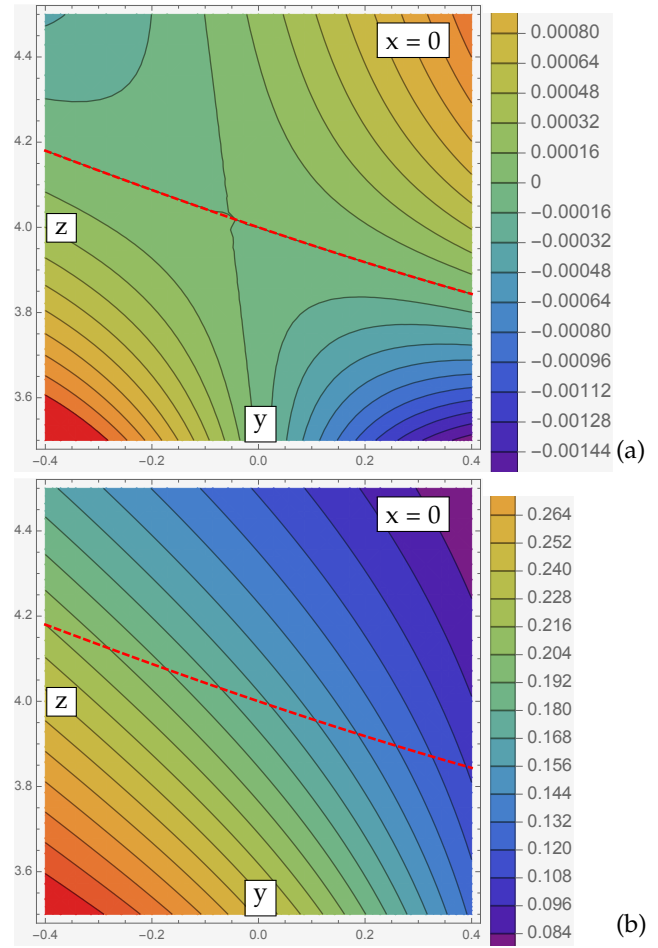
\begin{figure}
            \centering
            \includegraphics[width=0.7\columnwidth]{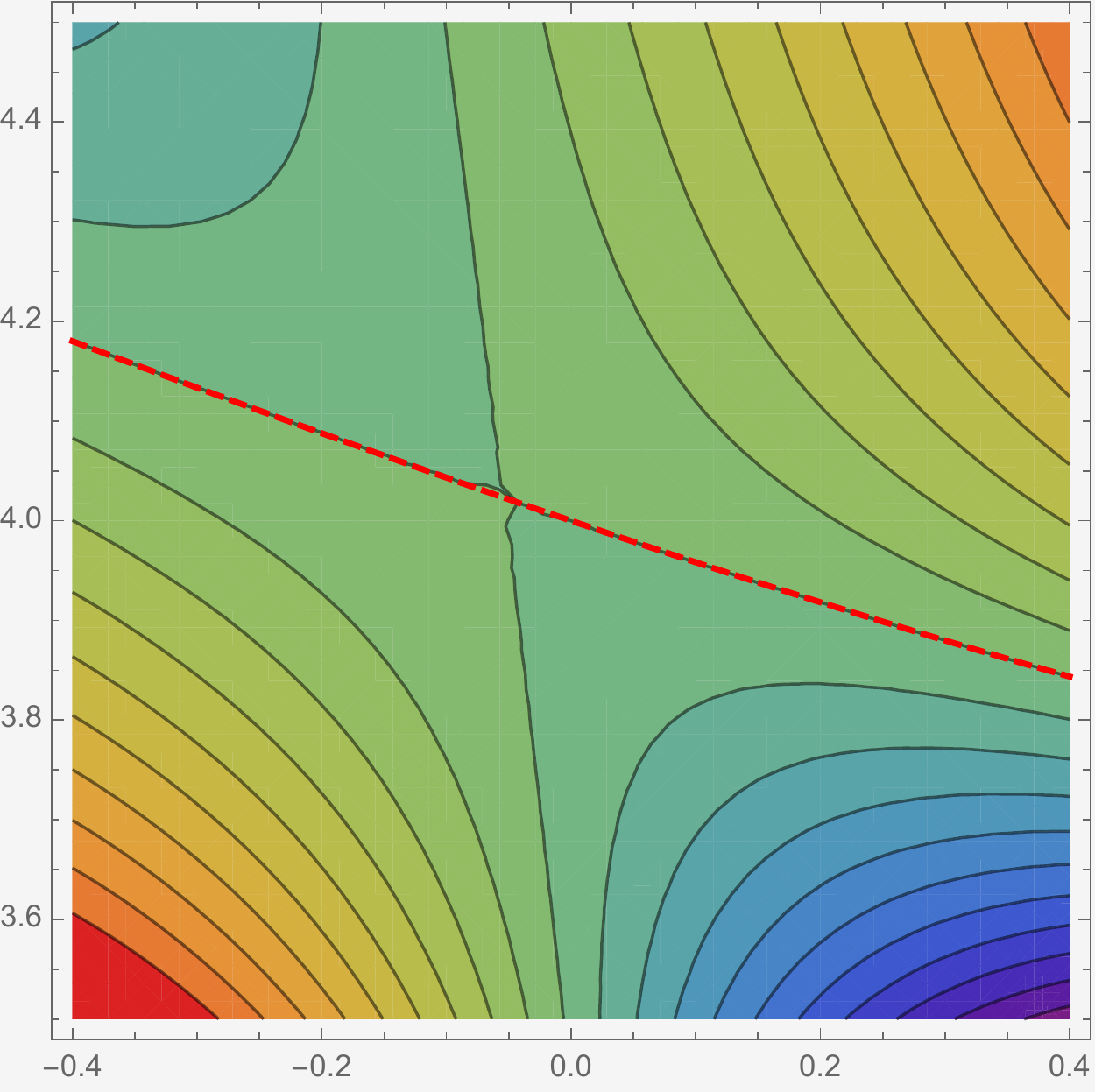}
            \put(-90,17){\fcolorbox{black}{white}{y}}
            \put(-165,90){\fcolorbox{black}{white}{z}}
            \put(-40,160){\fcolorbox{black}{white}{x = 0}}
            \includegraphics[bb=0 45 70 285, clip, width=0.2\columnwidth]{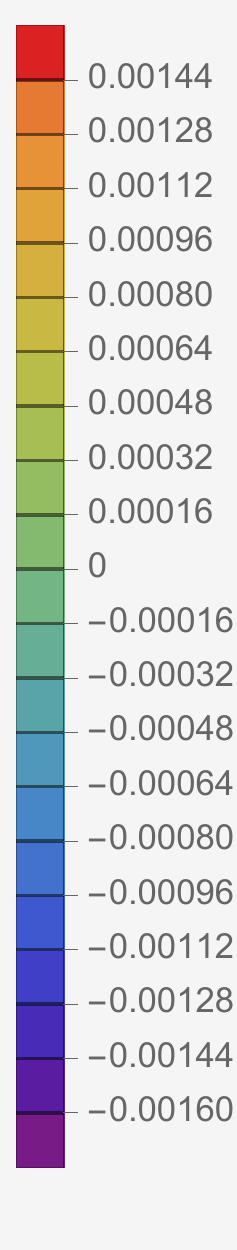}
            \put(0,5){(a)}
            \\ \vspace{0.1cm}
            \includegraphics[width=0.7\columnwidth]{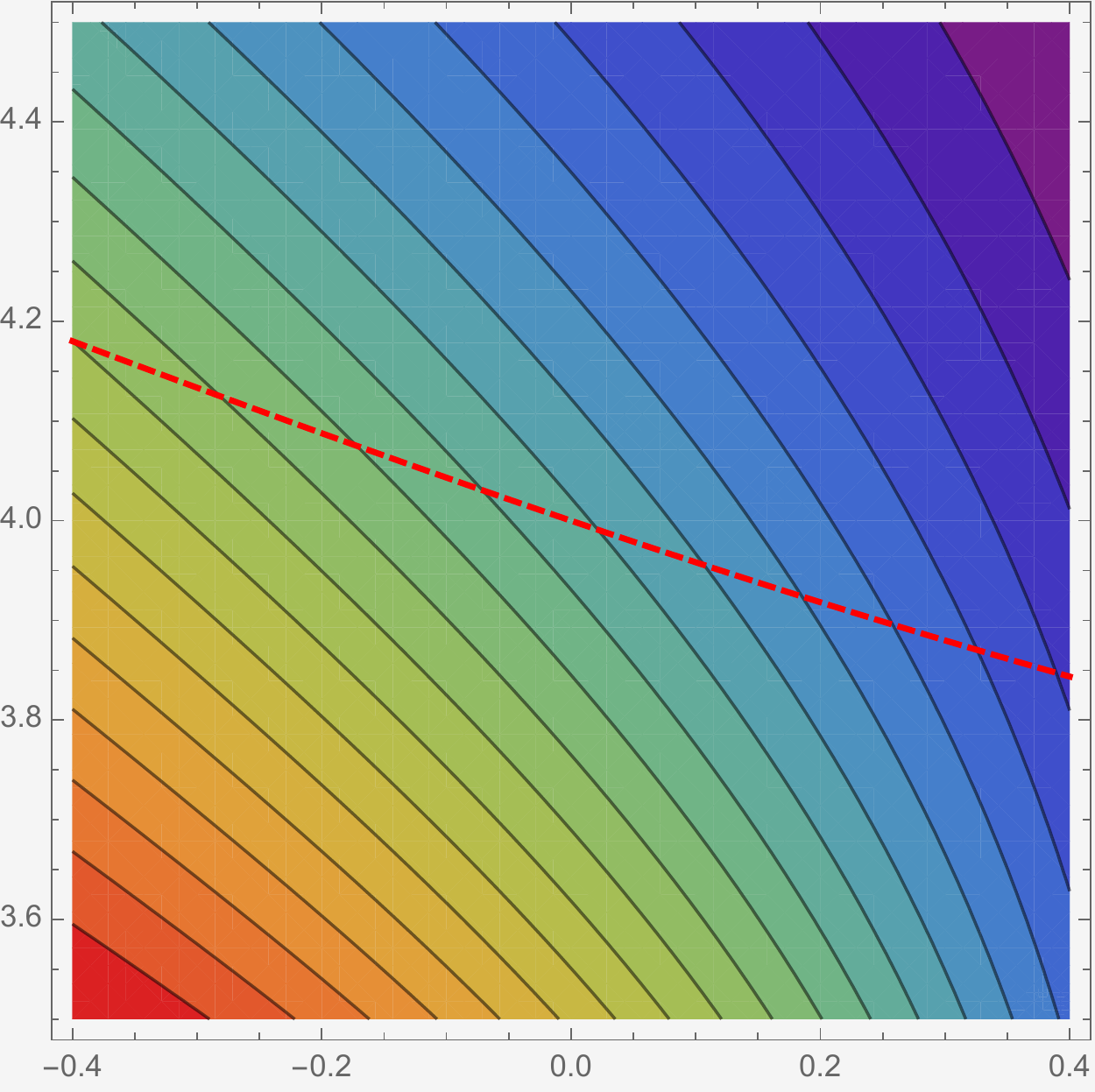}
            \put(-90,17){\fcolorbox{black}{white}{y}}
            \put(-165,90){\fcolorbox{black}{white}{z}}
            \put(-40,160){\fcolorbox{black}{white}{x = 0}}
            \includegraphics[bb=0 33 75 285, clip, width=0.2\columnwidth]{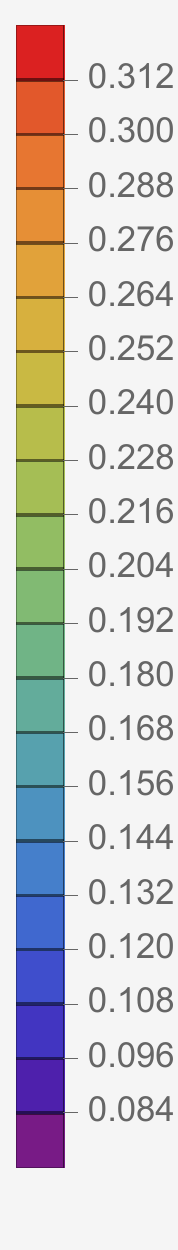}
            \put(0,5){(b)}
            \caption{Longitudinal component $\hat{n}^T(\vec{\nabla}\times\hat{n})$ (a) and the absolute value of the rotation $\|\vec{\nabla}\times\hat{n}\|$ (b) for the setup of Fig.~\ref{fig:geom}. The fields are evaluated over the plane $x = 0$ defined in the camera's system of coordinates. The red dashed line indicates the true position of the reflective surface. The coordinates and derivatives are defined in terms of the same length units. Note that $\hat{n}^T(\vec{\nabla}\times\hat{n}) = 0$ on the true surface.}
            \label{fig:rotn}
            \vspace{-0.6cm}
        \end{figure}

        Equivalently, the existence of a solution (a surface) at some point in space means that the longitudinal component of the normal field rotation $\vec{\nabla}\times\hat{n}$ must vanish~\cite{Pak.2014}:
        \begin{align}
            \hat{n}^T(\vec{\nabla}\times\hat{n}) &= 0,
            \label{eqn:integrability}
        \end{align}
        where we have suppressed the function arguments for clarity; Figs.~\ref{fig:no_long_rot_comp}
        and~\ref{fig:rotn} illustrate this result. (Note, however, that the rotation $\vec{\nabla}\times\hat{n}$ itself does not vanish in the general case.)
        
        Thus, the aforementioned "depth ambiguity" in DM is in fact spurious, and no integration should be necessary to recover the surface. In practice, however, normal fields induced by smooth objects are such that their longitudinal rotation components are numerically small even far away from the true surface (cf. the scale in Fig.~\ref{fig:rotn}(a)). The direct surface reconstruction then becomes very sensitive to noise, and the point-wise solutions -- unstable. As a remedy, one either needs to impose strong assumptions on the surface shape~\cite{Liang.2019} and/or combine Eq.~\ref{eqn:integrability} with some global optimization scheme for robustness~\cite{Zhao.2016, Graves.2018b}. We believe that Eq.~\ref{eqn:integrability} should best be used as an explicit regularization for some integration technique; the potential of such schemes for high-precision measurements is yet to be explored.
        
        \subsubsection{PDE-based reconstruction and regularization}
        \label{sec:regular}
        
        The surface reconstruction problem can in many ways be formulated as a system of partial differential equations (PDEs)~\cite{McGillem.1974, Freischlad.1992, Klette.1996, Elster.1999b, Elster.2000, Li.2004, Moreno.2005, Velghe.2005b, Agrawal.2006, Balzer.2008, Moreno.2013, Huang.2015b}. One simple approach~\cite{Balzer.2011, Werling.2011}, for instance, uses Helmholtz's theorem to decompose  a normal field as $\hat{n} = \vec{\nabla}\phi + \vec{\nabla}\times\vec{A}$, where $\phi$ and $\vec{A}$ are some scalar and vector potentials. Differentiating this relation, we obtain 
        \begin{align}
            \vec{\nabla}^T\hat{n} &= \Delta \phi
            \label{eqn:poisson}
        \end{align}
        (a Poisson's equation), where the left-hand side is derived from the data, and the right-hand side linearly depends on a scalar function $\phi$. After imposing some technical boundary conditions, we may solve Eq.~\ref{eqn:poisson} with e.g. finite element methods and find $\phi(\vec{p})$ in some volume of interest. Candidate solutions (reconstructed surfaces) can then be identified with isosurfaces $\phi(\vec{p}) = c$. In order to select a single solution from this family (parameterized by $c$) one needs additional information, known as {\itshape regularization}: e.g., a known point belonging to the true surface, or an observation by a different camera (stereo-deflectometry).
        
        Where does this ambiguity originate from? Inspecting Eq.~\ref{eqn:poisson}, we notice that by construction it ignores the curl component in $\hat{n}$ which is essential for Eq.~\ref{eqn:integrability} (i.e., strictly speaking, the found isosurfaces do not integrate the field $\hat{n}$).
        
        Other (first-order) PDE formulations that directly fit surface gradients formally remain sensitive to the field rotation and should therefore converge to a unique result. However, the respective cost functions for displaced surfaces are proportional to the longitudinal rotation components of the normal field, which, as discussed above, are often very small. The optimization then in practice must be additionally constrained (regularized) based on available auxiliary measurements or prior knowledge about the studied shape.
        
\begin{figure}
            \centering
            \includegraphics[width=0.9\columnwidth]{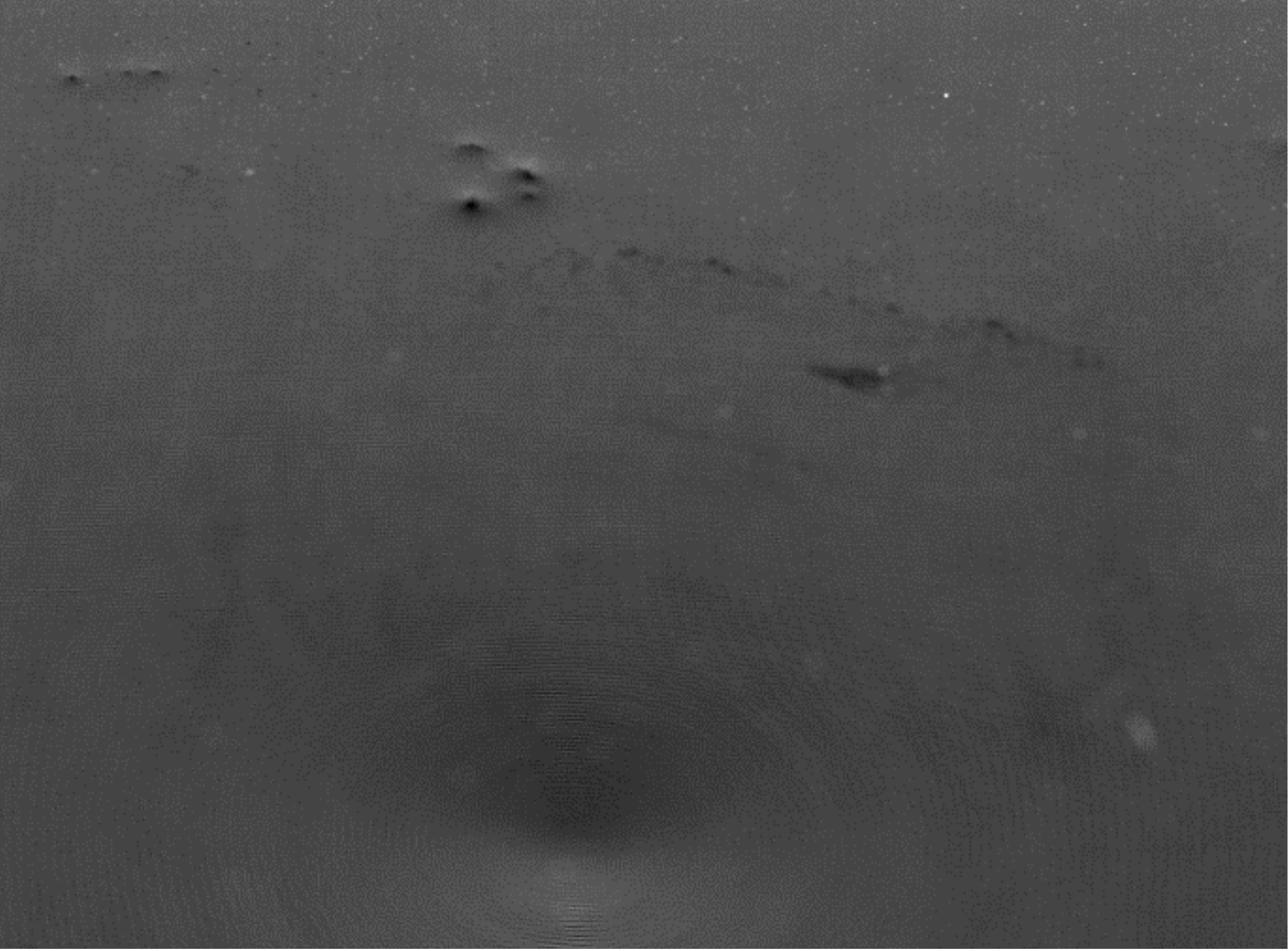}
            \put(-15,5){\colorbox{white}{(a)}}
          \vspace{0.1cm}
            \includegraphics[width=0.9\columnwidth]{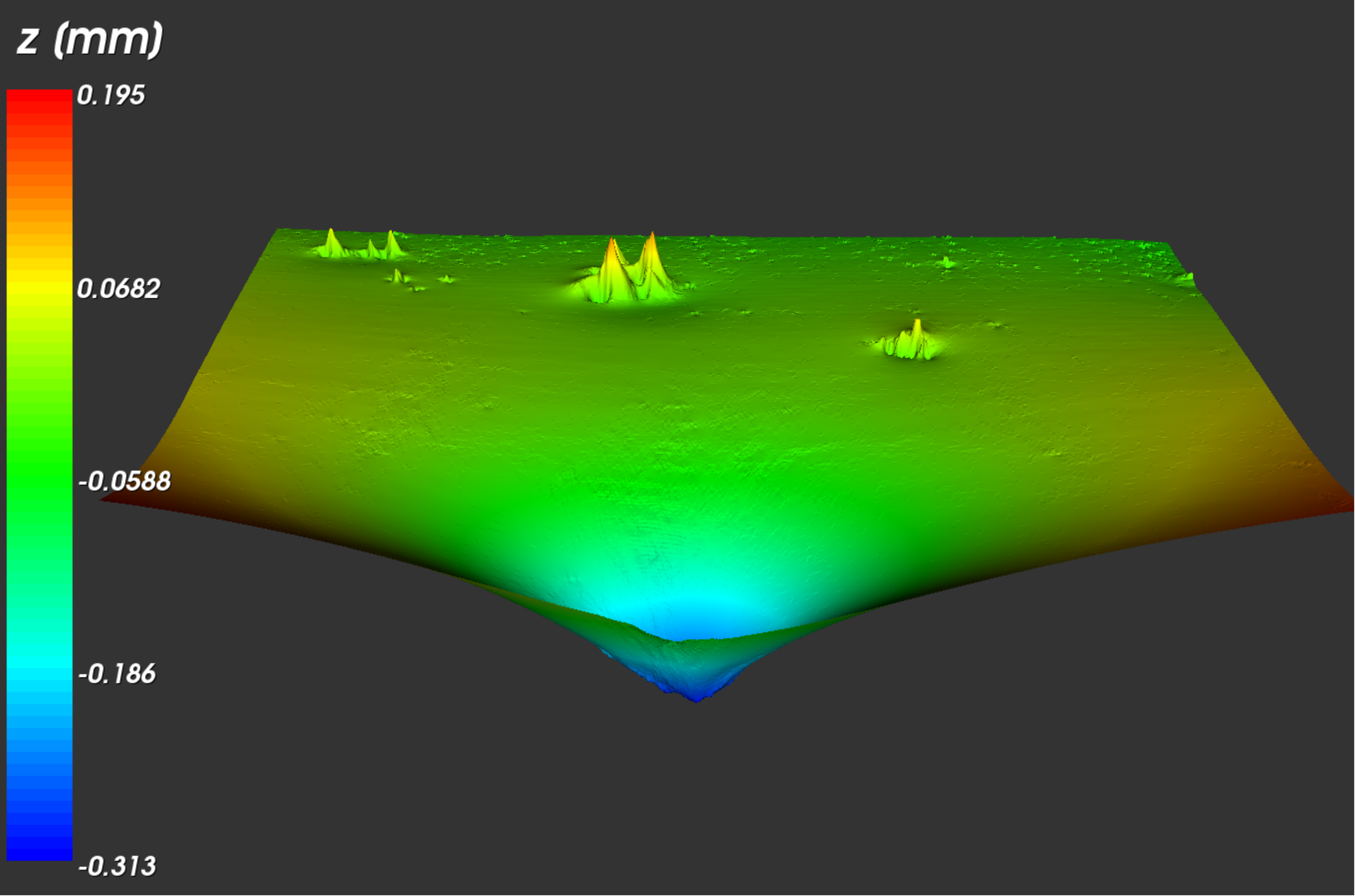}
            \put(-15,5){\colorbox{white}{(b)}}
            \caption{Local surface reconstruction. (a) Surface with defects; (b) local height map.}
            \label{fig:local_recon}
            \vspace{-0.4cm}
        \end{figure}

        If approximate geometry data are available, local reconstructions as in Fig.~\ref{fig:local_recon} are reasonably fast and efficient; however, since relevant defects are almost always specified by their lateral sizes, and almost never by their height or depth, reconstructions for semi-qualitative purposes or mere visualization are quite rare in practice.
        
        In what follows we briefly mention some other notable types of integration techniques used in practice. All of them to some extent share the general behavior discussed so far.
        
        \subsubsection{Transform-based reconstruction}
        
        Surfaces may be reconstructed in Fourier space~\cite{Freischlad.1986}, where it is also possible to formulate integrability requirements~\cite{Frankot.1988} (in this case, for SFS data). The integration itself may be iterative~\cite{Huang.2015} or deterministic~\cite{Bon.2012}. It has been observed that the discrete cosine transform alleviates many issues related to boundaries and/or incomplete data that are an obstacle for the discrete Fourier transform~\cite{Talmi.2006}.
        
        \subsubsection{Modal reconstruction}
        
        Constraining the surface to a combination of a few predefined shapes greatly accelerates computations. Depending on the ideal shape of the part, one typically uses Zernike, Chebyshev, or Forbes polynomials~\cite{Dai.1996, Li.2014e, Mochi.2015, Huang.2016, Aftab.2019, RamirezAndrade.2020}; in principle, any set of orthogonal functions can be used, which can be adapted to the aperture shape with suitable transformations~\cite{Ye.2015}.
        
        \subsubsection{Constrained zonal reconstruction}
        
        Tracking of small-scale features in the modal approach requires many coefficients, which may cause instabilities at the boundaries. A less rigid alternative is to use zonal techniques with adjustable stiffness/noise suppression, using e.g. radial basis functions~\cite{Lowitzsch.2005b, Ettl.2008, Huang.2013d, Alinoori.2016} or splines~\cite{Olesch.2007, Ettl.2007b, Huang.2017c}. An interesting new variety of this approach is emerging through the use of custom deep-learning network architectures utilizing information at multiple scales~\cite{Wu.2021, Dou.2022}.
        
        \subsubsection{Zonal reconstruction}
        
        Zonal techniques aim to find the best match of the reconstructed surface to local gradient maps using least-squares methods~\cite{Fried.1977, Southwell.1980}. As such, they suffer less from boundary artefacts, but are less resilient to noise, and are therefore often implemented as iterative schemes~\cite{Zou.2005, Huang.2015, Ren.2016, Li.2017}; mixed approaches have also been demonstrated~\cite{Espinosa.2010}.
        
        \subsubsection{Reconstruction with additional data}
        
        Frequently, the parts to be inspected may have specular and diffuse surface properties, either as a mixture on the same surface, or on different portions of the object. This makes it possible to add a fringe-projection measurement and thus to obtain significant extra information about the shape and location of the tested object~\cite{Sandner.2014, Sandner.2015, Wang.2021b}, although the calibration of such systems is fairly complex~\cite{Breitbarth.2009, Liu.2020}.
        
        \subsubsection{Integration/interpolation biases}
        
\begin{figure}
            \includegraphics[bb=88 0 610 509, clip, width=0.49\columnwidth]{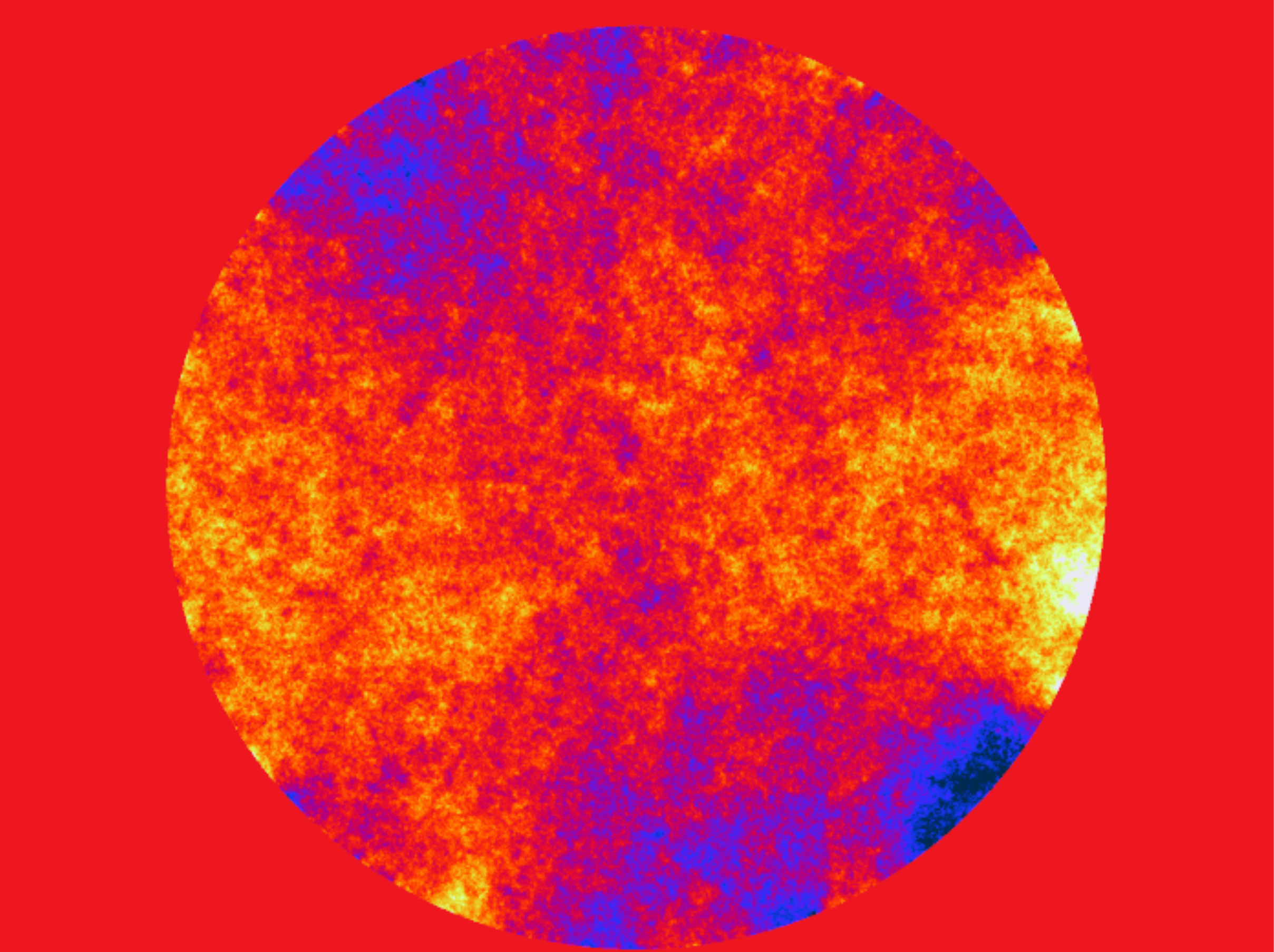}
            \includegraphics[bb=88 0 610 509, clip, width=0.49\columnwidth]{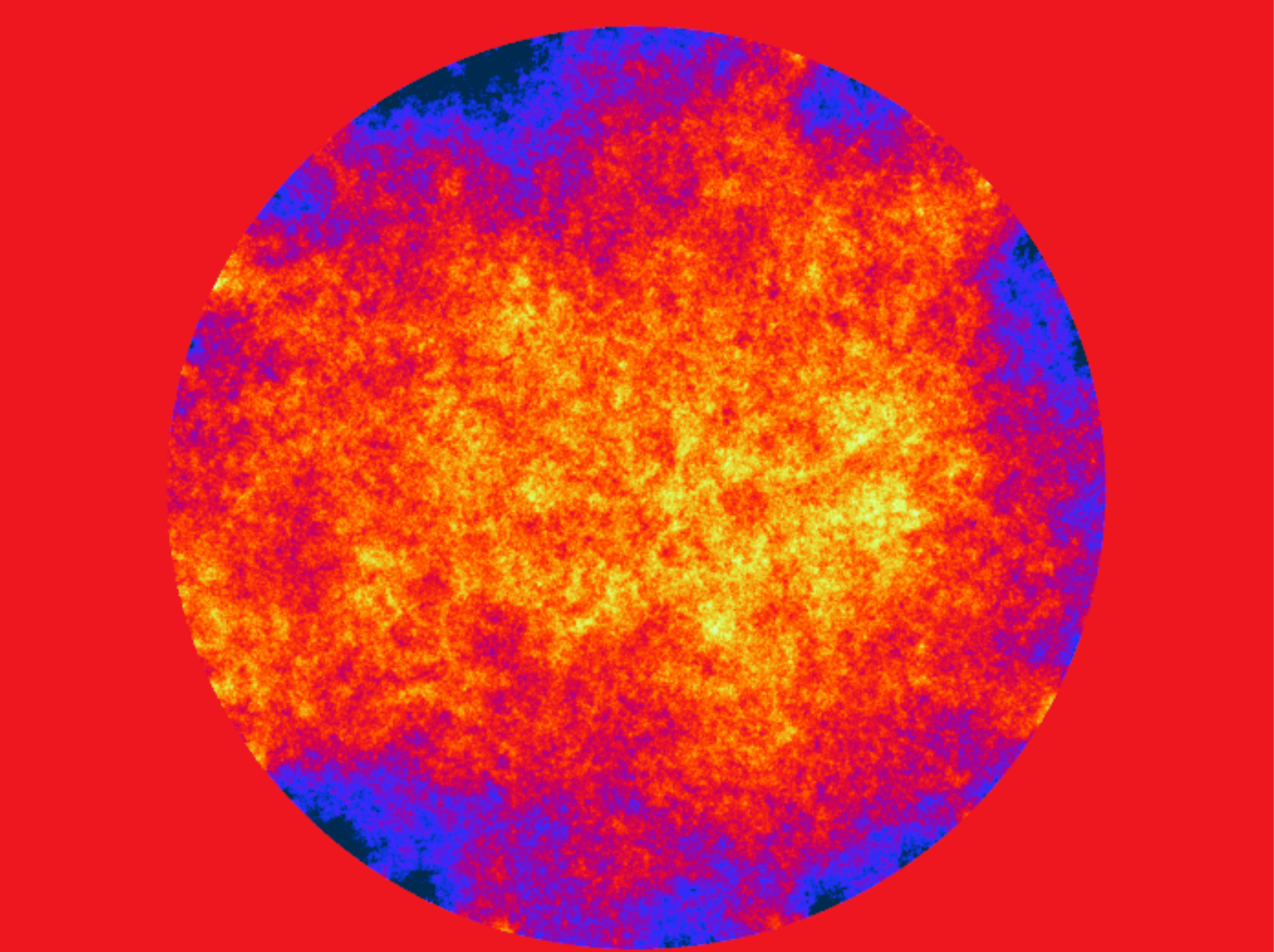}
            \vspace{0.075cm}
            \\
            \includegraphics[bb=88 0 610 509, clip, width=0.49\columnwidth]{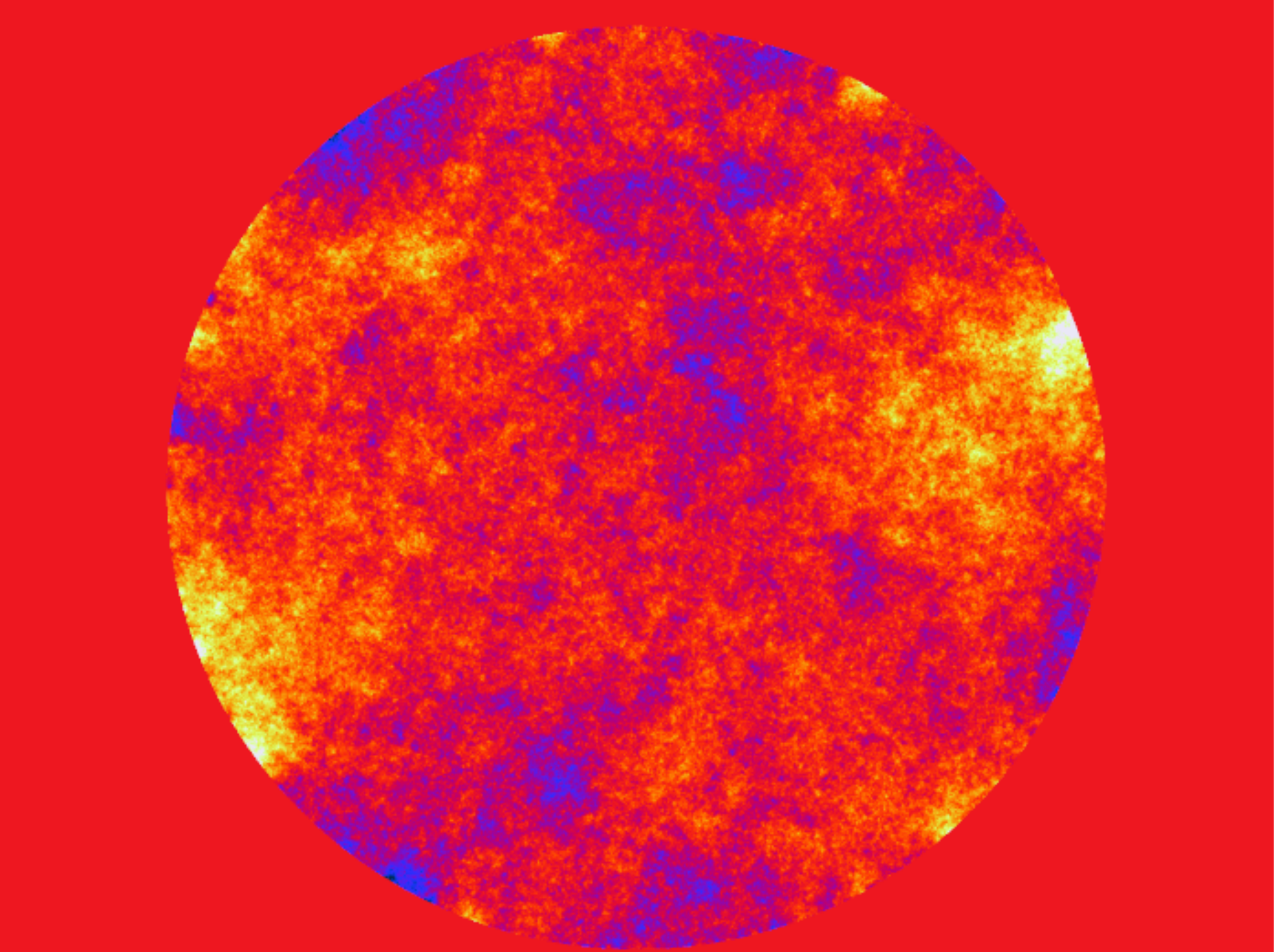}
            \includegraphics[bb=88 0 610 509, clip, width=0.49\columnwidth]{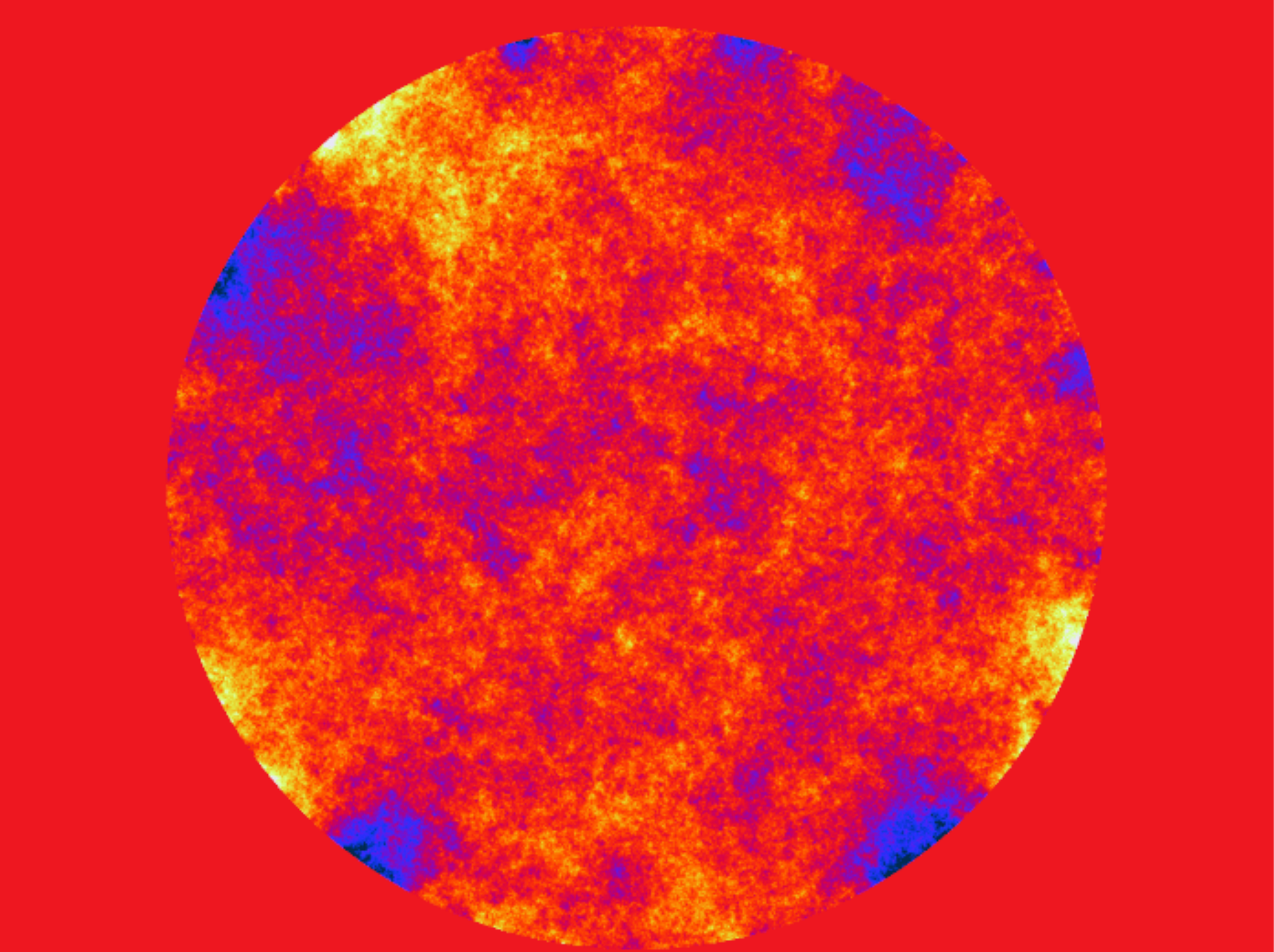}
            \caption{Reconstruction errors (reconstructed surface profile minus ground truth and tilt artefacts) in a simulated setup (object diameter 1420 mm, observation distance 4000 mm, resolution 1000x1000 pixels) for four realizations of random Gaussian deflection noise of 0.025 screen pixels RMS in the underlying registration maps. The deviation color scale is from $-50$ nm (blue) to $+50$ nm (yellow)~\cite{Li.2014e}.}
            \label{fig:int_noise}
            \vspace{-0.5cm}
        \end{figure}

        Any integration method effectively operates at the level of larger or smaller surface patches. Therefore, small errors in the DM or regularization data, sampling grid properties~\cite{Smith.2021}, or the influence of the tested surface itself~\cite{Zhang.2021b, Niu.2021} may have a global effect on the reconstruction outcomes, and deviations from the true shape end up correlated at different scales~\cite{Fard.2018b}. While we are not aware of a general method to predict integration uncertainties in DM, it is clear that simple quality metrics adopted e.g. in fringe projection or laser triangulation such as "RMS height error" do not capture the statistics of errors inherent for DM. Fig.~\ref{fig:int_noise} gives just one example: the possible effects of random noise on the reconstruction result. 
        
\section{System design}
\label{sec:design}

The apparent simplicity of DM systems and techniques (one just needs a light source, a reflecting object, and an imaging sensor) is deceptive. From the experience in our research projects, setting up a good deflectometry system and obtaining high-quality data takes weeks of training and months if not years of experience. Therefore, this section provides practical considerations for designing and using DM sensors and a closer look at how their components can be implemented.

    \subsection {Practical considerations and constraints}
    \label{sec:practical}
    
    Most deflectometric systems are custom-designed for specific applications: owing to geometric constraints associated with different specimen shapes, only nearly-flat objects can be assessed in a generic manner. Even with cylindrical (light tunnel) or hemispherical (light dome) illumination geometries that accommodate a wide variety of geometries, different objects typically need different measurement constellations and/or a customized sequence of object poses.
        
        \subsubsection{Lateral vs angular uncertainty trade-off}
        \label{sec:focus}
                
        Very generally, the product of lateral and angular uncertainties in a DM measurement is bounded from below~\cite{Hausler.2001, Ziebarth.2018}: in a high-precision measurement, decreasing one necessarily increases the other (although it it possible to image both the screen and the object sharply in certain cases~\cite{Li.2021}). In order to make a choice, one can adjust the camera focus to lie between the object surface and the light source. In case when high lateral resolution on the surface is not required, optics should be focused somewhere in the vicinity of the reference screen (not directly on it, as detrimental moiré artefacts will then appear). This reduces the resulting angular uncertainty of normal directions for two reasons: (i) the fringes on the reference screen appear sharper and can be made narrower, which reduces angle decoding uncertainties; (ii) a larger sampled area on the surface (circle of confusion on the object) sharpens the statistical distribution of the deflected rays.
                
        At the same time, once the focus moves away from the surface, the system is no longer in the "cat's eye" mode -- surface curvatures within the circle of confusion start acting as optical elements and may slightly bias the decoding results. In fact, similar considerations apply to the entire mirror image under any focus setting; in order to reduce these effects, one should place the camera far away from the foci or other caustics of the surface.
          
        \subsubsection{Surface roughness and partial specularity}
        \label{sec:roughness}
        
        The presence of roughness increases the diffuse and decreases the specular reflection of a surface; the diffuse background limits the achievable fringe contrast and hence the dynamic range of the measurement. (Painted surfaces are a common example: they are often glossy \textit{and} scattering.)
        
\begin{figure}
            \centering
            \includegraphics[width=0.5\columnwidth]{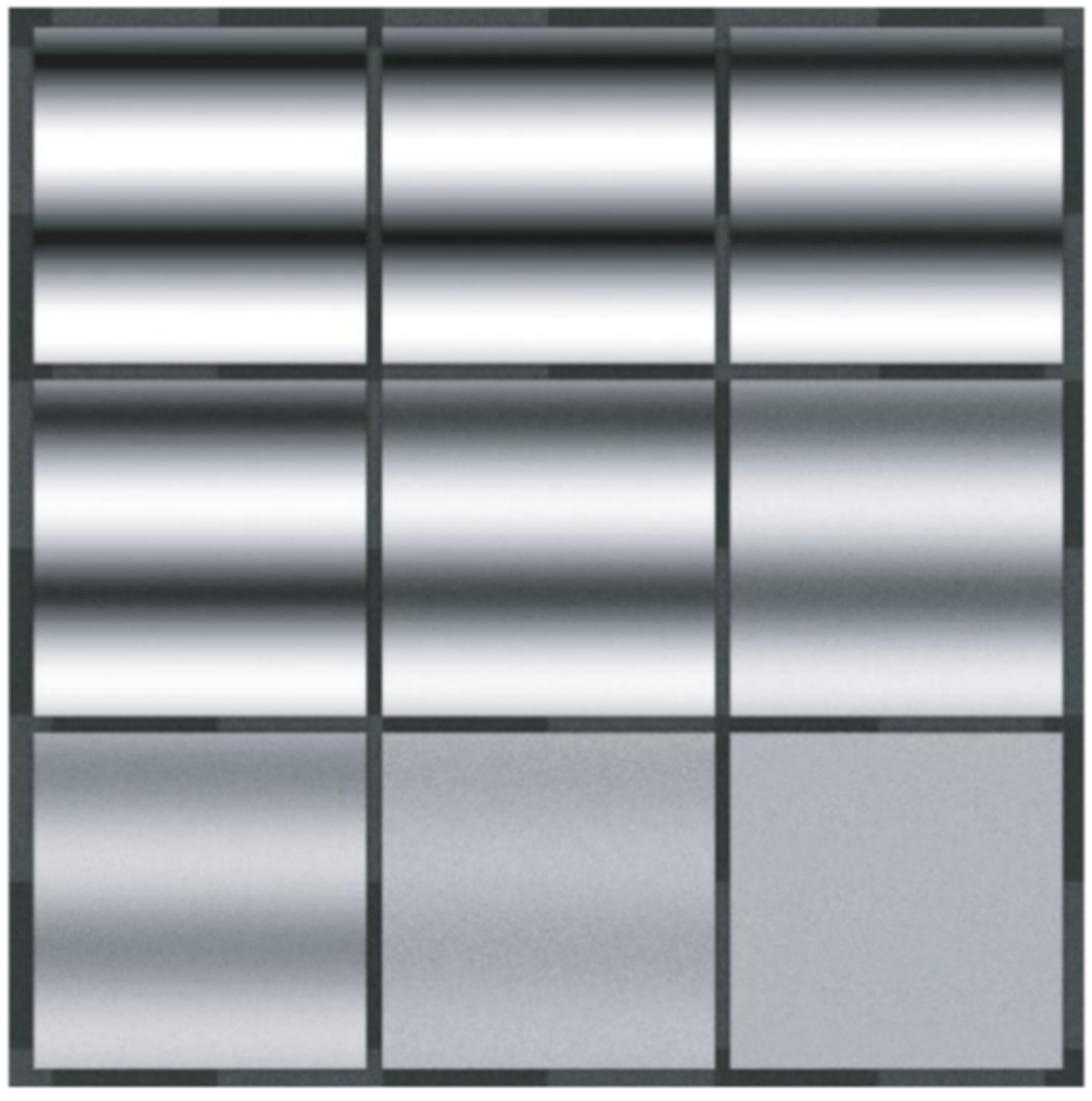}
            \put(-15,5){\fcolorbox{black}{white}{(a)}}
            \vspace{0.1cm}
            \includegraphics[width=0.98\columnwidth]{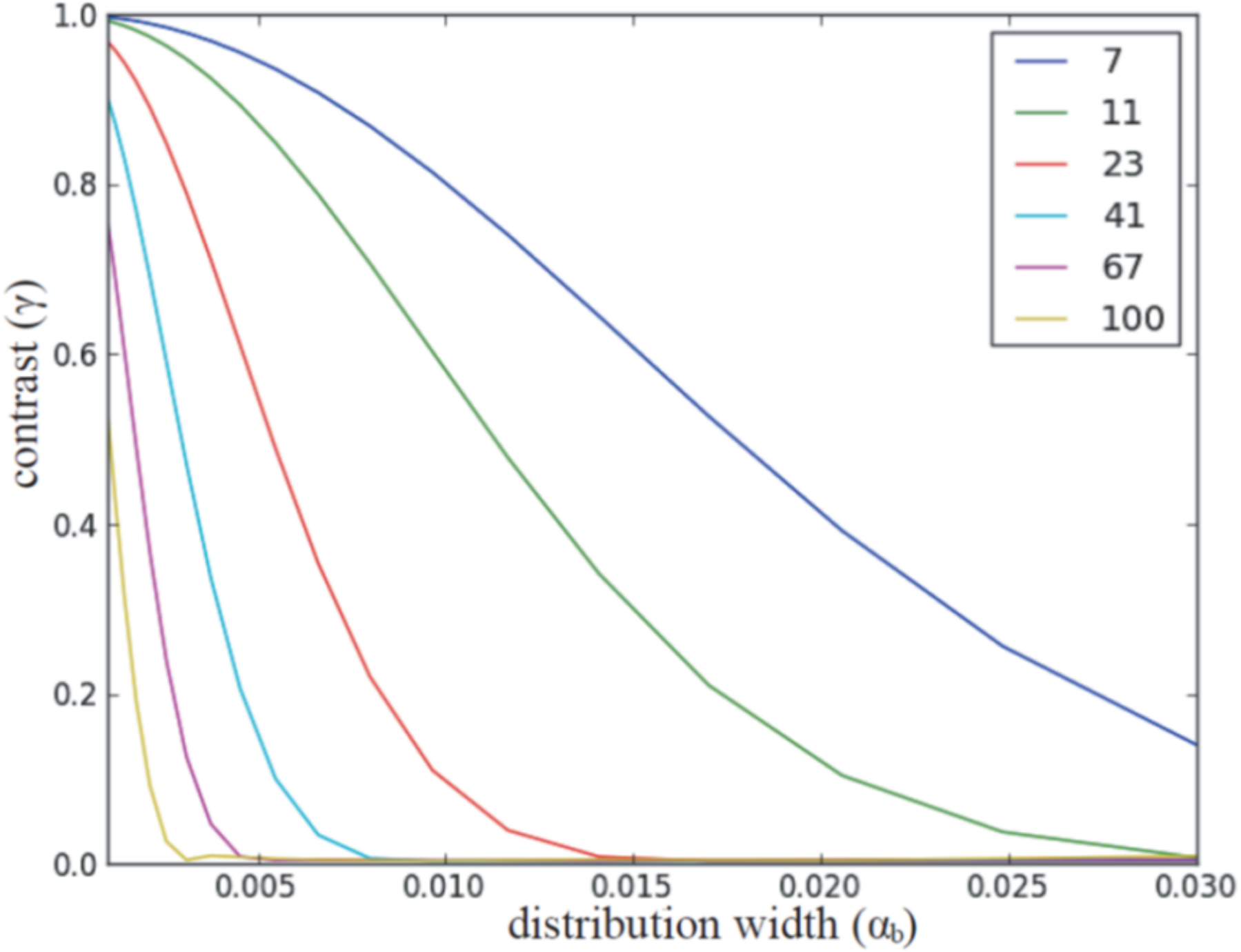}
            \put(-15,25){\fcolorbox{black}{white}{(b)}}
            \vspace{0.1cm}
            \includegraphics[bb=110 35 348 214, clip, width=0.48\columnwidth]{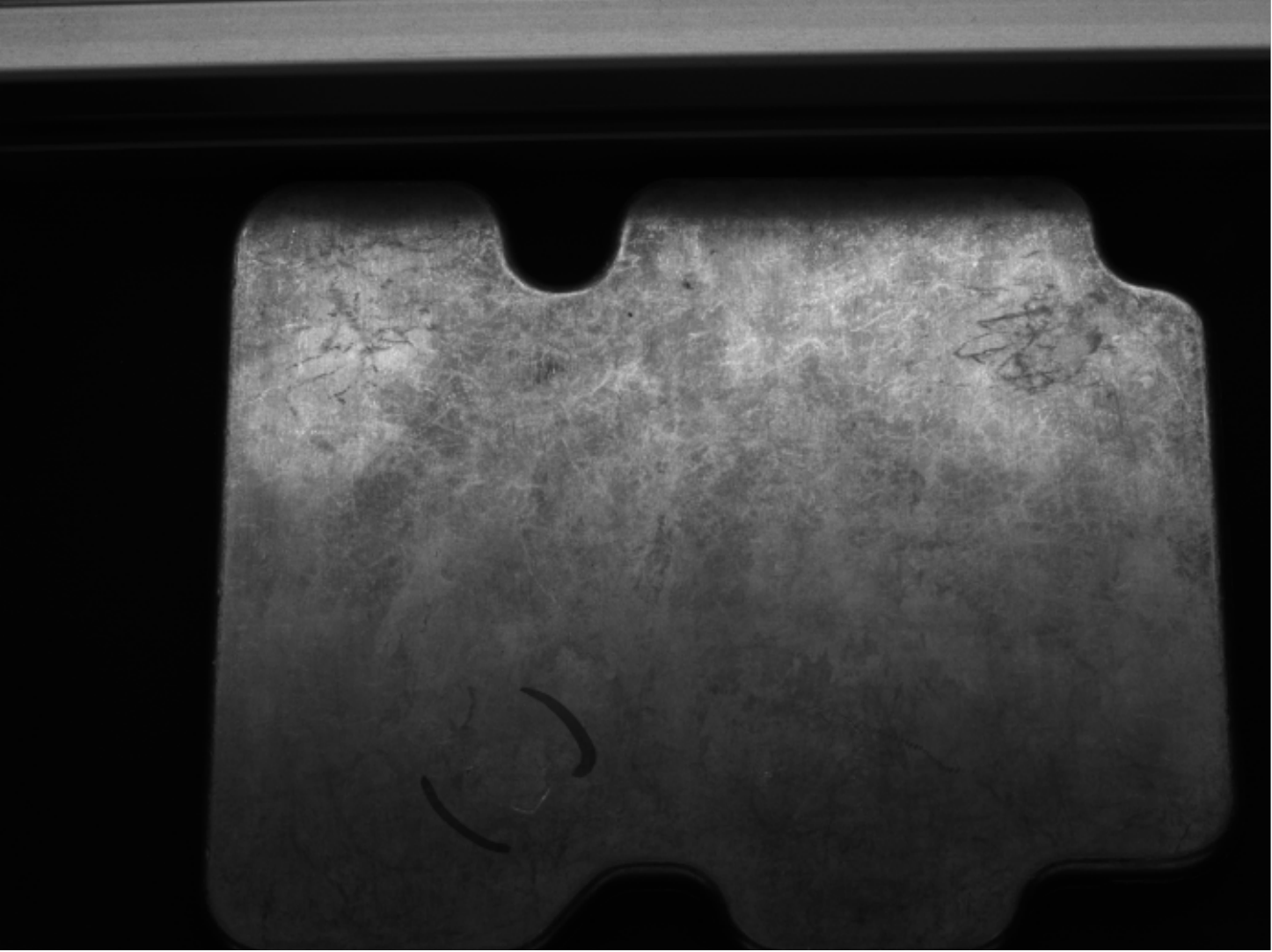}
            \
            \includegraphics[bb=110 35 348 214, clip, width=0.48\columnwidth]{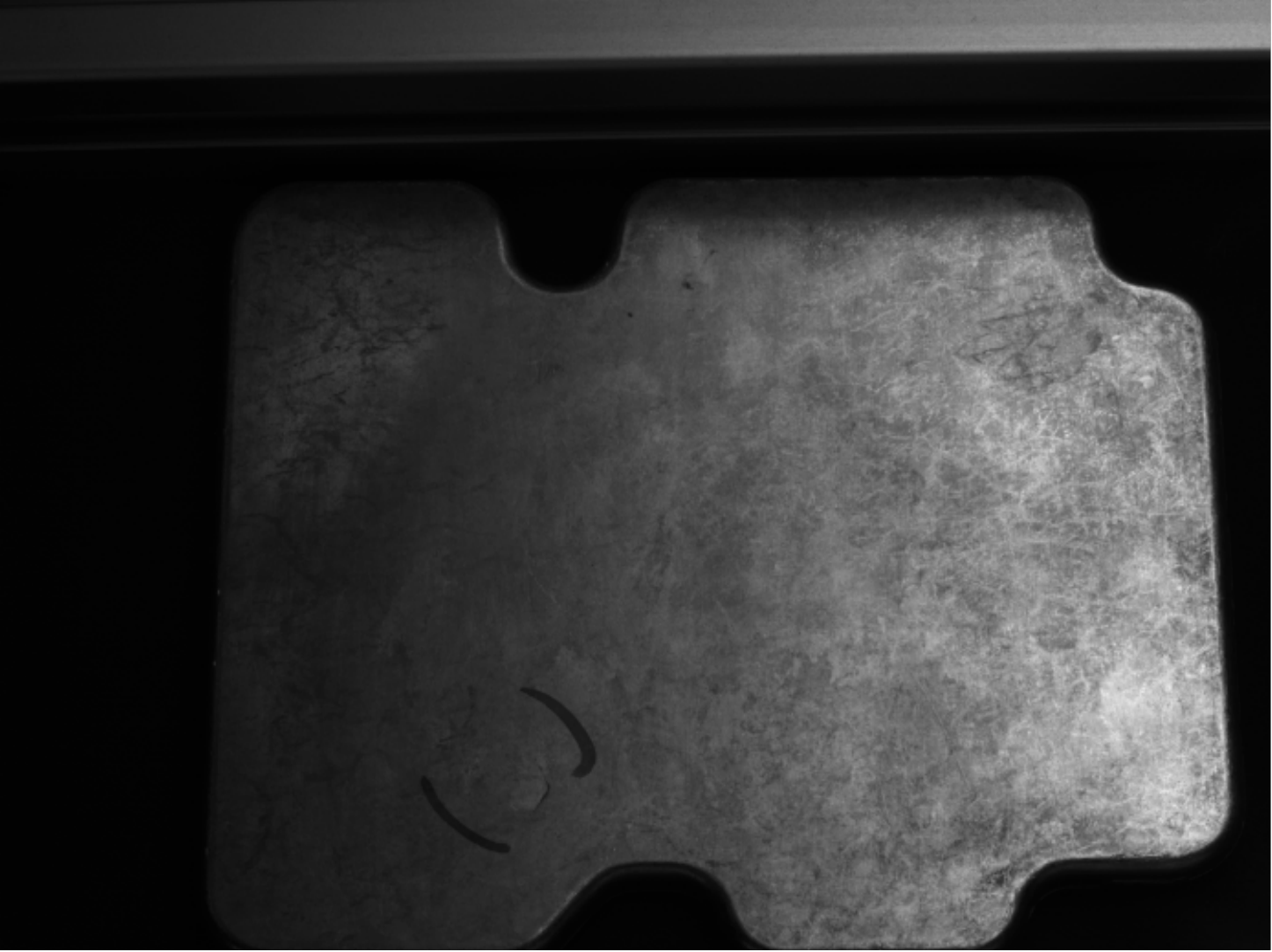}
            \put(-132,5){\fcolorbox{black}{white}{(c)}}
            \vspace{0.1cm}
            \includegraphics[width=0.75\columnwidth]{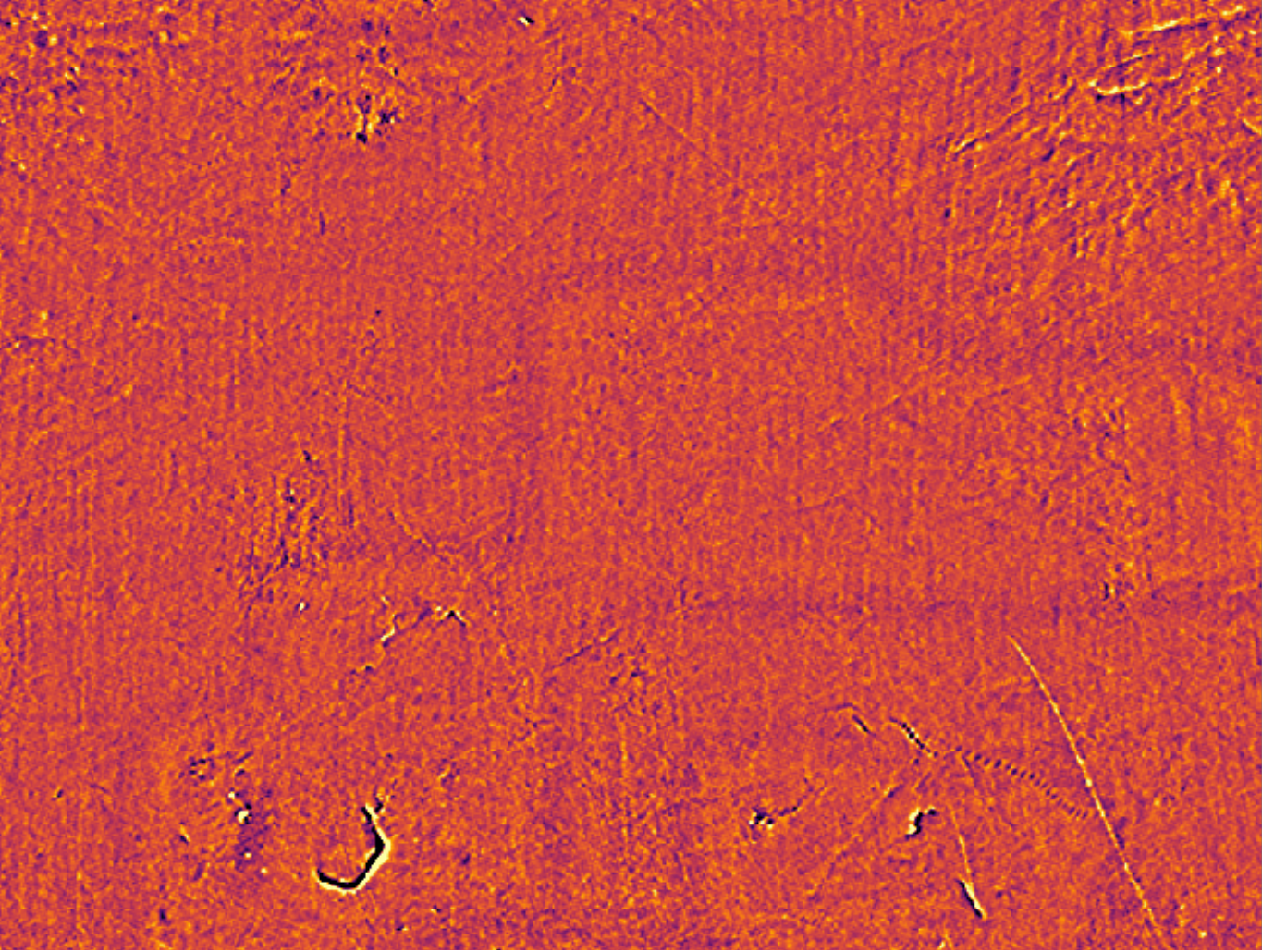}
            \put(-15,5){\fcolorbox{black}{white}{(d)}}
            \caption{Impact of surface roughness on the fringe contrast. The rendered image (a) shows fringe reflections from surface patches with varying distribution width ($\alpha_b$ = 0.001 to 0.03) at constant $L$ (11 fringes on the screen) for a microfacet scattering model~\cite{Hofer.2013b}. The plot in (b) shows the calculated fringe contrast as a function of $\alpha_b$  and for different $L$ (7 to 100 fringes on screen). (c) Two images from wide-fringe sequences (vertical and horizontal) used on a die-cast surface with low specularity; size approx. 60mm x 45mm. (d) Resulting curvature map in false color. Despite poor SNR, the decoding has succeeded: structure print-through from rear side and casting defects are clearly recognizable.}
            \label{fig:fringe_contrast}
            \vspace{-0.5cm}
        \end{figure}

        It is still possible to obtain valid deflectometric data with a rough surface as long as the reference structures/fringes are wider than the main scattering lobe from the surface and a usable contrast of the reflection can be obtained, but of course finer slope details will be lost in noise. (Note also that a sensitivity at the scale of nm would be meaningless on a surface whose roughness is in the $\mu$m range.) The practical limit is reached at an $R_q$ of $\sim$100 - 200 nm (see also  Fig.~\ref{fig:vis_vs_ir_spectrum}); Fig.~\ref{fig:fringe_contrast} gives an overview of this effect. Still, surprising detail can be reconstructed even from weakly specular surfaces, as Fig.~\ref{fig:fringe_contrast}(d) shows.
        
        Beyond that, one may continue with thermal infrared DM (IRDM, Sec.~\ref{sec:lightsources}), where rough metal surfaces are specular due to a smaller ratio of roughness scale to wavelength. Although the angular and lateral resolutions in IRDM are inferior to those in the visible light due to physical reasons, the approach is useful in the early inspection of raw or primed surfaces, on which defects would otherwise only become visible after painting.
                
        On the other hand, diffuse reflection enables the combinations of DM with other methods such as laser triangulation, SFS~\cite{Balzer.2010d}, or FP~\cite{Wang.2021b}. Further, roughness may be anisotropic: many technical surfaces have an increased roughness in one direction and remain quasi-specular in the other. Therefore, sometimes it is still possible to assess one slope component, or to rotate the reference fringes with respect to the surface, in order to maximize the signal to noise ratio. The optimal choices for such measurements have not been formulated theoretically since the relationship between roughness and fringe contrast has not yet been modeled in a satisfactory way.
                    
        \subsubsection{Calibration quality and stability}
        \label{sec:calibration_effort}
        
        When metric measurements are needed, the system must be calibrated. This is usually (but not always) done by optimizing unknown system parameters with a known reference object. As the parameter space of a DM sensor is relatively large, this process is often performed in smaller steps, when subsystems are calibrated separately (Sec.~\ref{sec:calibration}). Accurate calibration is particularly important when a surface is reconstructed by integration, where errors accumulate~\cite{Li.2012e}. Hence, a rigid construction and high thermal stability (ensured by e.g. low-thermal expansion materials) are important to reduce parameter drift between calibrations.
        
        These challenges are particularly important for large setups. For solar concentrators, testing often happens outdoors upon assembly or during maintenance and re-adjustment~\cite{Heimsath.2008b, Wang.2010}. In order to reduce the negative effects of wind buffeting it is important to record data fast. Different requirements apply to telescope mirrors: these are of course measured and polished indoors, but the slope/shape specifications are tighter by several orders of magnitude. Uncertainties here mainly depend on thermal stability; it may be necessary to continuously monitor the setup geometry with e.g. laser trackers~\cite{Huang.2015c}.
        
        When only slopes or curvatures are needed, these requirements can be relaxed. Under certain conditions, such maps can be used as a proxy for the surface shape~\cite{Dominguez.2012, Zhao.2013, Antoine.2019b}, e.g. for modal analyses with polynomials, as mentioned earlier. The misalignment effects then remain confined to local deviations.
        
        \subsubsection{Measurement speed}
        \label{sec:speed}
        
        Almost all industrial applications impose some time constraints on metrology; the prescribed production cycle dictates the rate of data acquisition and processing (or, alternatively, the number of inspection stations), whether or not the inspected surfaces are in motion. The following tricks may help find the balance between the measurement duration, costs, uncertainties, and defect detection efficiencies:
        \begin{itemize}
            \item Brighter light sources (maybe operating in flash mode) reduce the exposure time.
            \item Lower f-numbers of the optics reduce the exposure time; the required depth of focus puts a lower bound on the permissible f-numbers. As a rule of thumb, in a diffraction-limited lens the size of a focused point on the sensor in $\mu$m roughly equals the f-number of the optic~\cite{Reichel.2020}. Hence, increasing the latter in order to obtain a larger depth of field (which is often useful as most surfaces in DM are observed from an oblique angle) blurs the point image and reduces the effective image resolution (note that typical pixel sizes in modern sensors are in the single-digit $\mu$m range).
            \item Data processing can be accelerated by performing computations asynchronously, with distributed CPUs and/or GPUs.
            \item Non-demanding applications may use simpler coding methods than phase shifting (it yields unmatched sensitivity but needs many camera frames).
            \item A lower number of fringe sequences in phase shifting (Sec.~\ref{sec:cos_patterns}) may suffice. For continuous smooth surfaces one may use a single fringe sequence per direction in combination with a heuristical phase unwrapping technique (which may fail for less-trivial shapes).
            \item Similarly, fewer phase shifts per sequence (Sec.~\ref{sec:cos_patterns}) accelerate data acquisition. (An exact solution with no additional assumptions requires $N\geqq3$, but methods using more phases are significantly more stable in practice and reduce the effects of non-linearities in transfer functions).
            \item Multiple reference screens and/or cameras (in a single setup or as separate systems) utilize the available "light field" more efficiently.
            \item Single-shot methods (with static patterns that use spatial phase modulation or another principle) operate at the expense of spatial resolution and dynamic range (when multiplexed patterns are used).
            \item If the f-number is still too high, one may increase the camera gain as the last resort (this replaces digitization noise by electronic noise).
        \end{itemize}
                
        \subsubsection {Setup geometry and size}
        \label{sec:geometry_size}
        
        Finding the proper spatial frequency of fringes for reference patterns is straightforward for objects of quasi-constant curvature, but may be a challenge for complex shapes.
                
        Planar surfaces produce a largely undistorted mirror image of the reference screen. The latter then generally needs to be larger than the test piece (e.g., about twice as large in the basic geometry when the surface is placed at roughly the same distances from the camera and the screen). This situation may be compared to a fitting mirror that must be at least half the size of the person using it.
                
        Concave surfaces magnify the reference structure. Therefore, if the latter is placed in the vicinity of a caustic of view rays (but not too close to it), a relatively small screen may enable testing of very large parts~\cite{Su.2012} (another consequence of this magnification is, of course, the reduction of the source intensity recorded by the camera). Typical examples of large optics that can be tested with DM are solar concentrators and astronomical telescope mirrors. Their radii of curvature range from several meters to tens of meters, and in the most popular approach to test them, the reference screen and the detector are placed near the center of curvature. Therefore, such set-ups typically do not use large screens or many cameras, although multi-camera examples are known~\cite{Schulz.2011, Olesch.2014c}. As a drawback of this approach, for extremely large radii of curvature (for example, the E-ELT mirror facets have a curvature radius of about 69 m) it may be impossible to find or build testing chambers that are long enough.
        
        Convex objects are the most difficult case in practice. Many parts have edges and corners with small radii of curvature that strongly de-magnify the images of the structured light sources. In some positions even a large screen modulates only a tiny part of an edge in an object (and none of its remaining surface). In addition, since the entire screen image is compressed into a small area on the sensor, the fringes become too bright and often cannot be resolved. In such cases one may need to enclose the inspected object by reference structures~\cite{CamposGarcia.2011, Zhang.2017b, Graves.2019}. If large switchable structures are required, the method of choice is an array of projectors illuminating walls or scattering screens as in a "cave"~\cite{Balzer.2014}. In experiments, objects can be surrounded by screens on almost all sides; in practice, such systems are mostly built as tunnels or moving portals to increase the throughput.
        
        The inspection of pieces featuring a wide range of curvatures is a serious challenge even if the specular reflectance is uniform. If the object position is known, it may be possible in some cases to use inverse fringe or spot patterns~\cite{Perard.1995, DiazUribe.2000, Tosun.2007, Werling.2007c, Mai.2007, Liang.2016, CamposGarcia.2019, DelOlmoMarquez.2021, HuertaCarranza.2021}. Of course, sharper changes in curvature over the surface necessitate tighter constraints on the position of the sample with respect to \textit{all} degrees of freedom. The problem of shadowing that is important for fringe projection is less of an issue here, as shadowing is often associated with spurious double reflections on the edges anyway. The latter are a show-stopper in DM and must be avoided at the layout stage if possible. There exist geometries (e.g., funnel-like shapes) on which deflectometric measurements cannot be made unless the sensor and the reference structure are small enough (and are e.g. made to fit inside the cavities).
        
        For surfaces with steps~\cite{Karacali.2003, Xu.2022, Zhang.2022}, obscuration must be considered on the side of illumination and/or observation. Again, if possible, the optics must be modified in order to avoid this situation and ideally make the entire object surface accessible to testing. In one example, telecentric observation was combined with a coaxial structured light source~\cite{Werling.26.05.2014}. Note that this case is different from the situation when discontinuities occur in the \textit{slopes} of the surface~\cite{Durou.2009}.
                
        Instead of building a reference structure enclosing the object, it is possible to move e.g. a smaller screen around, so that the required solid angle is covered by its successive positions in a temporal sequence of measurements (Sec.~\ref{sec:planning}, ~\cite{Schwarz.2016}). By accounting for object and sensor locations, it is also possible to pre-distort patterns so that the resulting uncertainty is roughly the same as with a static sensor. In practice, though, this is a relatively complicated calculation. Instead, it is easier to adapt the highest spatial frequency in patterns (a single parameter) for each measurement.
        
        In general, the characterization of components in large-scale DM systems is challenging (Sec.~\ref{sec:calibration}). Efforts needed to calibrate them and ensure the consistency and the stability of parameters (inside or outside of a laboratory) increase with the number and sizes of the system components.
        
        \subsubsection {Low-cost and mobile sensors in deflectometry}
        \label{sec:lowcost}
        
        The increasing availability of high-quality screens and the miniaturization of electronics enables a range of new DM implementations with reduced system costs. The simplest example of a low-cost device suitable for DM is any laptop screen with a built-in web cam. Similarly, smartphones or tablets have become an ubiquitous example for the tight integration of sensors and displays in consumer electronics. They appear to be ideal platforms for rigid or hand-held DM systems, capable of qualitative measurements of small surface areas~\cite{Butel.2015, Trumper.2016}. With additional multi-view stitching tools, their field of view may be extended~\cite{Willomitzer.2020}. Despite a typically non-ideal geometry (e.g., concerning the camera alignment with respect to the screen), such systems demonstrate respectable results. In combination with sufficiently stable (e.g. 3D-printed) spacer frames, one can even ensure reproducible calibration uncertainties and measurements.
        
        \subsubsection{Other sources of measurement errors}
        \label{sec:uncertainties}
        
        Given the sensitivity goals for a setup (Sec.~\ref{sec:aesthetics}), one adjusts parameters in order to assess the largest possible surface area in the shortest possible time. Apart from calibration errors (discussed above), the resulting measurement quality also depends on the performance of the sensors. 
        
        For many surfaces, typical measurement distances are in the range between 0.3 and 1 meters. With typical LC displays as sources, the decoding errors (that determine the angular uncertainty of deflection maps) then significantly depend on the noise in the pixel values recorded by the camera. (This is less of an issue for the most demanding pieces measured with DM so far -- astronomical and synchrotron mirrors -- that have curvature radii of many meters and simple shapes and can be measured in very large setups with extremely high sensitivity.)
        
        The SNR in modern cameras is no longer limited by the electronic noise, but rather by the photoelectron collecting capacity of pixels, which is of order $10^4 - 10^5$~\cite{EMVA.2021}. In combination with the quantum efficiency level of sensors (that is no longer improving) this leads to the remarkable fact that the common 8-bit digitization (24-bit for color) is (coincidentally) well matched to the underlying physics. There may be cases with exceptionally poor illumination where digitization noise does play a role, but in practice one then takes measures to increase the photon count, or uses frame averaging for the same scene.
        
        As a direct consequence of this very well-predictable level of statistical noise, one may accurately state the uncertainty limits of phase-measurement methods~\cite{Surrel.1997, Horneber.2002, Fischer.2011, Faber.2012, Fischer.2012b, Hofer.2013b} and easily devise optimal encoding strategies at least in terms of reference fringe periods.
        
        Further, in defect detection tasks one needs to consider the spatial resolution. Remarkably, defects distinctly smaller than a pixel can still be detected by DM~\cite{Bothe.2004f, Petz.2018}. Thus, the well-known rule of thumb that a defect should be resolved by 2 or 3 pixels does not necessarily apply, and the specific requirements will depend on whether defects must be only detected or also classified by type.
        
        Apart from the systemic sources of uncertainty associated with the components and the trade-offs made in the configuration (Sec.~\ref{sec:design}), in practice one encounters a number of additional sources of error that interfere with measurement and classification tasks. In the remainder of this section we provide a few practical suggestions on how best to minimize or eliminate such errors and artefacts.
        
        Contamination of surfaces is an ever-present annoyance for visual inspection (even in clean rooms). Most often it has the form of dust or marks stemming from improper handling of an object (e.g., fingerprints). Since the detectability of minute contaminations much smaller than a pixel is the same as that of actual defects, no distinction or classification of these unresolved features of unclear origin is possible. Sometimes extra signals such as fluorescence under UV light can be used to detect contamination; however, the simplest strategy by far is to keep surfaces clean. Dust is most conveniently removed with a shower of ionized air which removes particles and static charge. After that, one has about 10-20 seconds to measure the part before it starts attracting dust again. Other contaminations, such as fingerprints, can usually be controlled by the appropriate equipment and procedures. Larger features can be distinguished to some extent by their signatures in phase and/or modulation maps~\cite{Burke.2017, Wu.2017}, but, of course, the error detection accuracy still suffers from any residual artefacts. 
        
        The stray light from external sources (e.g., sunlight, room lighting) or internal reflection on system components can introduce static or dynamic interference in the camera image. Measurements of surfaces with a relatively high diffuse reflectance are especially susceptible to surrounding light sources. In most cases, a lightproof housing and/or careful positioning of all components suppresses such interference. While it is not always necessary to cover all internal surfaces of the setup with matte black paint or foil, it is certainly a good idea to avoid glossy components. In cases where light contamination is unavoidable, solutions include constraining the measurement to a suitable undisturbed part of the light spectrum with custom light sources and spectral filters, e.g. in the near-infrared range~\cite{Chang.2020}, or recording the static part of the background intensity to subtract it from all recorded data later.
        
        Stable position and alignment of all system components is vital to minimizing uncertainties. This should be taken into account at all relevant time scales. Vibrations lead to blurred images due to movement during the exposure time. This can be a non-uniform effect, depending on the recorded fringe densities and object curvatures, so that at worst the fringe decoding may fail. An unstable object handling system or inadequate support might shift between the frames corresponding to a coding sequence. Even if the object support is highly reproducible, the measurement system must still be able to cope with variations in object shapes from part to part. From our practical experience, it is worthwhile to characterize the variations in object shapes before designing the layout and sensitivity of the inspection system.  
        
        Finally, long-term instability due to inadequate repeatability of the handling system or thermal expansion causes a mismatch between the calibration and the actual setup geometry. This affects the degree to which the measurements can be compared (e.g., when detecting defects with respect to a reference measurement) or combined for analysis.
        
        \subsubsection{Examples of practically achieved uncertainties}
        \label{sec:best_uncertainties}
        
        The measurement noise in the derivatives is a serious problem for surface reconstruction~\cite{Freischlad.1992, LegardaSaenz.2000, Elster.2002, Karacali.2004, Petz.2005c, Lowitzsch.2005b, Bonfort.2006, LegardaSaenz.2007b, Kolhe.2009, Fischer.2011, Li.2012e, Patel.2013, Hofer.2013b, Pak.2014, Dong.2014b, DeMars.2019} due to the low sensitivity of measured slopes to absolute surface height~\cite{Falconi.1964}. Today, only scanning techniques allow uncertainties commensurate with the tolerances of demanding optical surfaces (e.g., synchrotron mirrors)~\cite{Weingartner.2001c, Elster.2002b, Lammert.2004, Xiao.2011}.
        
        The state of the art sensitivity in DM is currently at the level of a few \si{\nano\metre} for small-scale defects~\cite{Kugimiya.1988, Hahn.1990, Beyerer.1997, Bothe.2004f, Juptner.2009, Burke.2019b}, several tens of \si{\nano\metre} for mid-frequency errors~\cite{Su.2012, Dong.2014b, Su.2014b, Coniglio.2021}, and about \SIrange{100}{200}{\nano\metre} for flat or low-curvature surfaces~\cite{Graves.2007, Ettl.2008, Sandner.2011, Li.2012e, Liu.2012b, Hofbauer.2013, Olesch.2014b, Ren.2015b, Li.2017}, with smaller parts yielding even better results~\cite{Su.2012b, Huang.2013g, Su.2013c, Bergmann.2015}. As mentioned above, for best results one may replace lenses with an actual pinhole aperture~\cite{Su.2013c}.
        
        The most straightforward use of DM is of course to map and evaluate slopes. While this approach cannot guide polishing or shaping processes based on 3D models, it has proven useful e.g. in characterization of solar mirrors~\cite{Ulmer.2007, Andraka.2009, Su.2010d}. Achieved slope sensitivities range from \SIrange{0.2}{20}{\micro\radian}, which is useful for solar mirrors whose performance specifications are typically in the range of a few mrad~\cite{Burke.2013}. In some cases, slope specifications can replace RMS wavefront specifications for precision optics~\cite{Burge.2010b, Li.2014f, Sironi.2016b, Antoine.2019b}; the required sensitivities are in the \SIrange{10}{1}{\micro\radian} range, which is achievable without too much effort. Higher demands, such as \SI{10}{\nano\radian}~\cite{Sironi.2016b}, are significantly harder to satisfy, and commensurate performance has not yet been demonstrated.
            
        Simple and accurate measurement of aspherics and free-form surfaces has been one of the hopes placed in phase-measuring deflectometry. With DM, one can indeed obtain valid data with less effort than with interferometry, but precision is so important in this discipline that interferometry or tactile methods, despite their own difficulties~\cite{Wiegmann.2011, Fortmeier.2016, Blalock.2017b, Pruss.2017}, have not been displaced except in niche applications.
        
    \subsection{Light sources for the reference pattern}
    \label{sec:lightsources}

        \subsubsection{Inspection in the visible spectrum}
        \label{sec:VIS deflecto}
        
        As mentioned above, computer monitors are convenient and affordable active reference devices. In what follows we discuss some practical details related to their use in DM.
        
        If the reflected view rays meet the screen at a shallow angle (which occurs frequently when convex objects must be placed close to the screen), it is important that the "view angle" (angular intensity spread by screen pixels) be as large as possible. With the improved availability of IPS (in-plane switching) displays, achieving a more uniform illumination has become easier. Still, if the distance between screen and object varies significantly across the rays captured, the illumination level and slope sensitivity is bound to be highly inhomogeneous for geometrical reasons alone.
        
        The screen's brightness is important for fast and low-noise measurements, and all flat-screen technologies keep improving in this regard. The brightest monitors (intended for window displays or outdoor use), however, typically use non-IPS matrices and cannot be universally recommended.
        
        Sometimes a fringe-phase measurement may succeed with a sufficiently small uncertainty using only Full-HD screen resolution. Even in such cases it may be beneficial to use monitors with more (and smaller) pixels, as the detrimental moiré structures are then easier to avoid. This becomes important when measuring complex surfaces whose geometry may create an undesirably sharp reflected image of the screen. For very demanding applications, one may consider medical monochrome monitors designed to display X-ray images; they have high brightness and dynamic range as well as a very high lateral resolution because they have no RGB matrix. Typically, these are available in useful square formats and are relatively expensive.
        
        Matte screens are better suited for DM than glossy ones. The scattering on the surface does increase the uncertainty of the measured ray directions; however, in practice this effect is less of an error source than the ghost images in the screen created via multiple reflections (especially when the screen is close to the object). Fortunately, large screens are primarily available with matte finish.
        
        Most screens are not designed for non-conventional mounting angles (e.g., parallel to the floor) and may deform or break under their own weight when tilted so. Certainly it is possible to mitigate such effects with extra support frames, but this may involve difficult and/or expensive modifications. The larger a screen, the more fragile it is; for e.g. moving systems on robotic supports, safe ranges of accelerations have to be established by trial and error.
        
        Systems where reference patterns are projected on some surface are significantly more flexible. While the geometry of such patterns is not nearly as easy to characterize as LCD screens (Sec.~\ref{sec:screen_models}), considerable freedom exists in placing and orienting projectors and projection surfaces. Similarly, pre-printed passive reference screens can be made in almost any shape and orientation, and a wide variety of illumination options exists, including very fast flash or strobe illumination. Uniform diffuse (isotropic) scattering by the material of such screens is beneficial for DM measurements.
        
        \subsubsection {Inspection with infrared light}
        \label{sec:irdm}
        
        In applications where surfaces have insufficient specular reflection to apply "normal" DM, light with longer wavelengths may help. This is the field of "infrared deflectometry" (IRDM), to be delineated against short-wave infrared deflectometry~\cite{Chang.2020} and thermal pattern projection~\cite{Landmann.2020, Landmann.2021}.
        
        While "infrared" (IR) commonly refers to the entire range from just beyond the visible (VIS) to millimeter-wave radiation, IRDM is most often implemented in the long-wave infrared (LWIR), or thermal spectrum. It corresponds to wavelengths of \SIrange{8}{14}{\micro\meter}, where affordable imaging sensors (thermal cameras) are available.
        
\begin{figure}
            \centering
            \includegraphics[width=1.0\columnwidth]{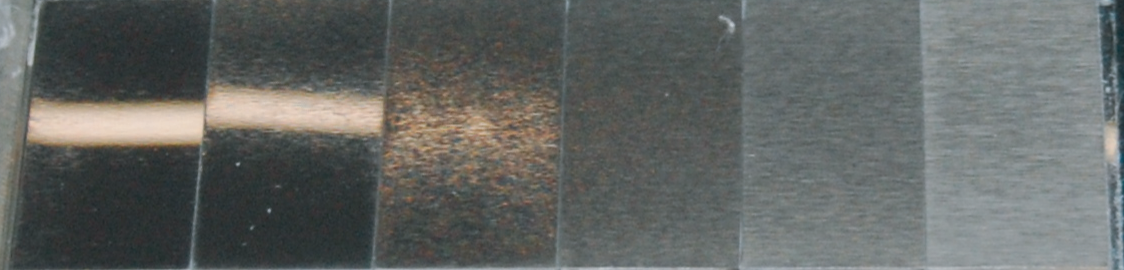}
            \put(-28,45){\colorbox{white}{VIS}}
            \put(-257,-13){\colorbox{white}{$\frac{R_{\mathrm q}}{\mathrm \mu m}$ = 0.05}}
            \put(-193,-13){\colorbox{white}{0.1}}
            \put(-151,-13){\colorbox{white}{0.2}}
            \put(-109,-13){\colorbox{white}{0.4}}
            \put(-67,-13){\colorbox{white}{0.8}}
            \put(-25,-13){\colorbox{white}{1.6}}
            \vspace{0.0cm}
            \includegraphics[width=1.0\columnwidth]{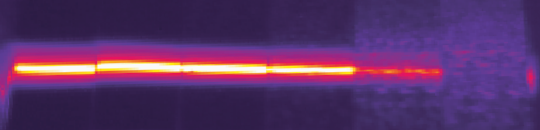}
            \put(-25,45){\colorbox{white}{IR}}
            \caption{Scattering by calibrated surface roughness standards in the visible ($\lambda = $ \SIrange{0.4}{0.7}{\micro\meter}) and long-wave infrared ($\lambda =$ \SIrange{8}{14}{\micro\meter}) spectra. The RMS surface roughness $R_{\mathrm{q}}$ is indicated for each sample. Images from~\cite{Hofer.2016c}.}
            \label{fig:vis_vs_ir_spectrum}
            \vspace{-0.6cm}
        \end{figure}
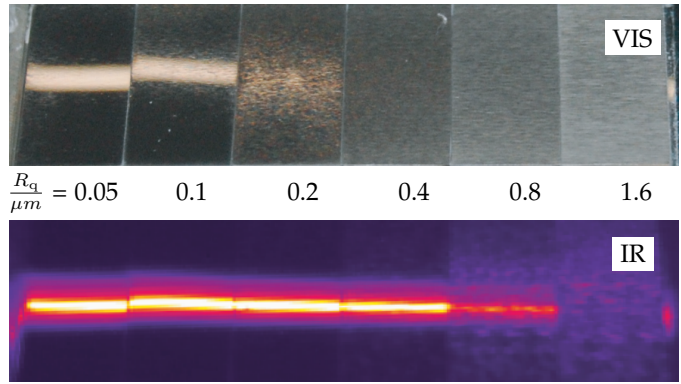

        As already noted for regular DM in Sec.~\ref{sec:fringe}, the IR radiation is {\itshape reflected} by the surface: no patterns are projected on it, nor is the material heated intentionally. Instead, IRDM exploits the wavelength-dependent scattering off the surface microstructure. The latter may appear as a result of machining or just be the raw state of the material's surface. While their typical scales are not resolved by DM, the roughness per se, nonetheless, affects scattering and leads to "blurring" of the reflected images, as shown in Fig.~\ref{fig:vis_vs_ir_spectrum}.
        
        As a rule of thumb, for the reflection to be considered "specular", the wavelength $\lambda$ of light and the root mean square (RMS) roughness of the surface $R_{\mathrm{q}}$ must satisfy
        \begin{align}
            \lambda &> 8 R_{\mathrm q} |\cos{\theta}|,
        \end{align}
        where $\theta$ is the incidence angle (in terms of Fig.~\ref{fig:reflection_components}, $\cos\theta = \hat{i}^T\hat{n}$). Therefore, IRDM extends deflectometry to surfaces with $R_{\mathrm q}$ roughly by an order of magnitude higher than those amenable to DM in the visible range (cf. Fig.~\ref{fig:vis_vs_ir_spectrum}).
        
        While thermal imaging cameras are available, the pattern creation remains a challenge in IRDM: there exist no off-the-shelf parts like monitors or projectors for LWIR similar to those operating with visible light. Starting from the initial experiments~\cite{Horbach.2005} to recent applications~\cite{Hofer.2016c, Su.2011b, Toniuc.2019}, one mostly uses static patterns. Harnessing dynamic thermal processes in order to generate patterns still remains a challenging engineering endeavor~\cite{Hofer.2013c, Hofer.2017c, Graves.2019b}.
        
        Known applications of IRDM usually focus on materials with "intermediate" surface properties, i.e. those that are too glossy for inspection with FP and at the same time too rough for regular DM. Prominent examples include the inspection of car body parts~\cite{Horbach.2005, Hofer.2016c, Hofer.2017c} in early production (before they are painted), or, in precision manufacturing, the measurement of large free-form optics during the grinding stage~\cite{Su.2011b, Su.2013d, Su.2014b, Lowman.2018, Graves.2019b}.
                
        Aside from creating specular reflections on rough surfaces, LWIR is also useful for some materials that are transparent in the visible light but opaque in the LWIR, such as glass or some plastics. By suppressing transparency and back-reflections, IRDM here facilitates the dedicated inspection of outer surfaces~\cite{Hofer.2016c}.

        \subsubsection{Inspection with ultraviolet light}
        \label{sec:uvdm}
        
        When inspecting glass optics (e.g. lenses) in visible light, the "parasitic" reflex from the back side may overlap with the front side reflection and corrupt the decoded data~\cite{Faber.2009}. The obvious solution -- covering the back side with non-reflecting paint -- is only practical in low-volume tests or for reference pieces~\cite{Schachtschneider.2018}, and is not suitable for serial production and/or finishing.
        
        Since most glass types are opaque in UV light, it has been suggested to suppress the back-side reflection in DM measurements by using UV illumination~\cite{Sprenger.2010}. As with IR, however, no cheap high-resolution displays are available for the UV spectrum. Therefore, static patterns (e.g., slit masks in front of a UV emitter) must be used with or without scanning. Note that the camera of course needs a lens that transmits UV and a detector that captures it (typically a "backlit" chip). UVDM systems are therefore slower, less versatile, and more expensive than the "standard" DM, and are only useful for special applications.
        
\begin{figure}
            \centering
            \includegraphics[height=2.7cm]{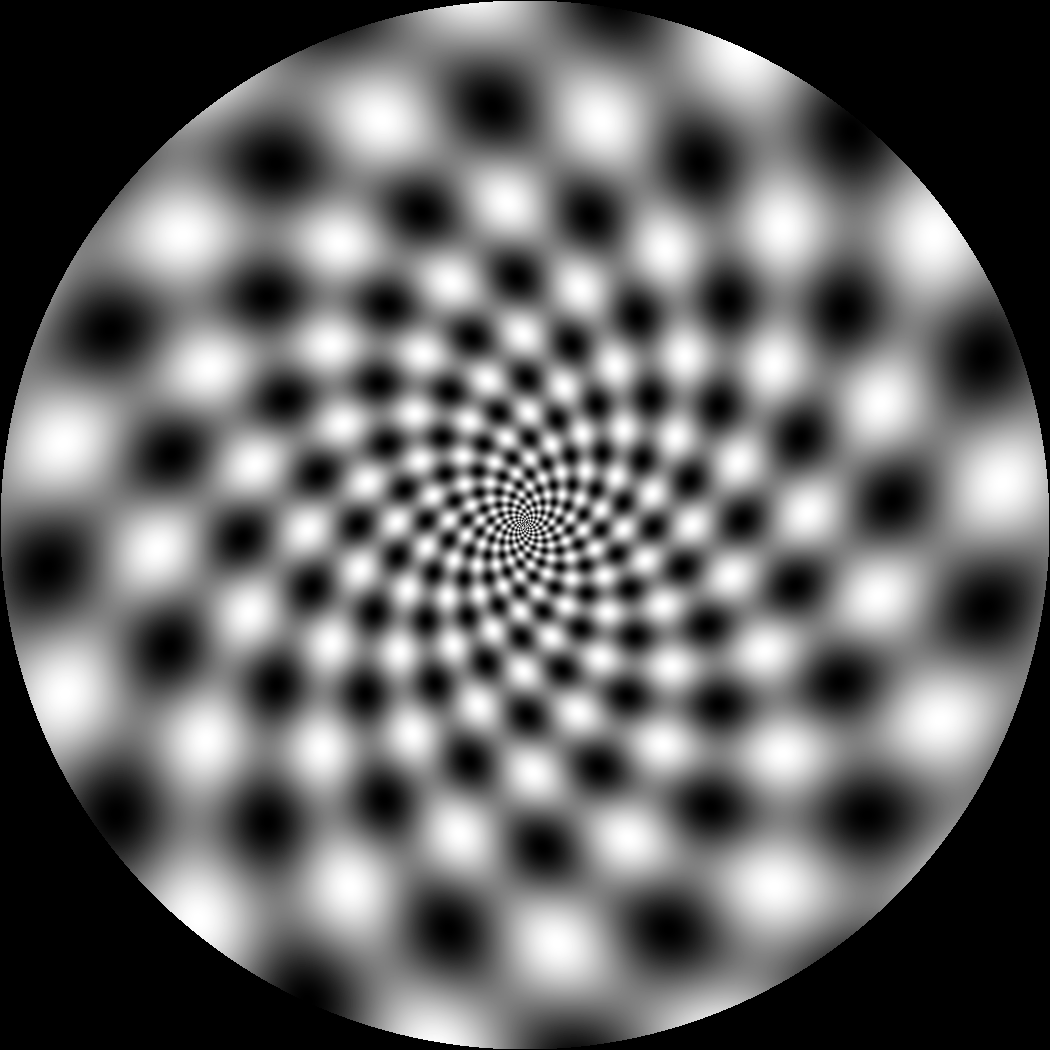}
            \put(-10,3){\colorbox{white}{(a)}}
            \hfill
            \includegraphics[height=2.7cm]{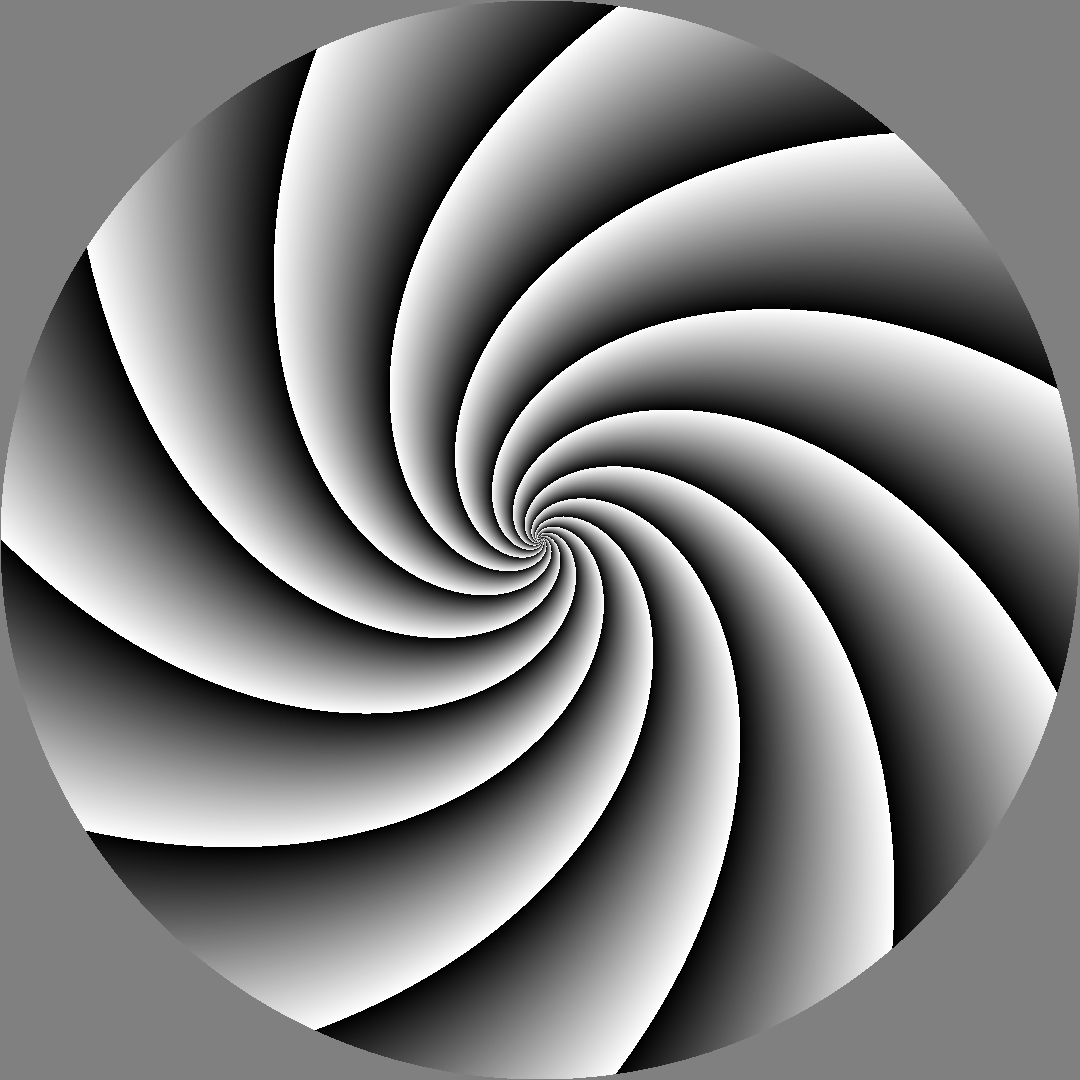}
            \put(-10,3){\colorbox{white}{(b)}}
            \hfill
            \includegraphics[height=2.7cm]{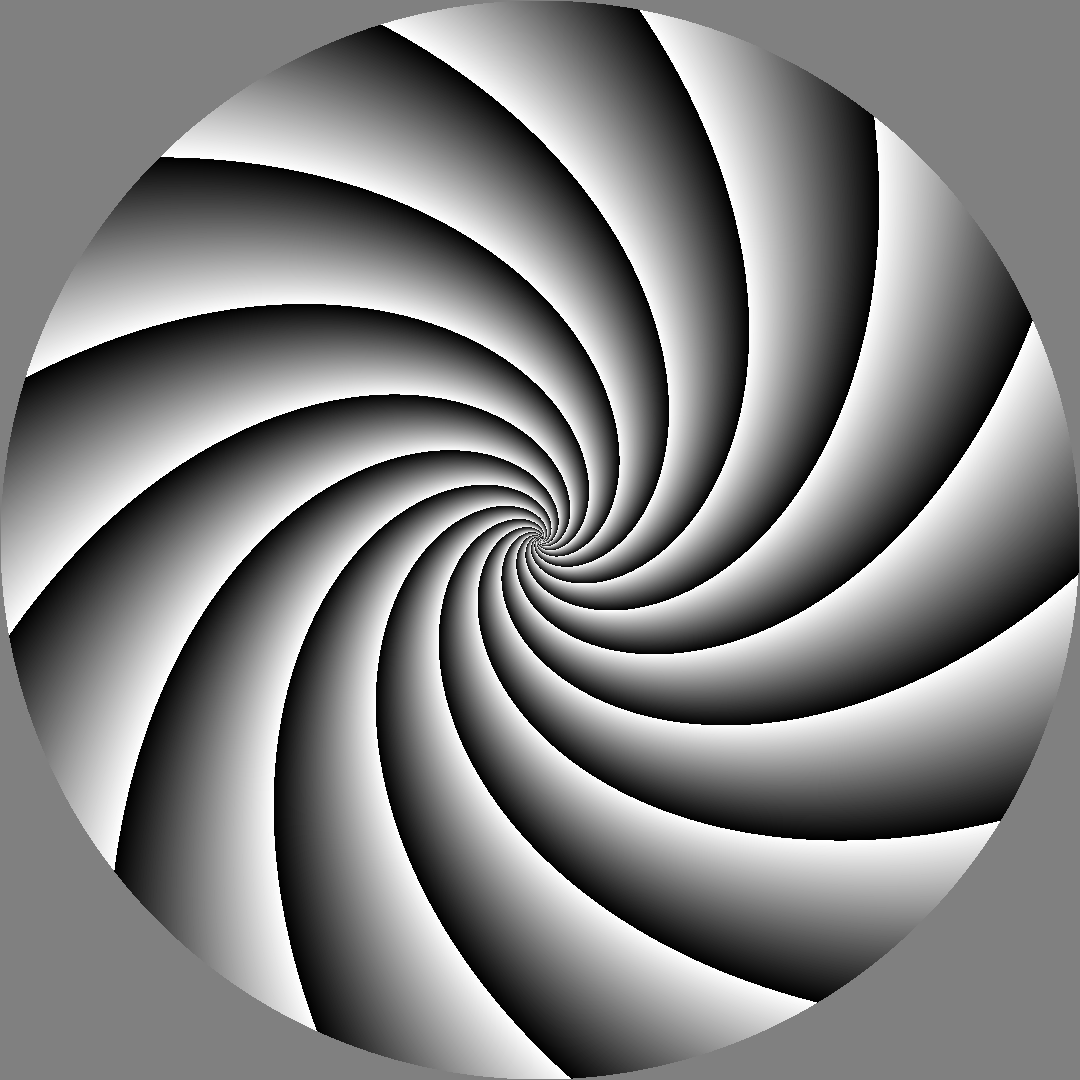}
            \put(-10,3){\colorbox{white}{(c)}}
            \caption{Using logarithmic spirals as coding patterns allows keeping the angle constant between the fringes and the radial/tangential directions. (a) Example with 11 fringes bending clockwise and 13 fringes bending counterclockwise. The natural widening of the fringes creates a (non-adjustable) form of pre-distortion for testing highly curved surfaces. (b),(c) phases of the respective fringe patterns. Note that both sets of fringes must have a radial component for the measurement to work by rotation of a static pattern.}
            \label{fig:spirals}
            \vspace{-0.6cm}
        \end{figure}

        Recently a UV-compatible scheme has been suggested for testing circular optical surfaces that can bring the strengths of phase measurement to bear in UVDM. It is based on multiplexed spiral reference patterns that can be evaluated spatially or (in case the pattern is a rotating mask) temporally; from the recorded phase data, one may decode spatial positions~\cite{Kludt.2018b}. Fig.~\ref{fig:spirals} gives a simple example.
        
    \subsection{Imaging systems}
    \label{sec:camera}
    
    As stated earlier, the photon accumulation by camera sensors is now very efficient and approaches the respective physical limits. When transporting photons to the sensor, most camera lenses designed for industrial vision have a remarkably accurate center of projection and low distortion; in our experience, the image geometry is seldom a cause for concern. On the other hand, the selection of lenses with appropriate resolution remains a challenge since MTFs have many parameters and are usually shown in data sheets only for selected settings. For sensors of ordinary sizes and aspect ratios (e.g., between 4:3 and 16:9), the statements by the manufacturers about the designed pixel size or the sensor resolution are reliable. However, even for high-quality lenses, the resolution is highest near the axis and decreases (sometimes to less than half the maximum) towards the field-of-view boundaries. We have found that many lenses work best at an f-number around 4, and attempts to increase the depth of focus by stopping down the aperture may lead to an unacceptable loss of resolution especially near the edges and corners of the image. No overviews nor plausible taxonomies of system layouts are available in the literature, possibly because one cannot study and compare different approaches with the same test piece, since -- once again -- the measurement geometry must be closely adapted to the geometry of the studied specimen.
  
        \subsubsection {Single camera, single screen}
        \label{sec:singlecam_singlescreen}
        
        This configuration is the entry point to DM; it has no moving parts and is easy to implement on the hardware level, but the concessions to be made may include sometimes insufficient surface coverage of complex objects and/or difficult or only approximate geometrical calibration.
        
        \subsubsection {Multiple cameras, single screen}
        \label{sec:multicam}
        
        There are several reasons to use many cameras in a DM setup, the most common being that several views are needed to cover a sufficient portion of the object surface. In a multi-view setup it is likely that the view fields overlap; such overlapping datasets can be matched correctly if all cameras have been calibrated and their relative orientations are known. For one object position, this is sufficient to synthesize a result; if data from several object positions are combined, relations between them must be known as well. Note that one cannot use common stitching methods in order to align datasets, since multi-camera observations do not automatically yield normal fields that can be compared and merged. In fact, every camera has its own sensitivity field due to varying distances and angles across its image field. If the overlaps between camera fields are large enough, they may be suitable for the application of stereo techniques~\cite{Wang.2021d}.
        
        In stereo or light-field deflectometry~\cite{Horneber.2001, Uhlig.2018, Ziebarth.2018, Xu.2018b, Meguenani.2021}, two or more cameras (or pixels) record the same part of the object (or its entire surface) from different locations and angles. Given the intrinsic and extrinsic camera parameters, it is then possible to reconcile the measured deflections from each camera in a so-called stereo regularization~\cite{Werling.2010}, find consistent surface normals, and reconstruct the surface in metric units~\cite{Kickingereder.2004, Werling.2011b}.
        
        The stereo positions do not necessarily have to be recorded with different cameras: it is possible to cover a surface with measurements from a single camera being moved from position to position and, if all its positions are known with respect to the object, to reconstruct the object surface with a so-called virtual stereo technique~\cite{Balzer.2011b}.
        
        \subsubsection {Single camera, multiple screens}
        \label{sec:singlecam_multiscreen}
        
        It is also possible to improve the surface coverage with more reference screens, either as an actual simultaneous set-up or sequentially by moving the screen to some pre-defined positions. (The latter option is also a relatively popular regularization method: since the screen is moved by a known distance, one may match the respective normal fields and select the unique surface position.) The challenges with several screens are similar to those associated with multiple cameras: one cannot always assume that they are identical, which complicates the calibration and data fusion. Using the same system for multiple positions partially avoids these problems but obviously incurs longer measurement times, and it is still necessary to combine individual measurements. Custom illumination devices such as arcs, tunnels or domes can be built to cover a much larger solid angle around the object than flat screens, but at a greatly increased expense for assembly and control.   
    
    \subsection{Surfaces in motion}
    \label{sec:motion}
    
    Sometimes it is desirable to scan the object in a linear motion even if it could potentially be covered with a sequence of static measurements, because (i) the sensor may be more compact and (ii) some production lines never stop. Tracking the relative positions of components (or exactly adhering to the design) is an important issue here, as an incorrect spatial assignment (especially in a cyclic motion) immediately causes artefacts in the results, which often take the form of temporal sequences of line data stacked into synthetic 2D maps as in Fig.~\ref{fig:ir_scan}.
    
    Unsurprisingly, the necessary amounts of data correlate with the required detection performance (in terms of sizes and the severity of defects to be identified). As an example, hail damage scanners are designed to find defects with lateral sizes of several cm, and depth in the mm range. These are easily detectable with one-dimensional stripe patterns, typically implemented as luminous portal structures. Phase shifting and evaluation of spatial derivatives here are unnecessary (i.e. data volumes can be reduced thanks to a-priori knowledge about the defect shapes). Moreover, the object may move past the sensor at an uncontrolled (but range-limited) velocity.
    
    Similarly, error detection is possible on moving objects in industrial inspection for production. The typical approach here is also to use a single bright line whose reflection is then captured on a scattering screen~\cite{Seulin.2001, Wedowski.2012c, Hugel.15.02.2016, Meguenani.2019}. The exact registration of positions is not very important here either, since the task is simply the detection of defects.
    
    In order to achieve a finer resolution on the surface and a more accurate localization of defects, the object movement must be controlled more tightly (e.g. with a conveyor belt), or alternatively the object should be stopped for a measurement (during which the reference structures move). This appears to be a practical way to obtain decent spatial registration, since the relative movement between the object and cameras introduces unwanted uncertainties. As a result, moving cameras are not used in any systems that we are aware of. Other solutions may involve semi-qualitative evaluation~\cite{Arnal.2017} as well as custom reference structures that allow two-dimensional decoding and even surface reconstruction~\cite{Penk.2020} from the temporal sequence of reflection data.
    
    Finally, robotic applications with a moving object and adapted active fringe patterns have been reported but they remain rare and relatively complex~\cite{Schwarz.2016}. 

    \subsection{Optimal inspection planning}
    \label{sec:planning}
    
    As discussed in Sec.~\ref{sec:geometry_size}, large or high-curvature convex surfaces may require multiple measurements with different placements of cameras and light sources in order to assess the entire surface. Finding the optimal placement manually is time-consuming and complex; automated optimization of DM setup parameters aims to mitigate this problem. Whether or not multi-pose procedures are the goal, starting with detailed simulations of the measurements~\cite{Li.2012c, Li.2014e, Huang.2016, Xu.2018b, NimierDavid.2019} is almost always a good idea. On the other hand, all simulations end up with more or less deviation from actual experimental data. Unfortunately, we are not aware of any systematic attempts to explore and compare the reliability of simulation approaches; this may be due to the role of specific implementations, which are typically not being published.
    
    The optimization complexity depends on whether the setup is static or dynamic. In a static setup, the placement of light source(s), camera(s), and the studied surface(s) is optimized at the design phase and remains fixed during measurements. In a dynamic setup, at least one component can be adjusted during the measurement. For example, an integrated sensor consisting of a display and a camera can be mounted on an industrial robot for a total of six "active" degrees of freedom. In this case one needs a plan -- a sequence of configurations for the controllable element.
    
    While static setups usually allow faster measurements (multiple cameras can be used in parallel to capture images), dynamic systems are sequential and require longer cycle times. On the other hand, depending on the available degrees of freedom, dynamic setups can be more efficient when inspecting complex shapes and multiple surface variants according to different respective plans.
    
    Only a few publications address the automated optimal design of DM constellations. Incipient stages of design optimization in DM can be found in~\cite{Kammel.2004} where a semi-automatic method is presented to position a camera and a display relative to the test surface for a single measurement. Iterations of bounding-box and fringe visibility computations are reported in~\cite{Xu.2012b}; \cite{Lobachev.2013} describes a device for the detection of hail dents on passenger cars and introduces a method to find the optimal camera positions for a static setup with fixed displays (projection screens). The algorithm minimizes a cost function that takes the visibility of multiple cars into account.
    
    While these works focus on the visibility (coverage) as the optimization criterion, \cite{Roschani.2020} uses probabilistic methods to optimize the measurement uncertainty. More specifically, the proposed algorithm minimizes the residual uncertainty of normal vectors or derived heights and attempts to reduce them below the level specified by the inspection task.
        
    All of the methods cited above assume that the reference surface is known. This is true in inspection tasks and considerably simplifies the planning process. In some applications, however, such information is not available. The optimization of DM inspection plans without prior knowledge of the reference surface or other simplifying assumptions about the studied object remains an important open problem.


\section{Data Acquisition and Processing}
\label{sec:processing}

We have already discussed the optical aspects of obtaining the reflected images of the reference screen and the integration of metric slope data. Here we concern ourselves with obtaining the slope maps from the captured information and with the alternative uses of that data when the 3D reconstruction is not needed. In most practical cases one can get away without it -- or conversely, since it is quite difficult, practice has evolved to avoid it where possible.

Depending on the nature and purpose of the measurement, there is an ever-widening selection of ways to acquire the necessary data, where of course the choice is affected by the available hardware for a given system design -- for it is the object's geometry that determines the possible system constellations. Once again due to the law of reflection, there is considerably less freedom for cherry-picking coding schemes in DM than e.g. in fringe projection~\cite{Gockel.2006, Salvi.2010, Zhang.2012d, Zuo.2016, Zuo.2018, Zhang.2018, Xu.2020b}. The choices that do exist can be weighted by a custom merit function as in the example in~\cite{Butel.2012}, and many coding methods developed and refined in fringe projection are equally as applicable in DM (which is why we present some methods here that have not yet been used in DM). Our categorization below is a best effort, but cannot be completely sharp because techniques are increasingly being combined as they evolve further.

    \subsection {Position correspondence coding}
    \label{sec:coding}
    
    In most cases, at least a qualitative description of relative ray directions is necessary to take advantage of the natural sensitivity of DM. As we will see, there are many ways to obtain this information.
    
        \subsubsection{Scanning methods}
        
        In Section~\ref{sec:scanning} we have already described point and line scanning techniques for deflectometry for which there is little alternative outside the visible range; in this context, let us point out that point-scanning techniques can yield one surface normal per measurement, whereas line-scanning techniques create ambiguity in the direction of the reflected line that must be resolved by scanning again in the orthogonal direction~\cite{Hofer.2016b}. In general, scanning techniques are useful where otherwise the evaluation may be impaired by the superposition of signals corresponding to multiple surfaces (or, when inspecting diffractive optics, to multiple diffraction orders~\cite{Chen.2020c}). The position in this case is often found as the centroid of the captured intensities for each location of the scanning unit; the latter must be sufficiently accurate in order to avoid wave-like artefacts on the reconstructed surface.
        
        \subsubsection{Discrete codes}
        
\begin{figure}
            \centering
            \includegraphics[width=0.95\columnwidth]{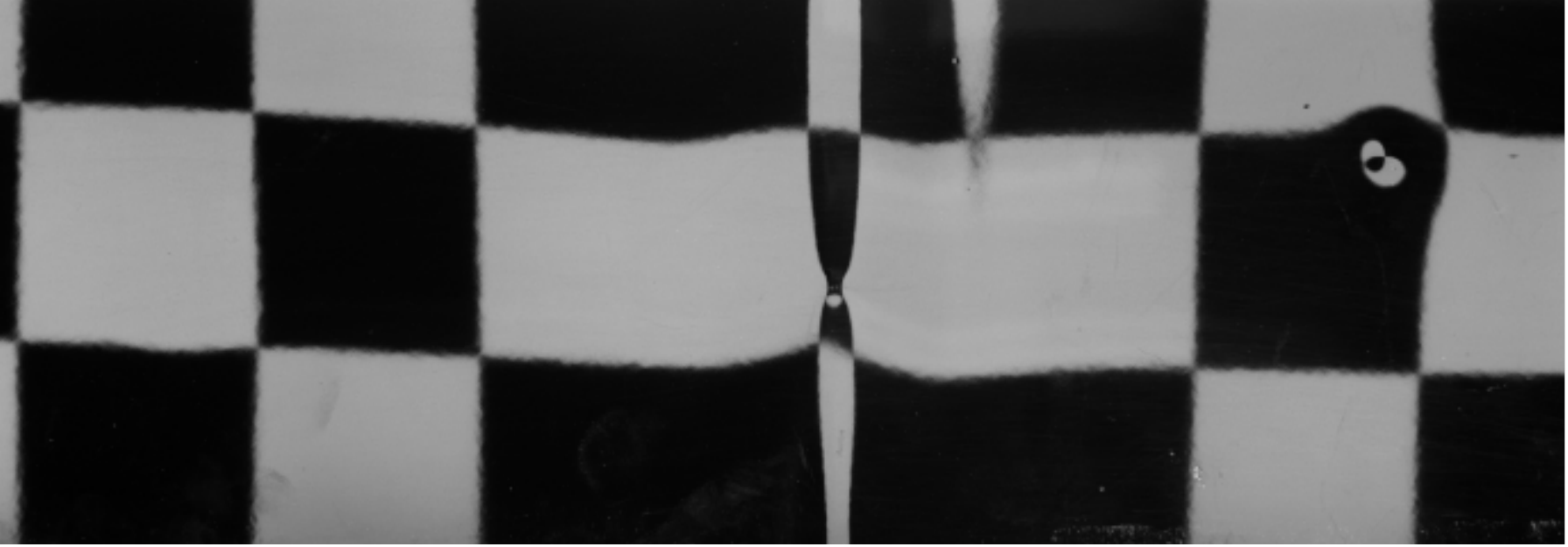}
            \put(-15,5){\colorbox{white}{(a)}}
            \vspace{0.1cm}
            \includegraphics[width=0.95\columnwidth]{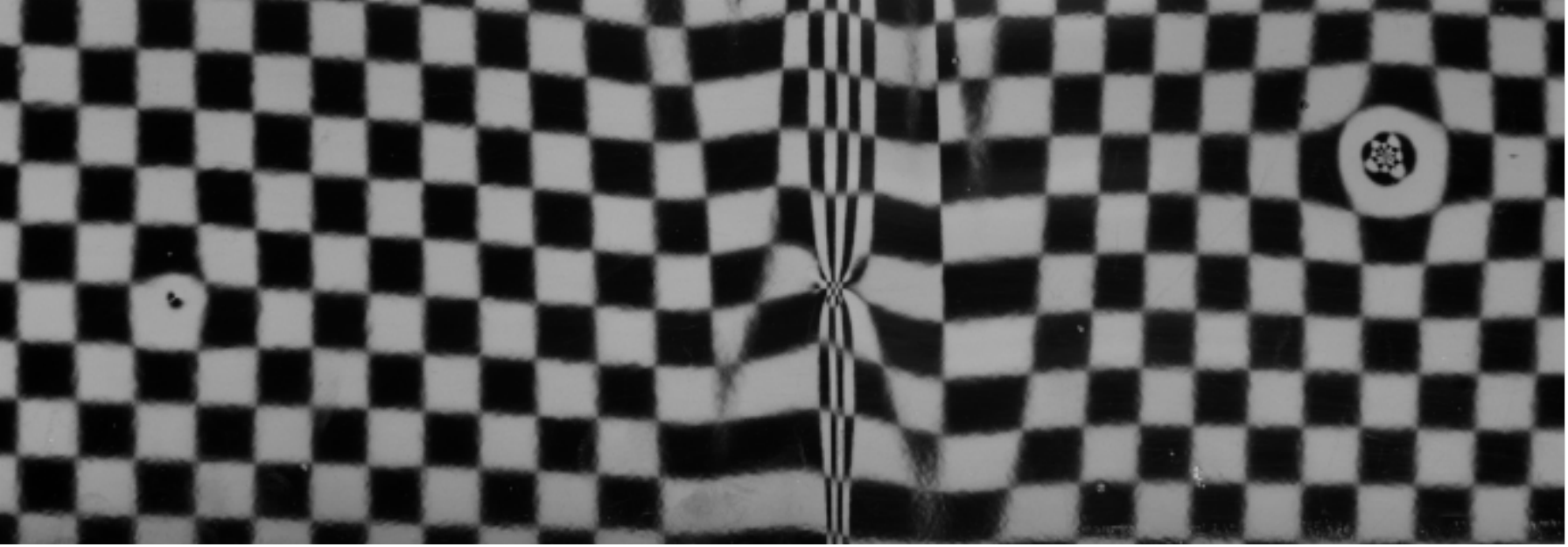}
            \put(-15,5){\colorbox{white}{(b)}}
            \vspace{0.1cm}
            \includegraphics[width=0.95\columnwidth]{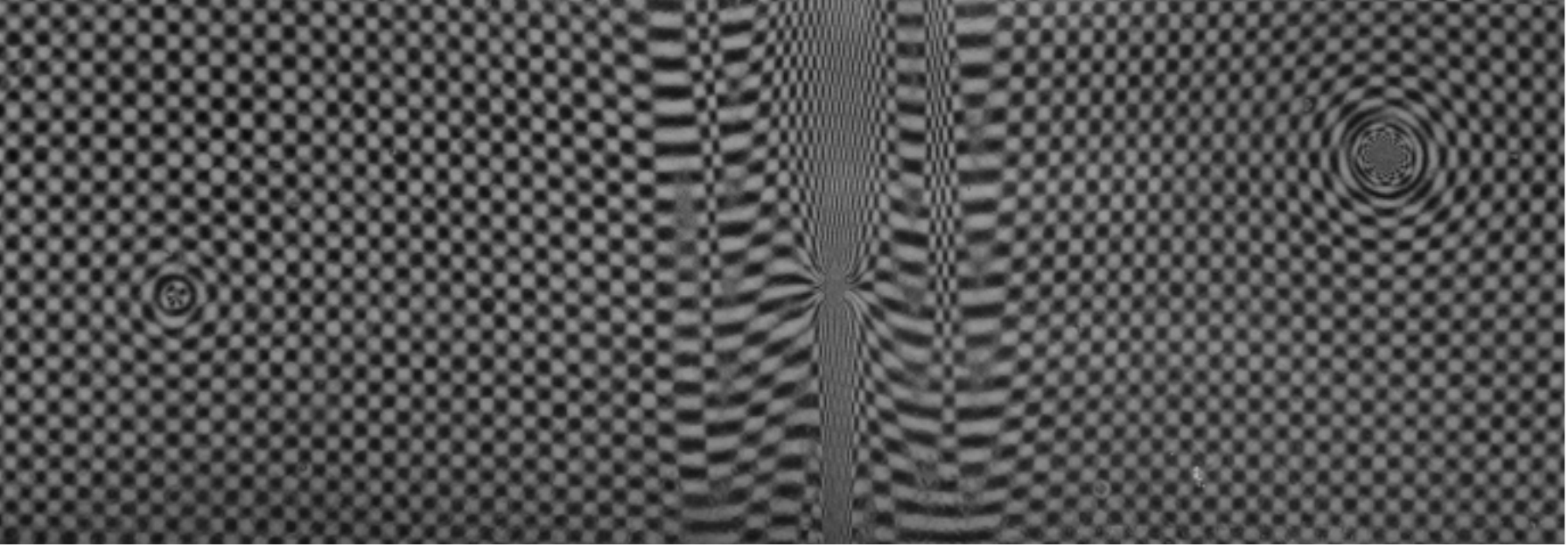}
            \put(-15,5){\colorbox{white}{(c)}}
            \caption{Reflections of checkerboard patterns from black painted metal sheet with kinks and dents; square sizes on screen are 240 pixels (a) to 15 pixels (c), in geometric progression. The dent on the right is always visible, but the reflections are difficult to interpret even in (a); the kink in the center demagnifies the squares strongly in the horizontal direction; the defect on the left is initially invisible; several smaller defects are hardly noticeable even with the smallest checkerboard pattern.}
            \label{fig:checkerboards}
            \vspace{-0.6cm}
        \end{figure}

        An early example of location coding is the family of binary (i.e. dark/bright) patterns known as Gray codes ~\cite{Gray.1953}, where the name refers to the inventor, not to the pattern. Originally designed as an error-suppressing data transmission scheme (where the original "reflected" binary approach indeed provides the best error suppression~\cite{Agrell.2004}), Gray codes can theoretically be used without gray-level calibrations. On the downside, finer spatial encoding requires a large number of frames and is bound to encounter difficulties when the reflected code pattern becomes blurred; for solutions to a subset of the issues, see~\cite{Butel.2014}. Nonetheless, binary patterns are no longer in common use in DM or fringe projection. Fig.~\ref{fig:checkerboards} demonstrates a set of binary images (but note that the checkerboard patterns as shown are not a coding sequence).
        
\begin{figure}
            \centering
            \includegraphics[width=0.95\columnwidth]{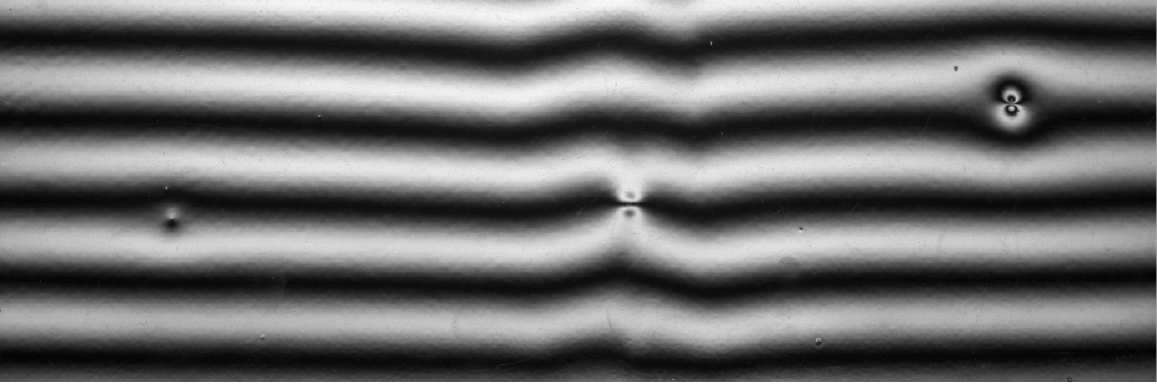}
            \put(-15,5){\colorbox{white}{(a)}}
            \vspace{0.1cm}
            \includegraphics[width=0.95\columnwidth]{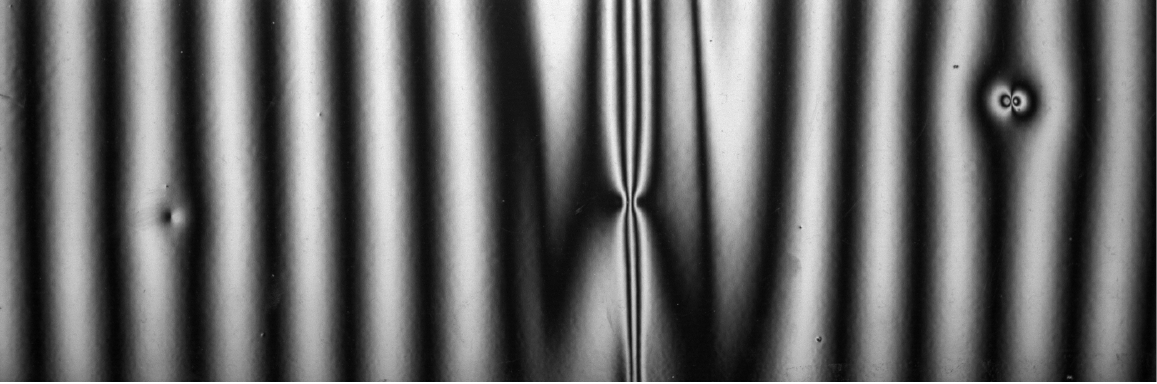}
            \put(-15,5){\colorbox{white}{(b)}}
            \caption{Reflections of cosine fringes from the same part as in Fig.~\ref{fig:checkerboards}; note how the smooth intensity profile of the reference pattern makes all kinds of defects more visible and thus more detectable.}
            \label{fig:cos_resolution}
            \vspace{-0.6cm}
        \end{figure}

        One possible upgrade has been the addition of dense phase data only in the last encoding stage, with coarser spatial assignments given by Gray codes ~\cite{Sansoni.1999, Yu.2018}. Fig.~\ref{fig:cos_resolution} gives an impression of the benefit of this approach by way of just one fringe pattern per direction.
        
        A comparison of the coding efficiency and error tolerance between Gray codes and phase measurement has been given in ~\cite{PorrasAguilar.2016}. Various attempts with ternary and quaternary Gray and gray codes have been reported~\cite{Zhang.2012f, Xu.2016, Zheng.2017b, He.2019, Yang.2021}, and it has been shown that the measurement benefits from ordering the codes for minimal spatial intensity gradients ~\cite{PorrasAguilar.2017b, Falaggis.2018}.

        \subsubsection{Aperiodic and pseudo-random codes}
        
        Besides deterministic code sequences, it is also possible to encode positions via pseudorandom patterns ~\cite{Sjodahl.1999, Wiegmann.2006, Heist.2016, Pak.2019b}. In comparisons with phase-shifting sequences~\cite {Kuhmstedt.2007c, Heist.2015}, it has been shown that the correspondence assignment with pseudo-random sequences can be almost as reliable~\cite{Lutzke.2013, Schaffer.2014} -- and potentially much faster -- if the pattern parameters are chosen correctly~\cite{Grosse.2013, Heist.2018}. The reported frame rates are in the kHz range and rely on fast custom projector units; with LC displays one must expect much longer measurement times since these are often surprisingly slow to switch from one pattern to the next one.
 
        \subsubsection{Phase-measuring methods}
        \label{sec:phase_measuring}

        The high resolution and low uncertainty of the phase-shifting method (Sec.~\ref{sec:cos_patterns}; see also Sec.~\ref{sec:history}) has unlocked the full potential of DM, not least because a defocused or otherwise smoothed cosine is still a cosine (albeit with lower contrast)~\cite{Huang.2014}. That means the decision where the camera is focused in the instrument will not affect the validity of the decoding results. Also, there are no undesired phase shifts once the fringe pattern is steady on the projector or LC display; thus the entire body of research dealing with phase-shift miscalibrations is unnecessary for DM, and a generic four-step phase-shifting method can usually be used with no penalties. However, when measuring partially specular surfaces, variations in background light or even re-distribution of scattering in the instrument due to switching fringe patterns may cause changes in the background intensity; hence, it may be necessary to compensate for background variations~\cite{Surrel.1997b}. Likewise, in most computer screens and projectors the relationship between the digital gray level and luminance depends on the graphics driver and its setting, is usually non-linear, and often does not span the full input or output range. These deviations may be interpreted as harmonics in the fringe profile, and comprehensive recipes have been given on how to suppress them by an appropriate design of the phase-reconstruction method~\cite{Surrel.2000, Burke.2012b}; however, typically the screen or projector is linearized by a look-up table or a fitted function before use~\cite{Ma.2012c, Zhang.2014k}. Thus, systematic phase errors can be removed almost completely.
        
        The remaining statistical errors mostly come from the camera as electronic and photonic noise~\cite{Rathjen.1995, Surrel.1997, Belsher.1999}; digitization noise is, in our practical experience, very seldom an issue, even for 8-bit cameras. Electronic noise only plays a role when the fringe modulation is very low (less than 3 bit); under practical conditions, photon shot noise is the main noise source ~\cite{Walkup.1973, Belsher.1999}. The physics of camera sensors has been thoroughly described and characterized in the EMVA 1288 guideline ~\cite{EMVA.2021} and has been utilized for understanding phase errors since ~\cite{Fischer.2011, Fischer.2012b}. Importantly, the measurement will benefit from choosing a phase-reconstruction method whose statistical error does not oscillate with the phase ~\cite{Bothe.2008}. Also, a higher static background intensity contributes extra photon noise, but no signal; therefore, measurements on partially specular surfaces are best carried out in the darkest possible environment.
        
        As opposed to the three previous methods, the fringe patterns used in phase-measuring DM are 2D and periodic. The maximum sensitivity corresponds to the smallest possible fringe period, which results in many nominally identical fringes being present in the captured images, and the necessity to "unwrap" the phase. For continuous surfaces, the simplest approach is to use localized intensity alterations as markers, in one or several locations~\cite{Zhang.2006j, Gai.2010b, Cui.2012, Liu.2014e, Xing.2017, Hu.2018}, or even on every fringe~\cite{Budianto.2014, Budianto.2016}, and apply spatial phase unwrapping from the reference point or line to recover the positions. In well-controlled geometries where the sample can be placed in the same position as the calibration surface, it is also possible to refer to calibration data for the correct geometry reconstruction.
        
\begin{figure}
            \centering
            \includegraphics[width=0.98\columnwidth]{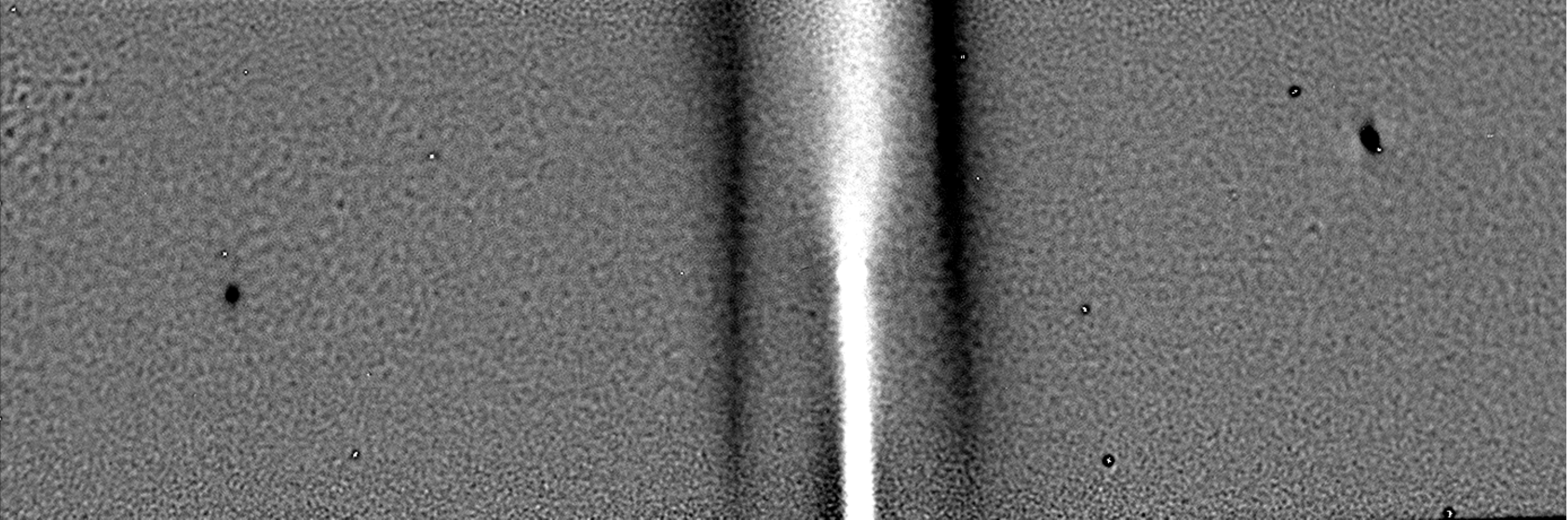}
            \put(-15,5){\colorbox{white}{(a)}}
            \\
            \vspace{0.1cm}
            \includegraphics[width=0.98\columnwidth]{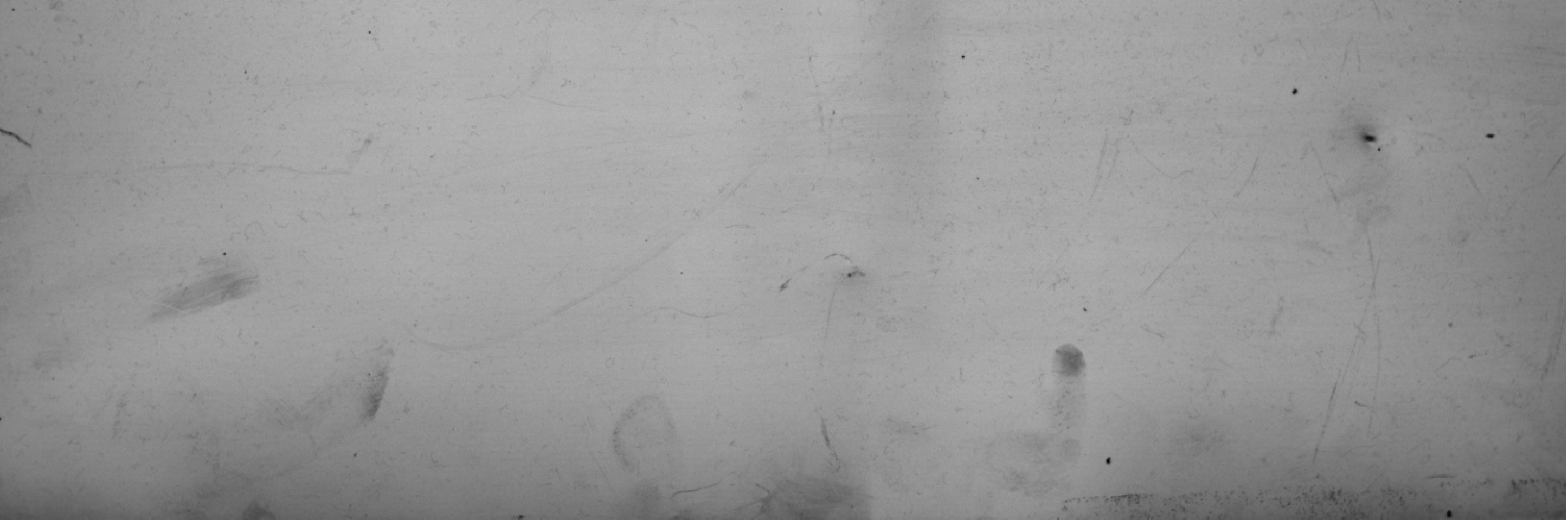}
            \put(-15,5){\colorbox{white}{(b)}}
            \caption{Test of the same surface as in Fig.~\ref{fig:checkerboards} with measurands derived from phase-shifted fringe patterns. (a) Curvature: kinks, dents, dust inclusions, and "orange-peel" paint waviness can be readily distinguished. (b) Fringe modulation amplitude as a proxy for glossiness; minor contamination not visible in Fig.~\ref{fig:checkerboards} or Fig.~\ref{fig:cos_resolution} is easily detectable.}
            \label{fig:multi_crit}
            \vspace{-0.6cm}
        \end{figure}

        The necessary sequence of fringe patterns is certainly a burden on the measurement time; but in contrast to binary sequences, the result is more than just ray correspondences: by virtue of Eqs.~\ref{eqn:encode} and ~\ref{eqn:decode}, one also obtains fringe-modulation and reflectivity data \textit{for each phase-shift sequence with $N\geqq3$}, so that various properties of the surface can be inspected at once. Fig.~\ref{fig:multi_crit} shows an example with curvature and modulation data from the same test piece.
        
        Spatial-only phase coding, and thereby one-shot recording of derivatives, requires de-multiplexing signals from crossed fringes (mostly by appropriate filtering in frequency space, but also wavelet transforms~\cite{Budianto.2014, Liang.2020}, regularization~\cite{Villa.2000} and deep-learning methods~\cite{Qiao.2020, Qin.2020}) and has been demonstrated mostly for flat mirrors or other optical elements that do not strongly affect the fringe spacing~\cite{Nguyen.2021}, or with pre-distorted fringe patterns~\cite{Massig.2001, Massig.2005, Surrel.2006, Liu.2009f, Huang.2011, Flores.2013, Liu.2013d, Liu.2014d, Liang.2016, Wu.2016, Trumper.2016, Nguyen.2019}. Again, the well-controlled geometry and/or differential nature of these measurements makes it possible to use only one coding frequency per direction. This approach should be preferred if possible, since superimposed reference patterns must share the dynamic ranges of screen and camera, which results in a lower modulation amplitude and thus SNR for each.

        All of these methods are inapplicable if the fringes cannot be tracked reliably, which happens e.g. either on surfaces with height steps, or (in the context of DM) in the presence of slope steps or peaks: large curvatures may create unresolvable or interrupted images of fringe patterns, which causes failures in decoding and unwrapping.
        
        Therefore, it is quite common to code positions with a sequence of phase measurements at different $L$ and then to use temporal phase unwrapping techniques for unique position recovery. Originating from two-wavelength interferometry, much research has been dedicated to finding the largest unambiguous range with the highest reliability. For integer ratios between the fringe periods, there exist number-theoretical tools that enhance the measurement range~\cite{Gushov.1991,Gushov.2013b}. Further research~\cite{Osten.1996, Surrel.1997c, Towers.2004c, Xia.2005, Falaggis.2011c, Falaggis.2013, Falaggis.2014, Li.2016, Xiao.2017, Wei.2017, Allgeier.2019, Petz.2020b} has provided an even better understanding of the available options~\cite{Falaggis.2018b}. An exciting new trend in multi-period coding is to use of the sum of individual phases (as opposed to the difference) to achieve narrow virtual fringe spacings that are no longer resolvable optically~\cite{Servin.2015, Xiong.2017, Wang.2018b, Servin.2018, DeMars.2019, Xie.2019b}. The actual narrow fringe patterns are best designed so that their period (and their advancement during phase stepping) is an integer number of pixels. This can help to avoid issues with another possible source of digitization noise: with only a few intensity samples per fringe period \textit{on the fringe display}, the modulation sequence computed and shown by the display can depend on the relative phase between the fringes and the pixel matrix.

        Temporal position coding is the safest approach for complex objects; if one can assume that the fringe patterns do not deform much, it is also possible to encode the required frequencies into one phase-shifting sequence in a so-called composite fringe pattern~\cite{Li.1997b, Guan.2003, Sansoni.2005, Su.2006, Kim.2009, Liu.2010, Kludt.2018b, Xie.2019b}. Above, we have discussed crossed fringes and how they must share the dynamic range, and the same is of course true of signals mixed for a pseudo-temporal phase shift. 
        
        The techniques described so far ensure either a unique coding for one spatial derivative, or coding for both directions that still needs to be regularized. Unsurprisingly, even more development was devoted to multiplexing all the information into one pattern at once. For fringe projection, the correspondence problem is easier to solve, and crossed fringes have been used in combination with advanced unwrapping approaches~\cite{Takeda.1997, Zhong.2001b, Choudhury.2002}.
        
        The addition of color information allows simultaneous three-step phase shifting~\cite{Tarini.2005, Castillo.2013, Flores.2015, Trumper.2016}, multi-frequency display~\cite{Zhang.2006l}, or direction separation by color~\cite{Liu.2009f, Wu.2016, Ma.2018, Xie.2019b}, including simple multi-frequency coding for easier unwrapping. Even discrete codes have been explored for use with colors~\cite{Xu.2016, Zhou.2017}, and a combination of color and direction coding has been presented as well~\cite{PoirierHerbeck.2021}. Color techniques of course work best on mirrors and glass, i.e. surfaces or media with no color of their own; but even then, one has to deal with the cross-talk between channels and the spectral dependence of the photon conversion efficiency. Moreover, on a patterned sensor, the color capability is traded for spatial resolution, while true three-color sensing amounts to tripling the number of sensors. On the other hand, the issue of limited dynamic ranges per signal is less important or can be avoided in this way; and if desired, object colors can be captured as well~\cite{Willomitzer.2020}. Dispersion effects and the sub-pixel location dependence of the color channels in LCD screens are always implicit, and this is true even when "white" light is used.

        \subsubsection{Inverse patterns}
        \label{sec:inverse_patterns}
        
\begin{figure}
            \centering
            \includegraphics[width=0.33\columnwidth]{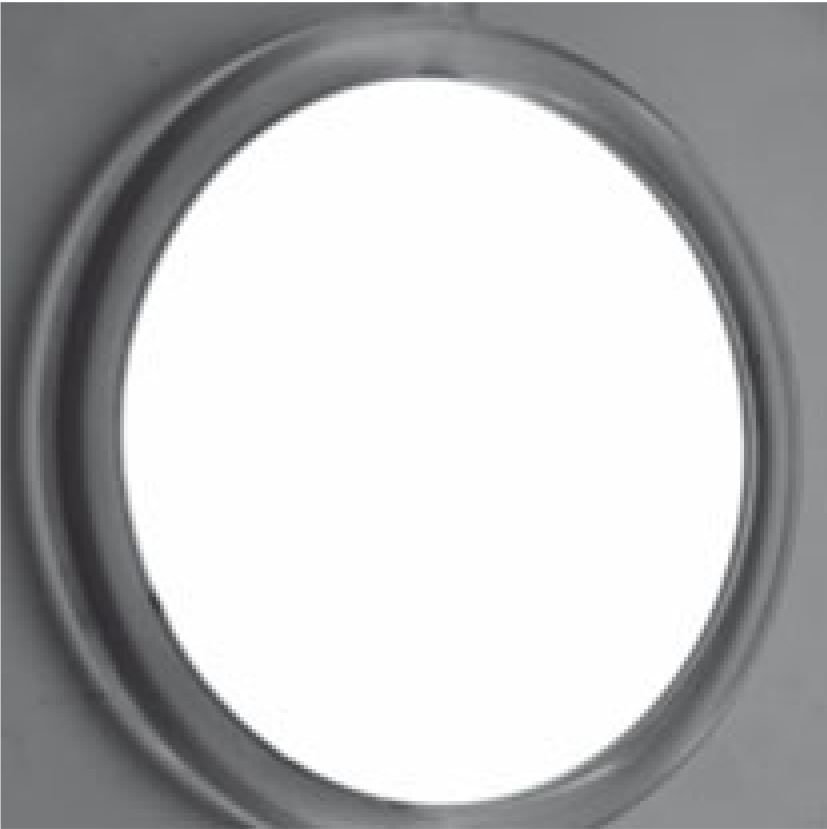}
            \put(-12,3){\color{white}(a)}
            \hfill
            \includegraphics[width=0.33\columnwidth]{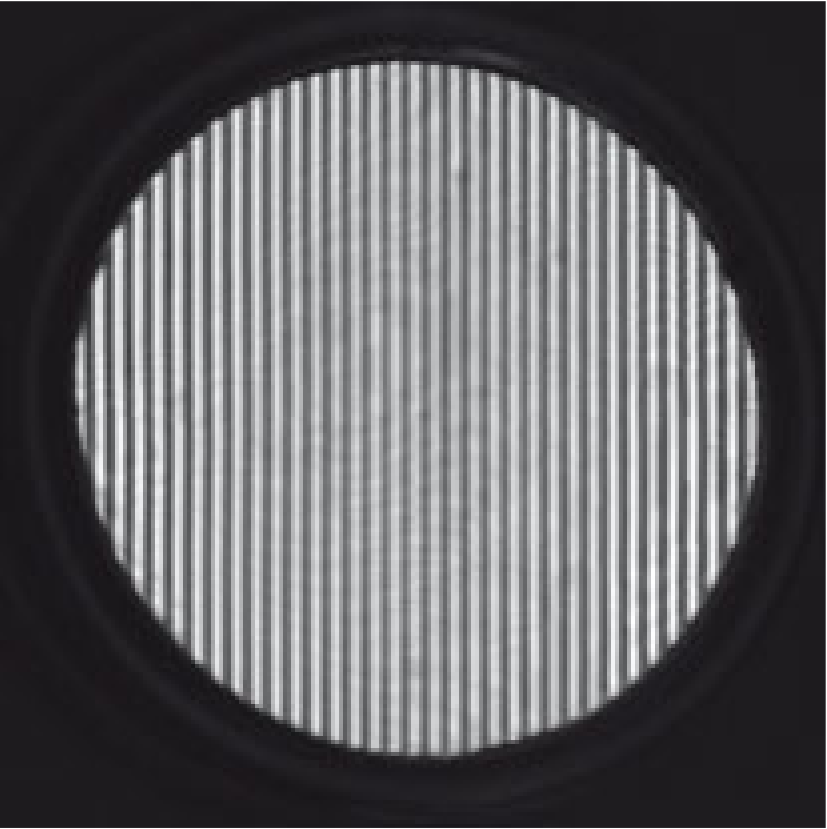}
            \put(-12,3){\color{white}(b)}
            \hfill
            \includegraphics[width=0.33\columnwidth]{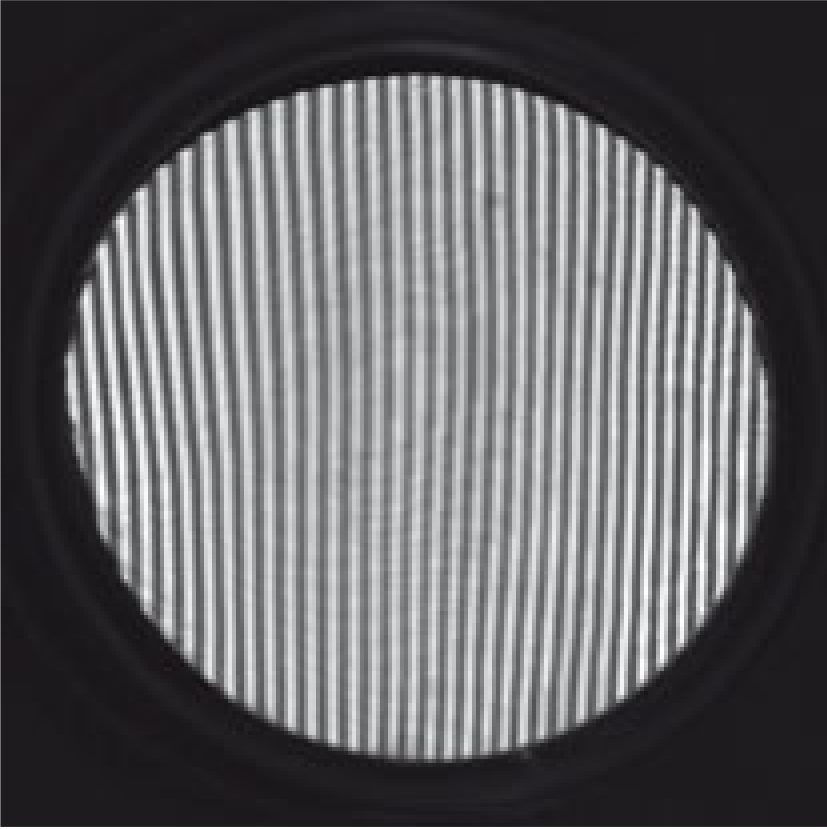}
            \put(-12,3){\color{white}(c)}
            \\
            \vspace{0.1cm}
            \includegraphics[width=0.33\columnwidth]{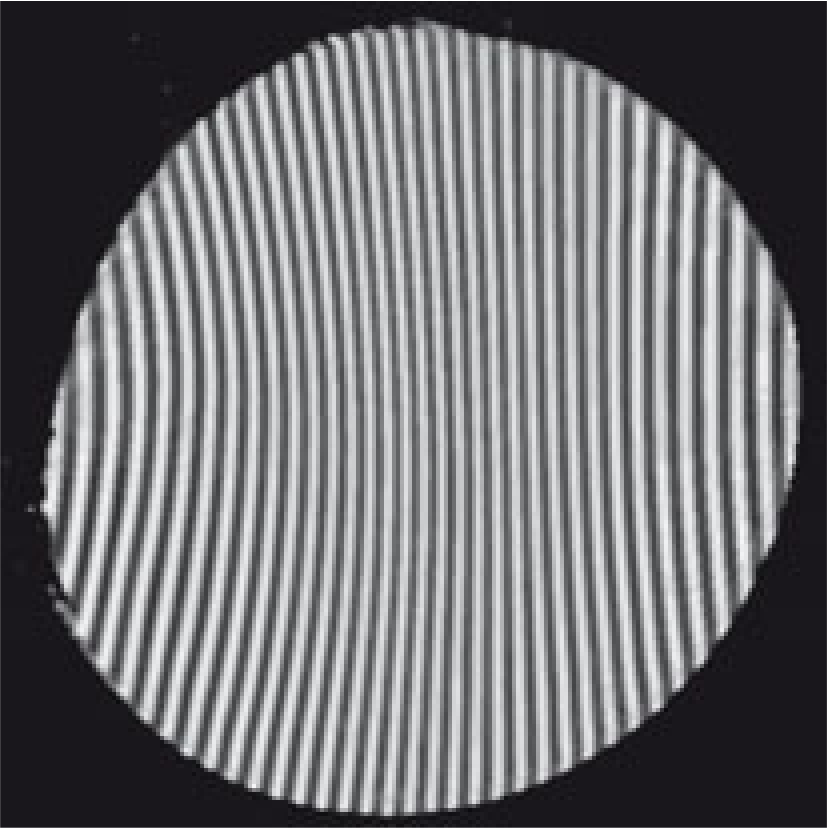}
            \put(-12,3){\color{white}(d)}
            \hfill
            \includegraphics[width=0.33\columnwidth]{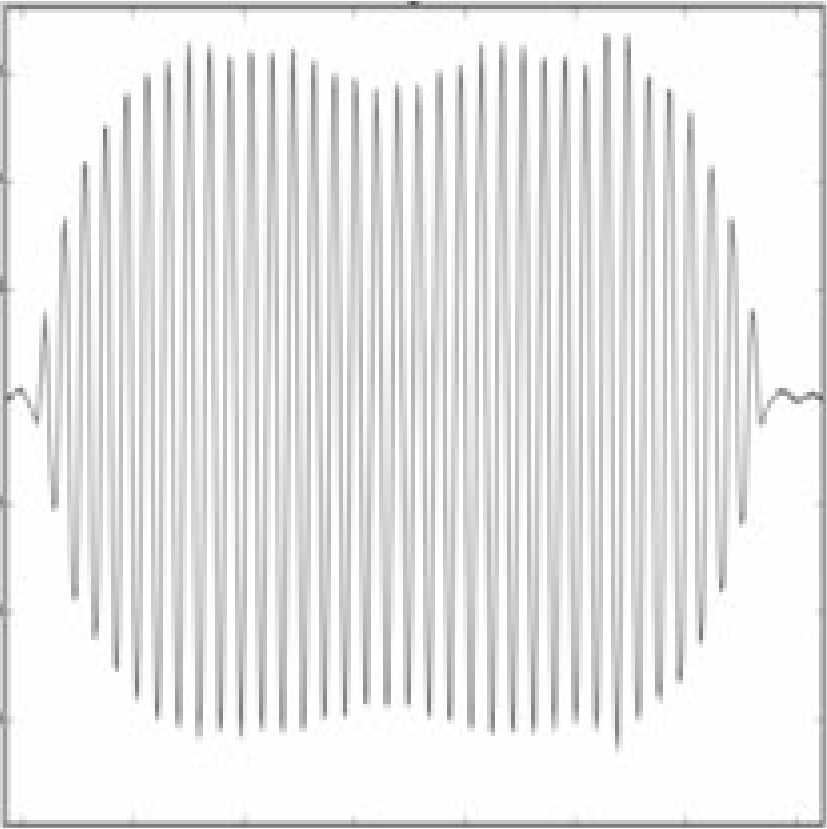}
            \put(-12,3){(e)}
            \hfill
            \includegraphics[width=0.33\columnwidth]{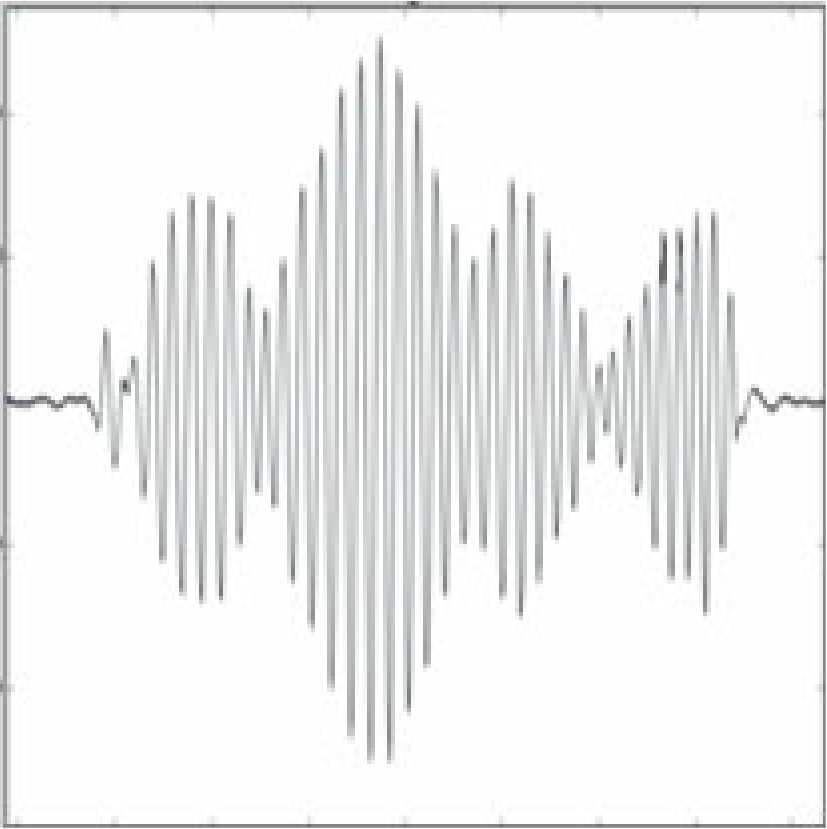}
            \put(-12,3){(f)}
            \caption{Inverse patterns for DM inspection of a magnifying mirror~\cite{Werling.2007c}: (a) reference object; (b) DM image of intact reference object; (c) DM image of deformed object, here a curved shaving mirror; (d) inverse pattern displayed on screen; (e) evaluation for reference object, vertical projection of (b); (f) evaluation for deformed object, vertical projection of (c).}
            \label{fig:inv_fringes}
            \vspace{-0.6cm}
        \end{figure}

        In serial tests of similar objects, one can alleviate the decoding difficulties that stem from excessive shape, modulation, or color-related effects on the reflected fringe patterns by pre-conditioning the patterns themselves. In other words, one generates "inverse patterns" such that they appear very regular upon reflection from an object of the prescribed (reference) geometry~\cite{Haist.2002, Li.2004b, Massig.2005, Werling.2007c, Tosun.2007, Caulier.2011, Liang.2016, Kludt.2018b, CamposGarcia.2019}. Although computationally demanding, such methods may provide easy visual cues and simplify automatic evaluation and/or error detection, as seen in Fig.~\ref{fig:inv_fringes}.
        
        \subsubsection{Handling multiple reflections}
        \label{sec:multi_reflection}
        
        Multiple reflections from the same surface, which cannot be avoided for some object geometries, are an unresolved problem so far -- unlike multiple reflections from different surfaces, such as two interfaces of a lens or another transparent object. Before the introduction of suitable decoding methods for mixed signals in DM, the obvious approach was to suppress the effects of transparency, either by eliminating the rear-side reflection~\cite{Synowicki.2008} or by using wavelengths for which the specimen is opaque~\cite{Sprenger.2010}. Known intensity differences between front and rear side reflection can be accounted for by intensity thresholding~\cite{Xu.2010, Xu.2012b}.
        
        As mentioned before, scanning techniques can be suitable to avoid superposition of signals, either as a single line or, if the object is sufficiently uniform, with the densest possible array of bright or (to enable phase measurements) sinusoidally modulated lines~\cite{Guhring.2000, Faber.2009, Wang.2019c}.
        
        Unless the object is almost flat on both sides, however, it is still difficult to avoid a superposition of reflections. In an interesting analogy to interferometry, where wavelength tuning was used to assess multiple surfaces at once~\cite{Okada.1990, Deck.2001b, Burke.2007}, it was found that the frequency-variation approach also works in DM. In this case, multiple measurements with varied fringe periods allow one to separate several mixed signals~\cite{Faber.2009, Huang.2012b, Ye.2021, Leung.2022}. Attempts have also been reported to use a model of the optical system's transfer function for signal separation~\cite{Wang.2018b, Tao.2019}. We are currently unaware of any efforts to implement the multi-frequency approach by means of any of the multiplexing schemes mentioned above.
        
\begin{figure}
            \centering
            \includegraphics[width=0.90\columnwidth]{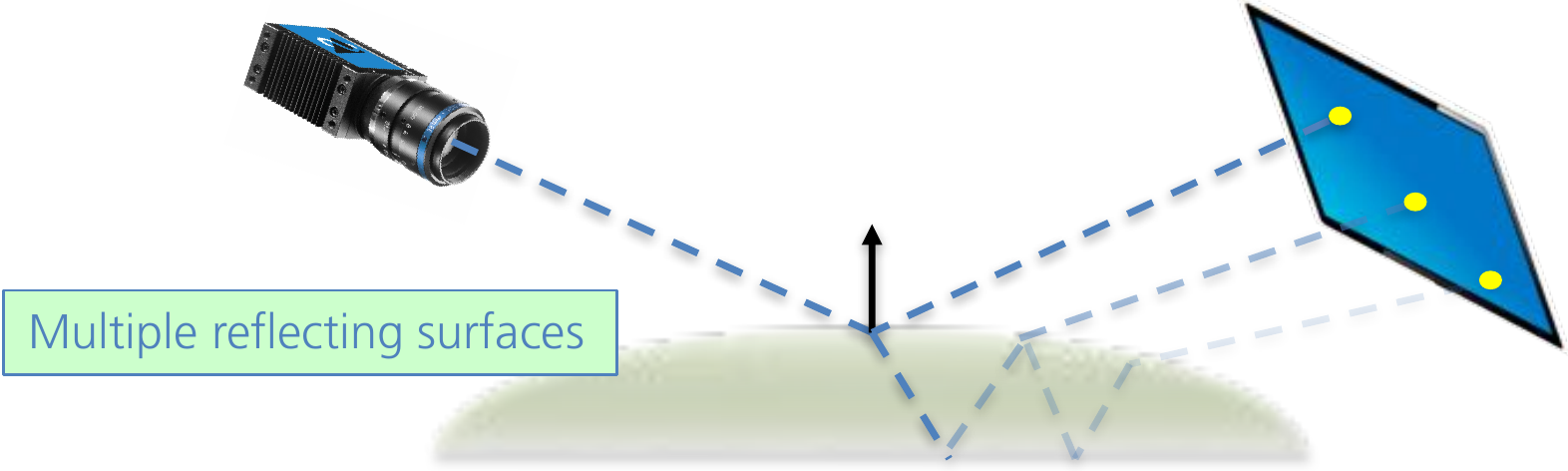}
            \put(-20,5){(a)}
            \\
            \includegraphics[bb=40 20 600 475, clip, width=0.9\columnwidth]{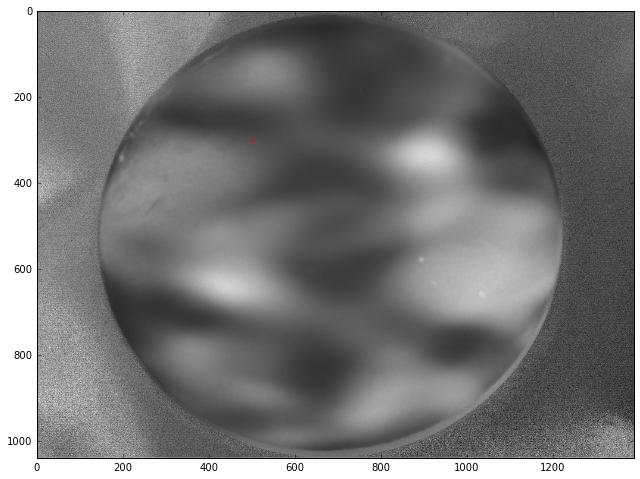}
            \put(-146, 126){\Large\color{yellow}+}
            \put(-20,10){\fcolorbox{black}{white}{(b)}}
            \\
            \includegraphics[bb=40 20 700 475, clip, width=0.9\columnwidth]{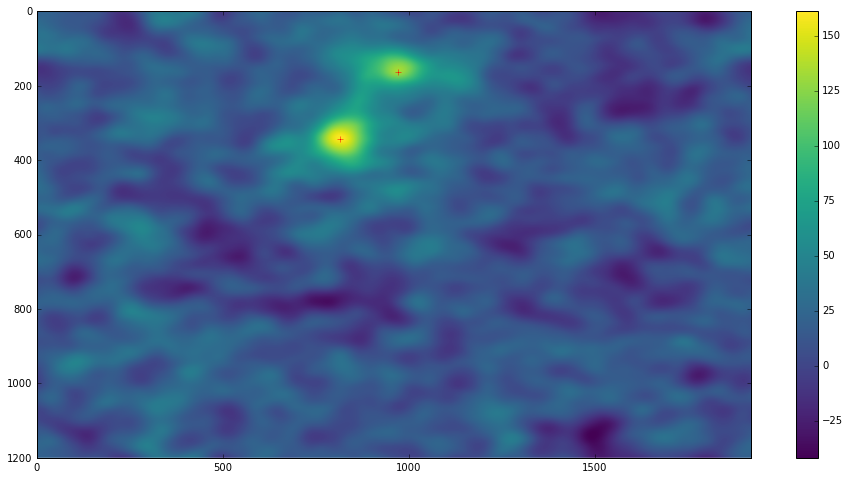}
            \put(-20,10){\fcolorbox{black}{white}{(c)}}
            \caption{Multi-valued deflectometric decoding for a plano-convex lens. (a) Scheme of ray paths during the inspection of the lens. (b) Camera image of a lens reflecting one stochastic pattern of the correlation sequence. The reflections from the front-side and the back-side superimpose; higher-order reflections are suppressed. (c) The 2D map of decoded correlations for a camera pixel indicated by the yellow cross in (b). The scale is from $-25$ (blue) to $+200$ (yellow). The coordinates in the map correspond to the coding screen positions; higher correlations denote higher probabilities that the signal originates in the respective screen pixel. The two peaks correspond to the two contributing paths~\cite{Pak.2019b}.}
            \label{fig:stochastic}
            \vspace{-0.6cm}
        \end{figure}

        Unique encoding can also be achieved by a sequence of stochastic band-limited gray-scale patterns~\cite{Pak.2019b}; in this case, the signals are separated by finding peaks in signal correlations with the coding sequences. Fig.~\ref{fig:stochastic} shows an example.
        
        \subsubsection {Structured backgrounds for specular flow}
        
        A generalization of all previous approaches, in the sense that deflectometry attempts to measure what humans perceive while moving past a specular surface, is the concept of specular flow (see also \ref{sec:sfsf}; here we focus on the encoding requirements relevant for SFSF methods). The properties of SF are quite different from those of diffuse OF~\cite{Lellmann.2008, Adato.2011c, Dovencioglu.2017}; a number of works investigate how the brain converts moving reflections into depth cues, i.e., how humans determine the shape of a specular object. Some good pictorial statements of the problem can be found in~\cite{Fleming.2004, Roth.2006, Doerschner.2011, Muryy.2013}.
        
        When using static backgrounds, there are currently no objective prescribed requirements on sufficient encoding. Quite generally, it is necessary for the background to have enough structure on multiple scales, so that the SF field can be assessed either by a motion of the observer or via a quasi-motion (implemented by several identical imaging systems located in close proximity that observe the same scene)~\cite{Wang.1993, Oren.1997, Savarese.2004b, Vasilyev.2008, Canas.2009b, Pak.2012, Pak.2013, Pak.2016b, Pak.2017}. Reconstructing object curvature and shape accurately then relies chiefly on a sufficiently distant background and/or assumptions/constraints on the smoothness and curvature of a surface. The attractiveness of this approach lies in the access to object shapes (including simple spheres) that can be evaluated only in small patches with backgrounds such as a screen or projector, which cover only a very small solid angle or must be arranged to surround the object on all sides~\cite{Balzer.2014}.
        
        If the restrictions are accepted, the usual coding screens can be used as moving backgrounds~\cite{Tarini.2005, Bonfort.2006b, Lellmann.2008, Nehab.2016}; the assumptions about a distant background are usually replaced by a careful geometrical calibration in this case, and unique position coding is again possible.
       
        \subsubsection {Polarization}

        It is tempting to try and use the strongly polarized light from LC screens to achieve some form of reflection angle coding via polarization effects; but according to some tests we did in 2013, the resulting polarization changes are very difficult to measure accurately enough to be of use in DM. Still, there is a body of work dedicated to the relationship of surface normals and polarization changes, applied mostly to computer vision ~\cite{Rahmann.2001, Rahmann.2003, Morel.2006, Garcia.2015}.

    \subsection {Measuring roughness}
    \label{sec:roughness_msmt}
    
    It is known to every practitioner that the fringe contrast is affected not only by the focus setting, but also by the properties of the surface. We have discussed partial specularity above; here we list a few attempts to use the fringe contrast as an indicator for surface roughness. The research started with moiré deflectometry~\cite{Glatt.1984}, and the trade-off between angular and spatial resolution was noticed very early on~\cite{Keren.1985, Keren.1988}, leading to the development of an instrument to quantify partial specularity~\cite{Keren.1990}. There has been much subsequent research on quantifying the surface roughness (particularly with speckle methods) that we cannot list here, so we restrict ourselves to a relatively recent comparison between deflectometry and speckle methods~\cite{Abheiden.2014} and an investigation of deflectometry for small-scale waviness characterization~\cite{Ziebarth.2014b}.
    
    One way to increase the amount of information gathered for this complicated problem is to use the wavelength dependence of scattering, and there have been a couple of attempts in this direction~\cite{Shafer.1985, Abheiden.2014}. When using ordinary LC displays for recording fringe patterns in different colors, the data are typically compromised by the relatively low brightness available in the blue channel. Presently, it does not appear that there are any more efforts underway to use deflectometry for surface roughness characterization.

    \subsection {Defect detection and classification}
    \label{sec:defect_detection}
    
\begin{figure}
        \centering
        \includegraphics[width=0.49\columnwidth]{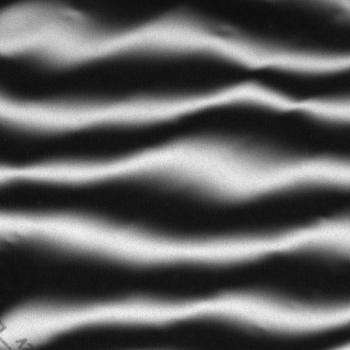}
        \put(-15,5){\colorbox{white}{(a)}}
        \hfill
        \includegraphics[width=0.49\columnwidth]{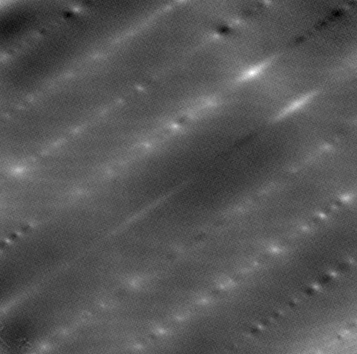}
        \put(-15,5){\colorbox{white}{(b)}}
        \caption{Measurements of a wavy surface (image dimensions approx. 46mm x 46mm). (a) Distorted single fringe image indicates waviness qualitatively; (b) false-color curvature map from phase-shifting DM reveals much more detail; the estimated waviness is in the 50 µm range.}
        \label{fig:fringes_vs_ps}
        \vspace{-0.6cm}
    \end{figure}

    In Section~\ref{sec:qualitative} we have already touched on some data-processing techniques to facilitate easier error detection; here we will compile some general guidelines and observations. Depending on the size and nature of defects to be found, the required encoding can be simple grid patterns ~\cite{Lippincott.1982, Rystrom.1987} (e.g., to detect dents in metal); for a wider range of defect amplitudes and sizes, once again phase-measuring methods or at least sinusoidal patterns are being used for very good sensitivity in absolute, "inverse-fringe" or differential measurements~\cite{Werling.2007c, Chan.2008b, Li.2014f, Macher.2014}. A quick comparison of two techniques is shown in Fig.~\ref{fig:fringes_vs_ps}.
    
\begin{figure}
        \centering
        \includegraphics[width=0.98\columnwidth]{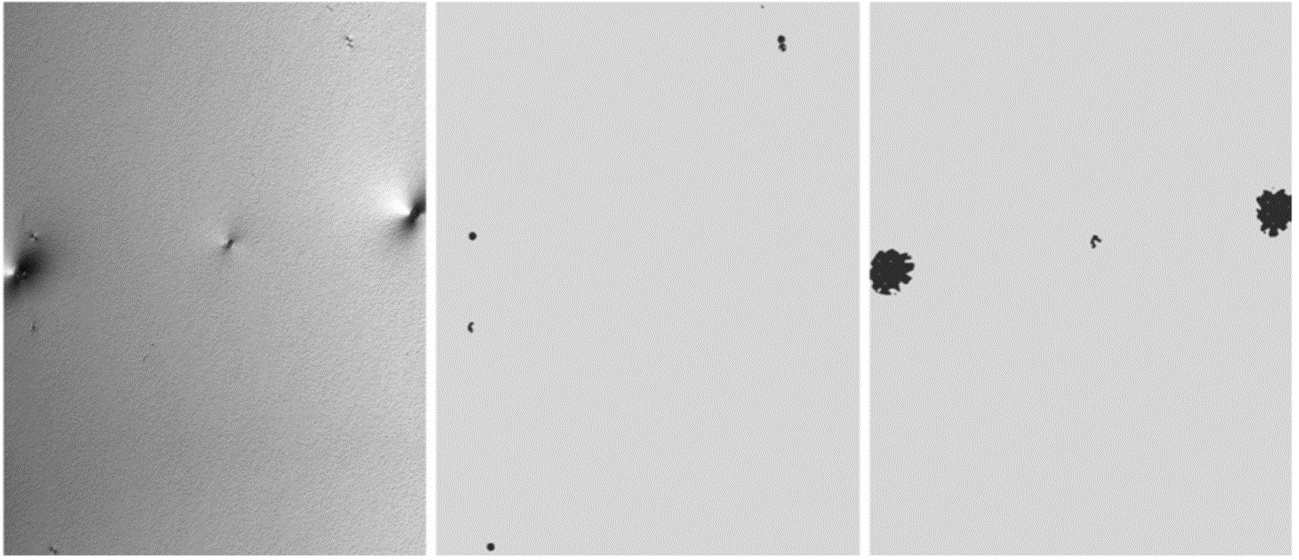}
        \put(-12,3){(c)}
        \put(-95,3){(b)}
        \put(-178,3){(a)}
        \caption{Example for detection and classification of surface defects with detection of dents with multi-scale wavelets; lateral extent of samples is approx. 100mm x 150mm. (a) One component of surface curvature; (b) automatic detection of dust inclusions; (c) automatic detection of dents. Classifier sizes can be scaled and adapted to respective lateral resolution~\cite{Greiner.2016}.}
        \label{fig:wavelet_classification}
        \vspace{-0.6cm}
    \end{figure}

    One difficulty with phase shifting is that defects featuring step-like changes in reflectivity induce artefacts in the measurement data that need to be accounted for to assess defects correctly~\cite{Guhring.2000, Macher.2014c, Burke.2017, Wu.2017, Tao.2019, Patra.2019}. For such defects, it is sometimes sufficient to observe the fringe modulation only~\cite{Petz.2018, Huang.2019}. In general, more complex surfaces tend to feature design elements that are hard to delineate against defects, "cross-talk" between measurands can occur, and easy recipes are elusive. Robust error detection has predominantly been demonstrated on simple surfaces under controlled conditions, and is an unsolved problem for the rest of the parameter space. One notable attempt besides classical image processing with steerable filters etc. explores the use of wavelets for size-sensitive error detection~\cite{Rosenboom.2011, Le.2013, Hahn.2013, Le.2016, Greiner.2016}. Fig.~\ref{fig:wavelet_classification} shows a result from one of these efforts.
    
    Another attempt to cope with the multitude of intentional structures and features of industrial mass products is to train and apply neural networks for error detection on DM data~\cite{Kuosmanen.2017, MaestroWatson.2018, MaestroWatson.2020}; first results look promising, but as mentioned above, no evidence exists yet as to the transferability of the results.
   
\section{Concluding Assessment and Prospects}
\label{sec:conclusion}

As we have seen, the variety of specular objects that one can test with DM, along with the wide qualitative and quantitative range of specifications, facilitates an astounding breadth of research activities and system designs.
Phase-measuring DM is an extremely versatile technique that offers very good sensitivity at moderate effort. Depending on the application, one may or may not need to perform a full geometrical calibration, registration and reconstruction of the rays and surface normals. We see three main research trends for the future.

For metrological tasks, one needs to bear in mind and use the strengths of deflectometry, which are the ease of application, high dynamic range, and excellent sensitivity in the medium to high spatial frequencies. Measurements of global shape and large and smooth surface features (corresponding to e.g. lower-order polynomial terms) should be conducted and interpreted with caution; so far no procedure has been demonstrated or accepted in practice that can seriously compete with interferometry~\cite{Hausler.2013}. The theoretical side of the problem appears solid, but in practice, the noise and calibration uncertainties are still impeding the path to even more accurate surface reconstructions. Still, DM has been successfully used for all but the most demanding metrological specifications. While further progress may still be possible, it remains to be shown that the associated practical efforts are manageable.

In defect detection, the high sensitivity of DM to small-scale features is used to its fullest, providing an excellent measurement technique to match and surpass the human eye. The challenge lies in finding defects in an inspection lasting a few seconds or minutes, where the consumer might take hours to go over each detail. Aesthetic surfaces often have design elements of high curvature and/or contrast, and distinguishing these from unwanted features or contamination is particularly challenging due to the very high sensitivity of DM to small and even unresolved features. The practical demands of DM data processing and the large number of parameters to optimize imply that it may be worthwhile to harness modern machine learning approaches to further improve the accuracy of defect detection.

Although the basics of image formation and processing are well understood, there is no measurement workflow that is optimal for more than one type of work piece. Today the design process is mostly based on customized simulations by experts. Especially for the larger parameter space of multiple camera/screen positions, much work remains to be done in devising and hopefully automating the shortest path to a quick and efficient measurement workflow.

To summarize, in this paper we have made an effort to describe today's deflectometric universe and to point out the "white spots" and possibilities on its map that we have noticed. Hopefully this will ignite discussions and stimulate further research.

\section*{Acknowledgments}

We thank all the inspiring and resourceful people in the deflectometric community for pushing the boundaries and sharing the fascination and passion with us; we have made our best attempt to do justice to all known efforts and achievements in this overview. Apologies for not listing our personal friends and acquaintances by name here: the list would be too long -- you know who you are. Also, we thank our industrial customers for constantly providing benchmarks for the practical challenges and difficulties. On the technical level that has made this paper possible, we thank the open-source communities, listing (non-exhaustively) Python, OpenCV, Mitsuba, Overleaf, TikZ, orcid.org, and doi.org; also, very importantly, our department for funding. Finally, thanks to our colleague Christian Kludt for providing the first fresh pair of eyes.

\printbibliography

\end{document}